\documentclass{article}

\usepackage{amsmath,amssymb,float,graphicx,setspace}
\usepackage[letterpaper, margin=2.5cm]{geometry}
\usepackage[colorlinks=true,citecolor=cyan,urlcolor=blue]{hyperref}

\title{The Stability of Transient Relationships}
\date{April, 2022}
\author{Valent\'{i}n Vergara Hidd\thanks{George Mason University. \href{mailto:vvergara@gmu.edu}{vvergara@gmu.edu}} \and
  Eduardo L\'{o}pez\thanks{George Mason University. \href{mailto:elopez22@gmu.edu}{elopez22@gmu.edu}} \and
  Simone Centellegher\thanks{Fondazione Bruno Kessler. \href{mailto:centellegher@fbk.eu}{centellegher@fbk.eu}} \and
  Sam Roberts\thanks{Liverpool Johns Moores University. \href{mailto:S.G.Roberts1@ljmu.ac.uk}{S.G.Roberts1@ljmu.ac.uk}} \and
  Bruno Lepri\thanks{Fondazione Bruno Kessler. \href{mailto:lepri@fbk.eu}{lepri@fbk.eu}} \and
  Robin Dunbar\thanks{University of Oxford. \href{mailto:robin.dunbar@psy.ox.ac.uk}{robin.dunbar@psy.ox.ac.uk}}
}

\begin{document}
\maketitle

\begin{abstract}
 In contrast to long-term relationships, far less is known about the temporal evolution of transient relationships, although these constitute a substantial fraction of people's communication networks. Previous literature suggests that ratings of relationship emotional intensity decay gradually until the relationship ends. Using mobile phone data from three countries (US, UK, and Italy), we demonstrate that the volume of communication between ego and its transient alters does not display such a systematic decay, instead showing a lack of any dominant trends. This means that the communication volume of egos to groups of similar transient alters is stable. We show that alters with longer lifetimes in ego's network receive more calls, with the lifetime of the relationship being predictable from call volume within the first few weeks of first contact. This is observed across all three countries, which include samples of egos at different life stages. The relation between early call volume and lifetime is consistent with the suggestion that individuals initially engage with a new alter so as to evaluate their potential as a tie in terms of homophily.
\end{abstract}

\section{Introduction}\label{sec:intro}
\begin{spacing}{1.5}
Humans are social animals and having strong and supportive relationships with others has large effects on both physical and mental health  \cite{holt2010, hawkley2010}. These social relationships are not static, but change over time due to two key processes. First, relationships have a natural tendency to weaken over time - to `decay' \cite{burt2000}. Indeed, if no effort is made to maintain relationships, the level of emotional closeness between two individuals will tend to decrease \cite{roberts2015} and the relationship will eventually drop out of the person's social network in terms of the meaningful ties they maintain with others \cite{burt2000}. Long-term studies of people's social networks (the set of relationships they maintain with their family and friends) show a degree of turnover in network members (alters), with some alters leaving the network and others joining \cite{mollenhorst2014, wellman1997}. Second, specific life events such as going away to study at university \cite{roberts2015,oswald2003}, entering a romantic relationship \cite{johnson1982, rozer2015, milardo1983}, having children \cite{munch1997gender, bidart2005}, or getting divorced \cite{milardo1987changes} can have an impact on the composition of social networks due to a decrease in the time available to maintain these relationships, or a change in the focus of attention (e.g. making new friends at university). What these various mechanisms highlight is that, despite the need and utility of stable relationships, over a period of time many relationships will cease to be active, i.e. many relationships are \textit{transient} (e.g. Wellman finds that over a 10-year period, only about $27\%$ of relationships remain active in Canadian adults~\cite{wellman1997}; see also~\cite{bidart20_livin} for a qualitative discussion).

Regardless of causal mechanisms, transient relationships form a considerable fraction of people’s communication (see our results). They are also ubiquitous, judging by the number of studies that report them even when the research is focused on other types of relationships~\cite{burt2000,mollenhorst2014, wellman1997,roberts2015,oswald2003,munch1997gender, bidart2005,milardo1987changes,saramaki2014persistence}. It is easy to appreciate that without them, adapting to the changing social needs of an individual would be impossible as this adaptation involves alters entering and leaving egos' social networks~\cite{saramaki2014persistence}. However, in contrast to long-term relationships, we know little about the amount of support they provide to a person, how many of them become long-term relations, or whether different individuals can handle more or less of them simultaneously. In summary, we do not have a good understanding of transient relationships as part of dynamic ego social networks.

From a theoretical standpoint, the existing literature does not readily offer a clear picture or definition of transient relationships (only transient romantic relationships seemed to have received systematic attention~\cite{buss2019mate}).
On the one hand, the literature on relationship decay has identified a gradual decline in emotional intensity before the end of a relationship~\cite{burt2000,roberts2015}, suggesting that transient relationships may display a gradually decaying volume of communication. On the other, some research suggests that communication is set to an amount appropriate to the perceived quality of relationships, with longer-lasting relationships receiving a greater amount of communication~\cite{MOK2007, wellman1997, dunbar2018anatomy}. In addition, the literature on homophily and friendship implies that an early and relatively fast assessment of relationships needs to take place in setting such a communication amount~\cite{McPherson,dunbar2018anatomy, Asikainen, Kossinets}. Under this picture, the volume of communication gauges the importance of a relationship~\cite{MOK2007,saramaki2014persistence} and the likelihood of the relationship ceasing after some \textit{lifetime}. Thus, whilst research on emotional intensity indicates that relationships are constantly and gradually degrading in the absence of active maintenance, other research on homophily and patterns of communication suggests a rapid evaluation followed by a pattern of steady communication. Further complicating the situation, there is some evidence that emotional intensity and communication volume are monotonically related (see e.g.~\cite{oswald2003,saramaki2014persistence}). This begs the question: \textit{are these pictures consistent or contradictory?}

Here, we study a variety of communication data sets, focusing on ties where measured communication is observed to cease, thus signaling a possible relationship hiatus or end. We call these relationships \textit{transient} because their communication is discontinued for a significant amount of time, perhaps permanently. As we show, transient relationships are not just a vanishing component of communication: in all our data sets, a substantial portion of phone calls is invested in transient relationships. By organizing the transient relationships of each ego into groups of similar lifetimes (actively communicating with ego for similar lengths of time), we find that egos display no dominant trends in their communication to such groups of alters; as a group, communication remains steady. This effect is present regardless of lifetime group. Such lack of trend in communication is in marked contrast with the steady decay that the literature reports for the temporal evolution of subjective measures of relationship intensity such as emotional closeness~\cite{roberts2015}. However, this effect requires lifetimes to exceed a minimum threshold that we characterize and measure. We also find that the call volume an ego invests in a transient tie during the initial weeks of relationship is an informative quantity in estimating tie lifetime. Our results are remarkably robust across cohorts in different countries, of various age ranges, and under different life circumstances. Beyond providing empirical understanding about an important and overlooked class of social relationships, our study suggests that a full understanding of transient relationships requires collecting both objective (e.g. contact events) and subjective (e.g. emotional score) measures of relationship intensity.

Previous research on ego communication patterns  has focused on a variety of related questions to the ones asked here such as overall properties of persistence and turnover in communication \cite{saramaki2014persistence, miritello2013time}, phone communication survival with individual alters~\cite{miritello2013limited,Navarro2017}, or link prediction in broader communication contexts~\cite{Almansoori2012,link-prediction}. Whilst this research has provided new insights into both the patterns and dynamics of social relationships, it has not offered specific information on the temporal regularities of communication to individual alters, particularly transient ones.

Although in many areas the study of dynamic networks has gained considerable traction~\cite{HOLME201297,Fu,Rand}, analytical convenience has meant that many studies in the psychology literature on network structure treat relationships as if they are stable over time. Yet, in fact, they are intrinsically dynamic~\cite{wellman1997,saramaki2014persistence}. This dynamic property arises partly as a result of changing friendship opportunities and partly as a result of adjustments that people make over time in the value they place on individual relationships. Constraints on the availability of social time result in networks having a layered structure~\cite{hill03_social_networ_size_human,sutcliffe2012relationships,maccarron16_callin_dunbar_number} between which individual alters are moved by increases or decreases in the time invested in them, including cases where communication virtually ceases leading to the effective removal of the alter from the layers. Understanding the processes involved in these decisions requires a better appreciation of the communication patterns involved.

It is important to note that the steadiness pattern we uncover here is not incompatible with the well-known burstiness of human communication~\cite{Barabasi2005}. Instead, while burstiness indeed plays a role, especially at short time scales when the contrast between activity or inactivity is clearly demarcated, at longer temporal scales such burstiness leads to overall activity levels that can have their own long term patterns such as seasonality and trends. In this study, we are interested in this longer time scale.

Before moving on to the body of the article, we summarize how our findings contrast with the possible hypotheses that the current theory on relationship subjective decay suggests about transient relationships. First, we find no gradual diminishing calling pattern trend. Second, the cessation of relationship communication is not generally presaged by reaching some low level of communication but, instead, is predicted by the volume of communication in the early periods of a relationship. Therefore, our results indicate that the view of transient relationships suggested by the literature on relationship emotional decay is incomplete.

Thus, the key aims of this study are to characterize the temporal communication patterns of transient alters, identify key variables and relations between those, and examine whether these patterns are consistent across different cohorts. We use three different mobile phone call data sets from the US, UK, and Italy, which include people of different ages, life stages, and cultural backgrounds. These data are from the time smartphones were not widely available in the respective countries and therefore do not suffer from the communication channel fragmentation of more recent data, where extensive use of multiple messaging services makes it more difficult to build up a complete picture of an ego's communication pattern to alters~\cite{bano2019whatsapp, phua2017uses}. As a parenthetical note, the remainder is exclusively concerned with transient relationships, but we occasionally simply call them relationships for brevity.
\end{spacing}

\section{Results}\label{sec:results}
\begin{spacing}{1.5}
Consider an ego $i\in\mathcal{E}$, where $\mathcal{E}$ is one of the cohorts we study (a data set or subset thereof). The set of alters of $i$ is denoted $\mathcal{A}_i$. To develop a clear picture of how an ego-alter relationship evolves over time, we focus on two quantities: the first is the \textit{observed lifetime} $\ell_{i,x}$ of the relationship, i.e. the number of days, reduced by $1$, alter $x\in\mathcal{A}_i$ remained in ego $i$'s network from their first until their last observed phone call. The second is the \textit{observed elapsed duration} $a_{i,x}$ of the relationship at the time of a phone call, i.e. the number of days between the first and a subsequent call between $i$ and $x$, where the first call is defined to occur at $a_{ix}=0$. By definition, $0\leq a_{i,x}\leq\ell_{i,x}$. To refer generically to the elapsed duration and observed lifetime of relationships without specifying the ego-alter pair, we simply use $a$ and $\ell$ without subindices. For ease of reference, the symbols with their definitions and terms used in this paper are summarized in Table~\ref{tab:symbols}.

Since we are interested in studying relationships in which contact stops for a sufficiently long time that one can assume that the communication has either ceased or become dormant, in all our cohorts we eliminate from consideration ego-alter pairs that have contact with each other within a time window of $\Delta t_{w}$ days before the last day $T_{\mathcal{E}}$ of data for cohort $\mathcal{E}$; the larger $\Delta t_{w}$, the more stringent our filter is in terms of which relationships we select as having ceased. Note that many relationships that cease communication may be dormant for a considerably longer time than $\Delta t_w$, as they may stop communication well before the end of a data set approaches; $\Delta t_w$ is therefore a lower cutoff of the duration of time without contact (over all our datasets, on average transient relationships cease communication $238$ days before the end of their studies). Our method follows a similar logic to~\cite{miritello2013limited,Navarro2017}, and although a small percentage of relationships could become active again as indicated in these references ($3\%$ after 6 months), the level of error this induces is very small; note that, since longitudinal data is always limited, other criteria to determine tie end is very difficult, or even impossible, to apply. Finally, for each cohort $\mathcal{E}$, we limit the relationship lifetimes we study to a maximum value $\mathcal{L}_{\mathcal{E}}$ to avoid issues of poor sampling (see details in Supplementary Information, Sec.~S1). These filters lead to three cohorts for the UK, Italy, and the US; the Italian cohort is filtered one more time for additional analysis (so called IT${}_n$ subcohort, see Sec.~\hyperref[sec:data]{Data} below, as well as Supplementary Information, Sec.~S1.4.2).

To provide a sense for the magnitude of communication volume to transient alters, we note that for $\Delta t_w=60$ days, each cohort exhibits large proportions of activity dedicated to transient relations. For ties that involved more than just casual exchanges (defined here as at least $3$ calls): i) in the UK cohort they take up $\approx 45\%$ of overall communication, ii) in the US cohort they receive $\approx 27\%$ of overall communication, and iii) in the Italy cohort they take up $\approx 17\%$ of overall communication.

\begin{figure}[h!]
\centering
\includegraphics[width=\linewidth]{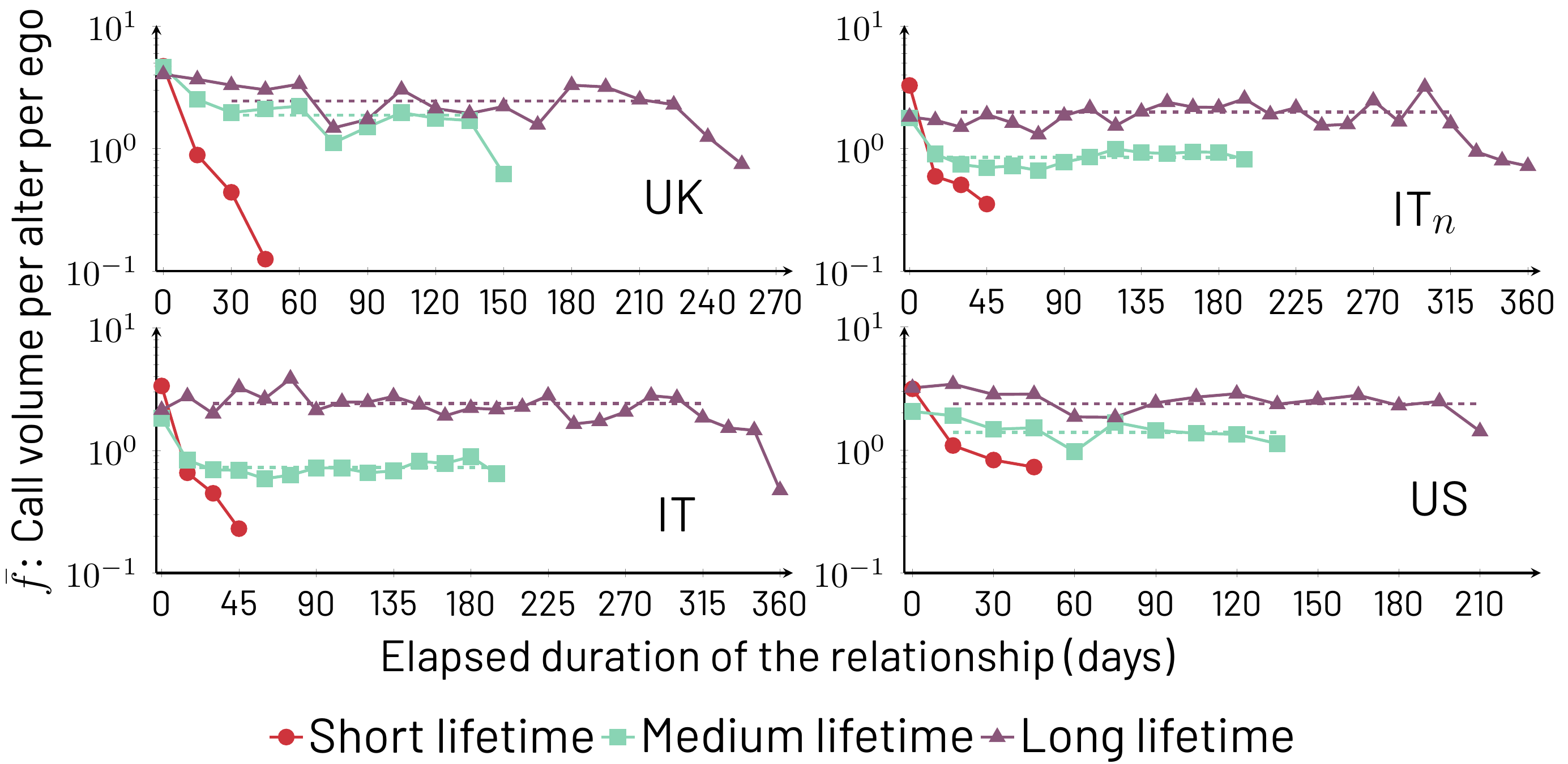}
\caption{Average per alter per ego phone call volume $\bar{f}(a,\ell)$ as a function of elapsed relationship duration $a$, binned with $\Delta a=15$  and $\Delta\ell=50$ for the four cohorts. The lifetime groups correspond to $\ell=0$ (short), $\ell=\lfloor(\mathcal{L}_{\mathcal{E}}-\Delta \ell)/2\rfloor$ (medium), and $\ell=\mathcal{L}_{\mathcal{E}}-\Delta\ell$ (long). To calculate the exact $\ell$ per country, as stated in Fig.~S1 of the Supplementary Information, we use $\mathcal{L}_{{\rm UK}} = 270$; $\mathcal{L}_{{\rm IT}_{n}} = 365$; $\mathcal{L}_{{\rm IT}} = 365$; and $\mathcal{L}_{{\rm US}} = 220$.  The transient condition is $\Delta t_w=60$ days, and for cohort IT${}_n$, the gap between the entry of an ego and the acceptance of an ego-alter pair is set to $\Delta t_s=50$ days. The number of resulting ego-alter pairs induced by our selection criteria is reported in Table~\ref{tab:numegoalter}. Robustness checks with different values for parameters $\Delta \ell, \Delta a,\Delta t_w$, and $\Delta t_s$ are shown in the Supplementary Information, Sec. S3. The curves are stable for medium and long lifetime groups. For curves displaying stable regions, we show a dashed line that represents $b(\ell)$, the average number of phone calls to alters of a given $\ell$ during the stable regime of communication.}
\label{fig:exb_vertical}
\end{figure}

\subsection{Stable volume of calls during the relationship}\label{sec:stablevol}
In order to study the evolution of attention allocation from ego to its alters, we focus on call volume as a function of the elapsed duration and observed lifetime of relationships. Specifically, we measure for each ego $i$ the quantity $\bar{f}_{i}(a,\ell)$, namely the per alter average number of phone calls to alters whose lifetimes fall within $\ell$ and $\ell+\Delta \ell$ when the elapsed duration of the relationship is between $a$ and $a+\Delta a$ (for definitions of $\Delta a,\Delta \ell$, see Sec.~\hyperref[sec:methods]{Methods}). If communication volume exhibits any general trend over the duration of ego-alter relations, $\bar{f}_i(a,\ell)$ would reflect such trend (in the Supplementary Information, Sec.~3.7, we show that another possible way to measure communication, time spent talking, is highly correlated with the number of calls).

To aid in our study of $\bar{f}_i(a,\ell)$, and because any single ego $i$ has few alters with a given combination $a,\ell$, we also measure $\bar{f}(a,\ell)$, the average of $\bar{f}_i(a,\ell)$ over egos with $a,\ell$ (using the same $\Delta a$ and $\Delta \ell$ as $\bar{f}_i(a,\ell)$). Intuitively, $\bar{f}_{i}$ and $\bar{f}$ capture stable estimates of the communication volume (attention allocation) egos invest per alter.

We first focus on the UK cohort (as described in greater detail in Materials and Methods and Supplementary Information, Sec.~S1) which is extracted from a study of students in their last months of secondary school and their first entire year of university study~\cite{roberts2011}. From this study, we form a cohort comprised of the transient relationships that egos form with alters after they transition to university ($6$ months from the start of the study)~\cite{roberts2015}, and that also satisfies the transient relationship filter explained above.

\begin{figure}[!h]
    \centering
    \includegraphics[width=0.95\textwidth]{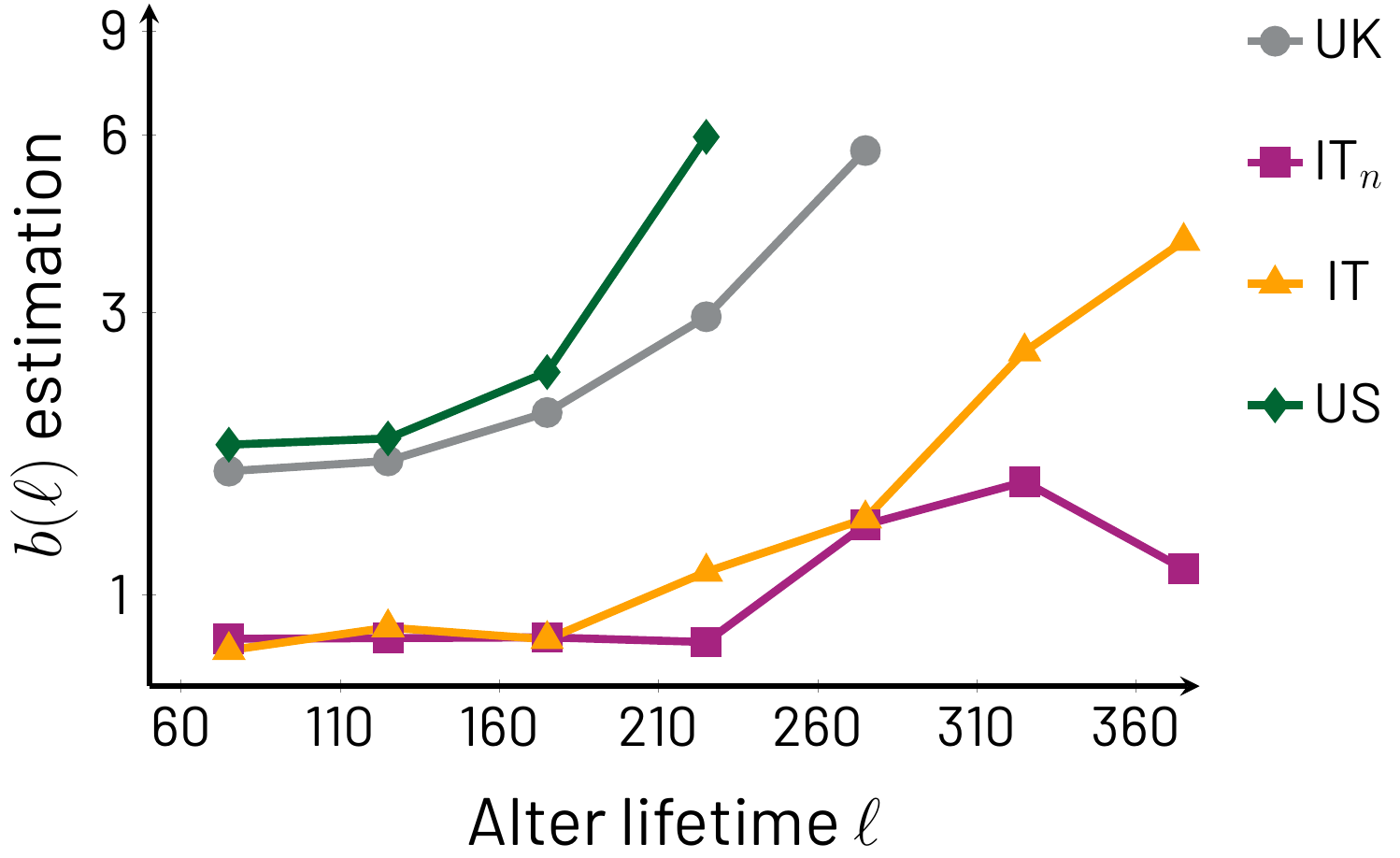}
    \caption{$b(\ell)$ as a function of $\ell$ obtained through the stable region average method. The vertical axis is in logarithmic scale. Clearly, $b(\ell)$ has an increasing trend with respect to $\ell$, with minor exceptions. The UK and US cohorts display a faster increase than IT and IT${}_n$. This could be a consequence of specific differences between details of the cohort participants, such as country, age, and/or personal circumstances of the participants; for example, since the Italian cohort is focused on adult parents with pre-teenage children, these participants may have less available time to invest in phone communication.}
    \label{fig:bestimation}
  \end{figure}

The new alters that emerge after $6$ months of the start of the study are almost certainly new social relationships for the egos, as prior research has shown that almost no relationships survive after $6$ months without communication~\cite{dunbar2018anatomy,miritello2013limited}. In this cohort, $a$ and $\ell$, respectively, approximate very well the \textit{actual} duration and lifetime of transient relationships.
In Fig.~\ref{fig:exb_vertical} (UK), we present $\bar{f}(a,\ell)$ for three groups of transient relationships based on their lifetimes: short starting with $\ell=0$, medium starting with $\ell=\lfloor(\mathcal{L}_{\mathcal{E}}-\Delta \ell)/2\rfloor$, and long starting with $\ell=\mathcal{L}_{\mathcal{E}}-\Delta \ell$; in all cases, $\Delta\ell=50$, and $\lfloor\rfloor$ represents the floor function. We standardize these ranges for this and subsequent analysis of $\bar{f}(a,\ell)$ and $\bar{f}_i(a,\ell)$ to avoid idiosyncratic choices, but see our comments about lifetime ranges in the discussion of Fig.~\ref{fig:bestimation}. First, we note that alters with longer lifetimes receive a greater volume of calls (i.e. $\bar{f}(a,\ell_1)>\bar{f}(a,\ell_2)$ if $\ell_1>\ell_2$). Second, lifetime groups exhibit an initial period of slightly elevated activity up to an elapsed duration we label $a_s$ and, after this period, medium and long lifetime groups exhibit $\bar{f}(a,\ell)$ that stabilize with respect to $a$, remaining close to constant for a long range of values of $a$, or
\begin{equation}\label{eq:bell}
    \bar{f}(a,\ell)\approx b(\ell)\quad[a_{s} \lesssim a \lesssim \ell; \ell \gtrsim \ell_s],
\end{equation}
where $\ell_s$ is the value of lifetime when the steady behavior sets in (see below). In other words, for $\ell>\ell_s$, $\bar{f}(a,\ell)$ approaches an $a$-independent value $b(\ell)$ from about $a_s$ (which corresponds to a value of 3 days, as described in the Supplementary Information Fig.~S8) to just before the observed lifetime ($a \gtrsim \ell$). Both $b(\ell)$ and $\ell_s$ are determined by finding the range of $a$ where, respectively, $\bar{f}(a,\ell)$ and $\bar{f}_i(a,\ell)$ become steady. Note that $\ell_s$ marks the upper bound for another type of transient relationship with $\ell<\ell_s$, one that is too short and ephemeral to achieve any stability; in Fig.~\ref{fig:exb_vertical}, all short lifetimes correspond to this type. We estimate $\ell_s$ as explained in the Supplementary Information, Sec.~S5.4, and find that, depending on the estimation technique, the average value for all the cohorts studied here ranges from $\approx 56$ to $62$ (values for individual cohorts are similar, and are reported in Supplementary Information, Table~S2). In this study, we do not pursue this line of inquiry further.

The UK cohort, while highly informative because of being constituted almost purely of new relationships (transient or long-lasting), is limited by its size ($30$ egos) and represents only one example of the behavior shown. To strengthen our results, we introduce the US and Italy data sets, which have a larger number of egos (for details see Sec.~\hyperref[sec:data]{Data} and Supplementary Information,  Sec~S1)~\cite{aharony2011social,centellegher2016mobile}. The Italian data set in particular has both a large sample and a longer duration, allowing us to construct two different analyses to support our findings. Further, whilst the UK cohort was specifically recruited to capture a period of transition in the egos' social networks~\cite{saramaki2014persistence}, the Italian and US data were collected for egos under steadier social circumstances which could, in principle, lead to different characteristics of transient relationships. Therefore, studying transient relationships across these three cohorts provides a test of the robustness of the findings of stability of communication in transient relationships, using egos at different life stages.

Whilst in the UK, $a$ and $\ell$ accurately reflect actual elapsed duration and lifetime of transient relationships, respectively, for the Italian and US studies these measures become approximate as the precise start of a relationship cannot be guaranteed to occur after the study was initiated. In order to provide a second test that transient relationships in other contexts have the behavior observed in the UK, we create a subcohort IT${}_n$ out of the Italian data in which, for an ego, we restrict the ego-alter pairs to those that satisfy both the transient criterion ($\Delta t_w$) \textit{and} begin at least $\Delta t_s$ days after the entry of the ego into the study. Beyond providing a cross-check for the UK results, in this subcohort $a$ and $\ell$ accurately reflect actual elapsed duration and lifetime. Fig.~\ref{fig:exb_vertical} shows the equivalent analysis of the UK subcohort, now for IT${}_n$, with remarkably consistent results.

As we show next, the robustness of the behavior of transient relations is such that even a more approximate measurement of $a$ and $\ell$ continues to be informative. In the two bottom panels of Fig~\ref{fig:exb_vertical}, we present $\bar{f}(a,\ell)$ for both the Italian and US cohorts still restricted to transient relationships but without restricting the timing of the entry of ego-alter pairs.
The communication patterns in these cohorts are once again consistent with those of the UK and IT${}_n$. This should not be surprising because, given that one is selecting for transient relationships, the properties they possess lead to the same qualitative patterns (steady $\bar{f}(a,\ell)$ with a growing tendency as a function of $\ell$). The nature of the approximation in using these cohorts is reflected in the measurement of $\ell$, particularly if it is to be interpreted as \textit{actual} lifetime of a relationship. If we define $\hat{\ell}$ and $\ell$ as, respectively, the actual and the observed lifetimes, then the Italian and US cohorts can have examples of $\hat{\ell}>\ell$ for particular relationships, whereas for the UK and IT${}_n$, one expects $\hat{\ell}\approx\ell$. In reality, only a fraction of ego-alter pairs in the unrestricted Italian and US cohorts are affected by this, because many relationships indeed start a considerable amount of time after an ego enters a study (average entry day per cohort: $119$ UK, $287$ IT${}_n$, $292$ IT, and $283$ US; complete distributions found in Supplementary Information, Fig.~S2). Below, we take advantage of the robustness with respect to the measurement of $\ell$ to perform the analysis leading to Figs.~\ref{fig:survival} and~\ref{fig:contour} with the UK, Italy, and US only since they provide larger statistical sampling.

\begin{figure}[h!]
\centering
\includegraphics[width=0.95\textwidth]{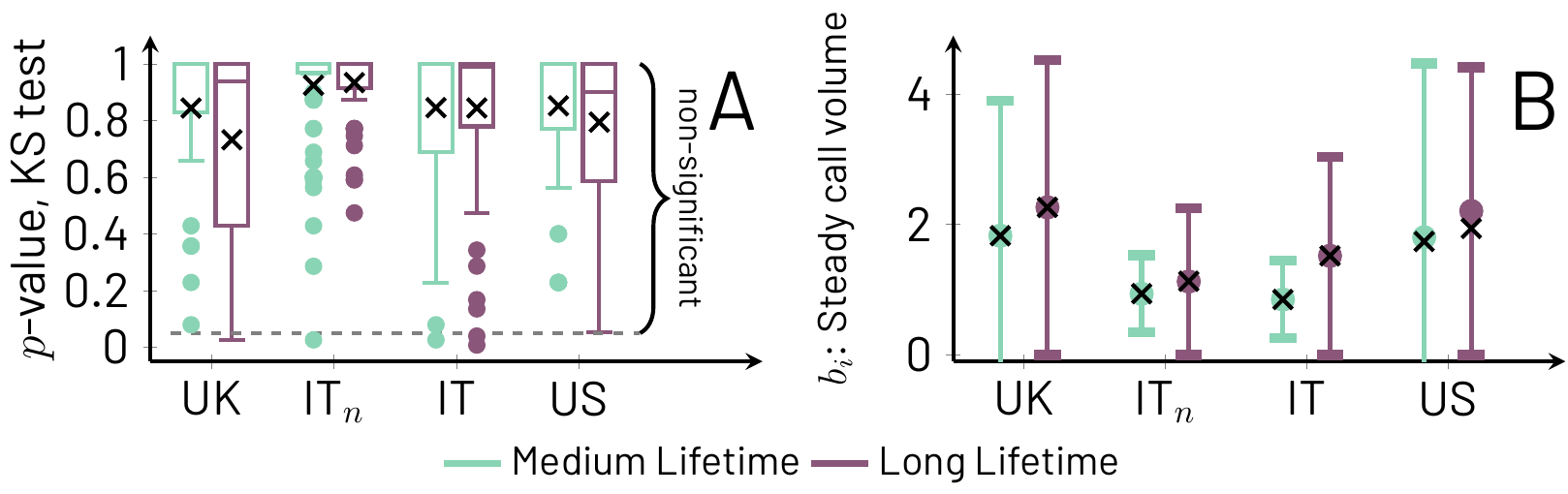}
\caption{Panel A: Box plots for all cohorts using the $1.5$ interquartile range convention for $p$-values from Kolmogorov-Smirnov tests for egos in medium (teal) and long (purple) lifetimes in all cohorts. The per alter call averages $\bar{f}_i(a,\ell)$ are divided into two equally-sized ranges of $a$, the early range $\Delta a\leq a< \lfloor(1/2)\left(\lfloor\ell/\Delta a\rfloor -1\right)\rfloor\Delta a$ and the late range $\lfloor(1/2)\left(\lfloor\ell/\Delta a\rfloor -1\right)\rfloor\Delta a\leq a\leq\lfloor\ell/\Delta a \rfloor\Delta a-\Delta a$). Large $p$-values mean that the early and late ranges of $\bar{f}_i(a,\ell)$ are not distinguishable, and thus, show no trend with $a$; small $p$-values mean there is a trend in $a$. We draw a dashed line at the $0.05$ significance threshold and the averages over all egos are represented with the symbol $\times$. As it is clear from the plot, the vast majority of egos show no trend with $a$. Panel B: Average values of $b_i(\ell)$ (circles) and standard error of the means (whiskers) for medium (teal) and long (purple) lifetimes for all cohorts. Superimposed to each circle and associated whisker is a symbol $\times$ that represents the value of $b(\ell)$ for the corresponding cohort, which matches well the averages of $b_i(\ell)$ across cohorts and lifetimes.}
\label{fig:boxplots}
\end{figure}

As noted above, $b(\ell)$ is observed to increase as a function of $\ell$. To provide further evidence for this observation, we present Fig.~\ref{fig:bestimation} which systematically displays this relation. The fact that $b(\ell)$ increases with $\ell$ highlights that our selection of the medium and long lifetimes used in Figs.~\ref{fig:exb_vertical} and~\ref{fig:boxplots} (below) does not affect the conclusions we draw about the behavior of $\bar{f}(a,\ell)$; in other words, one can work with values of $\ell$ from $\ell_s$ and up. From Fig.~\ref{fig:bestimation} we also note that, while the trends of $b(\ell)$ are increasing, there are differences among the cohorts, with the US and UK showing a more rapid growth than the Italian cohorts, which start roughly steady and then begin their marked increase at larger values of $\ell$ (the IT$_n$ cohort shows one decaying point for the largest $\ell$, due to sampling issues, as discussed in the Supplementary Information, Sec.~S3.2). This may have implications in terms of how effectively one can distinguish medium lifetimes in Italian ego-alter pairs in comparison to the other cohorts on the basis of early phone call activity.

While $\bar{f}(a,\ell)$ allows us to describe the temporal patterns of communication more easily, this is an average quantity over egos and therefore may not be representative of $\bar{f}_i(a,\ell)$. However, it is the latter quantity that genuinely interests us because it captures a more accurate picture of how each ego generally behaves with its alters, i.e., what are the trends in communication over time. To examine $\bar{f}_i(a,\ell)$, we carry out two analyses. The first one consists of determining the level of steadiness of $\bar{f}_i(a,\ell)$ as a function of $a$. This is done ego by ego, taking for each time series $\bar{f}_i(a,\ell)$ two parts of equal duration in $a$ around the mid-point of the series and excluding the first ($a=0$) and last ($a=\lfloor\ell/\Delta a \rfloor\Delta a$) points  (details found in~\hyperref[sec:methods]{Methods}). The two ranges of elapsed duration generate for each ego two samples of $\bar{f}_i(a,\ell)$ at points in $a$ within each of the periods, and we perform a Kolmogorov-Smirnov test to determine if the values of the two samples come from the same distribution. Fig.~\ref{fig:boxplots}A captures the results of the test. Overwhelmingly, the test shows non-significant differences between the values of $\bar{f}_i(a,\ell)$ before and after the mid-point, ego by ego. Moreover, the average $p$-values of the tests over each and every $\bar{f}_i(a,\ell)$ are actually quite high (see symbols in Fig.~\ref{fig:boxplots}A, values range between $0.73$ and $0.94$, and are reported per cohort and lifetime in Supplementary Information, Fig.~S15), not merely rejecting the possibility of change, but confirming a high probability that communication volumes remain largely unchanged between time periods. In other words, $\bar{f}_i(a,\ell)$ remains steady between the first and second periods of the lifetime. The second analysis pertains to the robustness of $b(\ell)$ as a good approximation for $\bar{f}_i(a,\ell)$ or, more precisely, that each individual $\bar{f}_i(a,\ell)$ does not deviate much from $b(\ell)$. We test this by calculating $b_i(\ell)$ for each ego and form its distribution over $i$ (see Fig.~\ref{fig:boxplots}B). The results show that indeed the values of $b_i(\ell)$ are typically close to those of $b(\ell)$ and, therefore, can be treated as approximately equal, i.e. $b_i(\ell)\approx b(\ell)$.

The results illustrated by Figs.~\ref{fig:exb_vertical},~\ref{fig:bestimation}, and~\ref{fig:boxplots} together support the following interpretations. First, the pattern of communication that each ego maintains with its transient alters does not exhibit systematically increasing or decaying trends, that is, no trend is dominant (unless $\ell<\ell_s$, in which case there do not seem to be stable relationships). This steadiness due to the absence of trends is strongly supported by the lack of statistically significant results, and indeed large $p$-values approaching $1$, from the Kolmogorov-Smirnov test comparing the first and second time periods of each ego's call volumes $\bar{f}_i(a,\ell)$. The steadiness is a surprising result that indicates that communication related to transient relationships does not tend to gradually fade away in parallel with measures of emotional closeness~\cite{roberts2015}; when communication ceases, it appears to do so without warning. Second, the similarity between $b(\ell)$ and the set of $b_i(\ell)$ (that is, $b_i(\ell)\approx b(\ell)$) shows that the $b_i(\ell)$ follow a growing trend with $\ell$. This trend, displayed in Fig.~\ref{fig:bestimation} for $b(\ell)$, also means that the definitions of medium and long lifetimes used in Figs.~\ref{fig:exb_vertical} and~\ref{fig:boxplots} can be changed without affecting our conclusions. Third, the fact that the behavior of various cohorts is in agreement means that the variables $a$ and $\ell$ capture useful measures of transient relationship duration and lifetime even if the start of a relationship has not always been observed in a study. Fourth, in Fig.~\ref{fig:exb_vertical} a number of curves begin with an elevated volume of communication and rapidly settle to their steady long-lasting behavior.

\subsection*{Survival of alters}\label{sec:survival}
The increase of the $b_i(\ell)$ with respect to $\ell$ suggests that it may be possible to estimate $\ell$ for transient relationships on the basis of the communication volume they maintain. Note that while $b_i(\ell)$ is not specific to a given relationship of ego $i$, it is nevertheless formed by the aggregation of ego $i$'s communication with alters of lifetime $\ell$ and therefore each individual relationship's communication volume is likely to be of a similar scale as $b_i(\ell)$.

Let us define $g_{i,x}(a_o,a_f)$ as the number of phone calls ego $i$ places to alter $x$ when their relationship is between observed elapsed durations $a_o$ and $a_f$ ($g_{i,x}$ is an $\ell$-unrestricted version of $f_{i,x}$). The increase of $b_i(\ell)$ with $\ell$ suggests that, for a randomly chosen $x\in\mathcal{A}_i$, $g_{i,x}$ is likely to increase with $\ell$. To confirm this, we define the probability $P(a \mid a_o,a_f,g)$ over a set of egos (cohorts or combinations thereof) and their transient alters with lifetimes $\geq a_o$ that one of those alters, randomly chosen, with call volume $g$ within the window $a_o\leq a\leq a_f$ is still active for elapsed durations $a> a_o$ (note that $a$ can be smaller or larger than $a_f$). The intuition of this quantity is that if we take, for example, the number of calls $g(a_o,a_f)$ placed by an ego to one of its alters in a given period of the relationship (when $a$ is between $a_o$ and $a_f$), the probability that the relationship will still be active for $a>a_o$ would grow with the number of calls $g(a_o,a_f)$ received by the alter; in other words, the more calls received, the longer the lifetime. The period comprised by $a_o\leq a\leq a_f$ can be chosen with some level of flexibility, but if it corresponds to an early period in the observation of the relationship (for example, the second complete month of activity), it may provide an early forecast for the lifetime of the relation. Due to the discreteness of the $g$ and the finite sample size, we slightly modify the probability we study to include a range of values of $g$, and represent the quantity by $P(a \mid a_o,a_f,\gamma)$, where $\gamma$ characterizes a range of values of $g$ (specifically, $\gamma$ is defined as the exponent characterizing the bin $3^\gamma\leq g<3^{\gamma+1}$).

\begin{figure}[!h]
\centering
\includegraphics[width=0.7\textwidth]{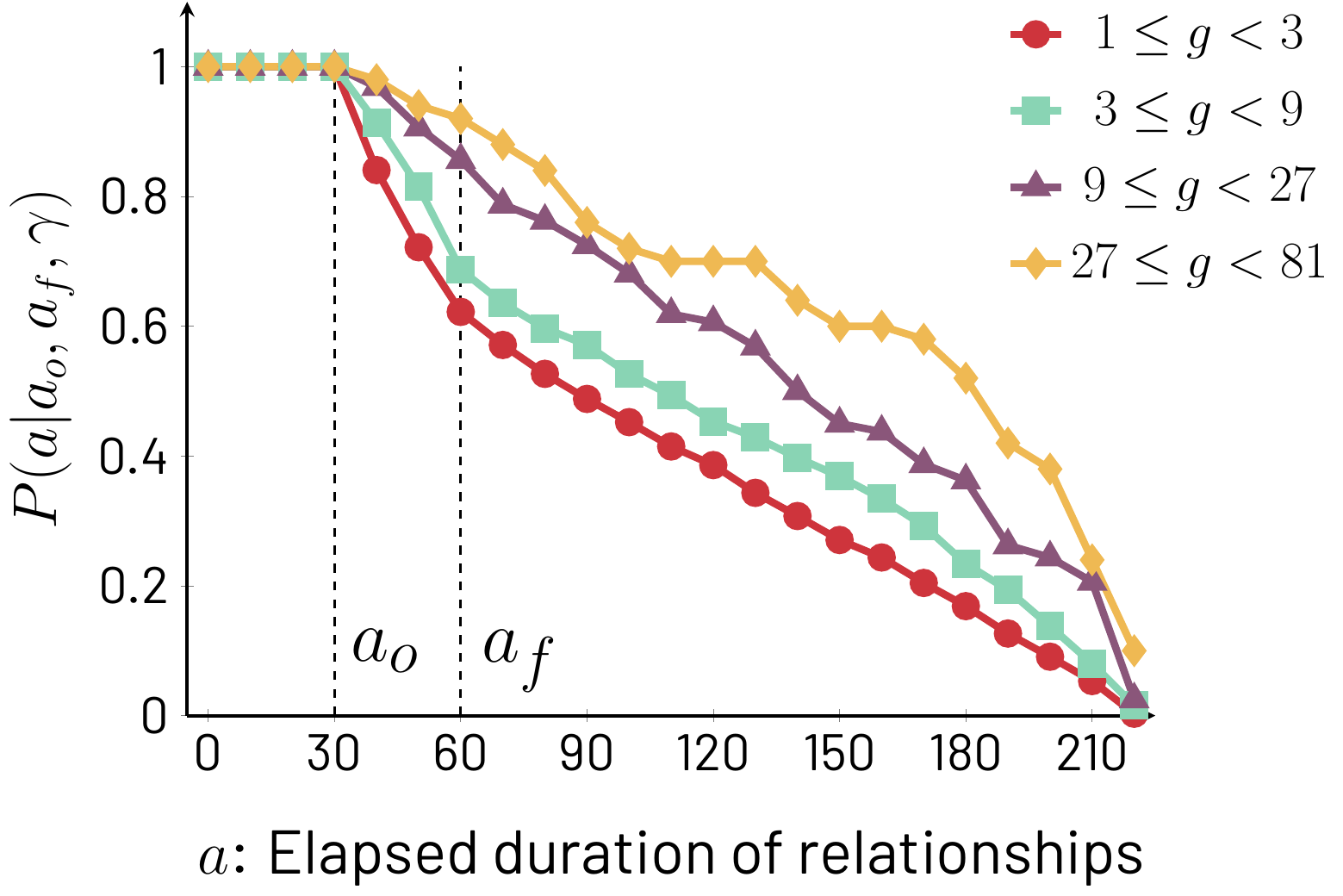}
\caption{$P(a\mid a_o,a_f,\gamma)$ of transient alters to duration of at least $a$ for different bins $\gamma$ of amount of mobile phone calls between $a_o=30$ and $a_f=60$ days. We use the combined data for UK, Italy and US, and therefore, we only look at relationships active for $\ell<\mathcal{L}_{{\rm US}}=220$ days or less, in order to include data for all three cohorts. The bins represented by $\gamma$ as the exponent in $3^{\gamma}\leq g<3^{\gamma+1}$ are $\gamma=0,1,2,3$. As $\gamma$ increases, the probability of survival also increases, i.e. for $\gamma'>\gamma$, $P(a\mid a_o,a_f, \gamma') > P(a\mid a_o,a_f,\gamma)$ which is equivalent to saying that $P(a\mid a_o,a_f,\gamma)$ decays more slowly in terms of $a$ as $\gamma$ increases. See Supplementary Information, Fig.~S19, for various combinations of $a_o,a_f$.}
\label{fig:survival}
\end{figure}

In Fig.~\ref{fig:survival}, we combine the UK, Italy, and US cohorts to show that there is a monotonically increasing relation between $P(a\mid a_o,a_f,\gamma)$ and $\gamma$, i.e.
that the survival probability of a specific alter in an ego's network grows based on the number of calls ego makes to alter between days $a_o$ and $a_f$ (here taken to be $30$ and $60$, respectively) of the observed relationship. The monotonic behavior is robust to different choices of parameters and cohorts (see Supplementary Information, Fig.~S18). Note that we deliberately used an extremely simple test that captures an early period of relationships, even including the challenging choice of $a_o,a_f<\ell_s$ which means that many alters we consider do not reach steadiness. Nevertheless, the measurement clearly shows the monotonicity of $P(a\mid a_o,a_f,\gamma)$ with $\gamma$. Since this survival analysis is meant to illustrate the relation between $\ell$ and $g$, we refrain from developing this point further, as a more precise prediction of the continuation of relationships may require the use of additional variables beyond call volume. A selection of such variables may be informed by several considerations, including other work that has explored the related (but not identical) question of alter persistence~\cite{Navarro2017}.

\subsection*{Relation between early call volume and relationship lifetimes}\label{sec:callslifetimes}
The results displayed in Fig.~\ref{fig:survival} demonstrate that knowing $g_{i,x}(a_o,a_f)$ for relationship $i,x$ provides information about $\ell_{i,x}$. Next, we perform two analyses that further illustrate and quantify this.

First, we present the \textit{symmetric uncertainty} $U(\ell, g)$ between the two random variables $\ell$ and $g$ measured for each ego-alter pair in each cohort as well as for all unique cohorts combined. This quantity ranges from 0, when $\ell$ and $g$ are independent, to 1, when $\ell$ gives complete information on $g$ and \textit{vice versa}. Concretely, $U$ is a monotonically increasing function of how tightly interdependent two variables are to each other and it is therefore a function of the joint distribution of the variables. Symmetric uncertainty is a normalized version of the more well-known concept of mutual information $I(\ell,g)$ (see Sec.~\hyperref[sec:methods]{Methods}). Combining all cohorts, $U(\ell,g)=0.09$, while separate cohorts yield $U(\ell, g \mid {\rm UK}) = 0.3632$, $U(\ell, g \mid {\rm IT}) = 0.1044$, $U(\ell, g \mid {\rm IT}_n) = 0.0998$, and $U(\ell, g \mid {\rm US}) = 0.1597$. Although these values are not near $1$, they are nevertheless quite significant, and to interpret them we must take into account that $g$ is measured very early in relationships, ignoring other variables related to the value of $\ell$~\cite{Navarro2017}.

The interpretation of Fig.~\ref{fig:survival}, along with the consistency of the results over various cohorts, suggests another interesting possibility: by quantifying the behavior of one cohort, one may be able to predict the behavior of another. In our final analysis, we examine whether $P(a\mid a_o,a_f,\gamma)$ calculated from the combined cohort made of the US and UK data sets can predict the behavior of the Italian cohort.

\begin{figure}[h!]
\centering
\includegraphics[width=0.7\textwidth]{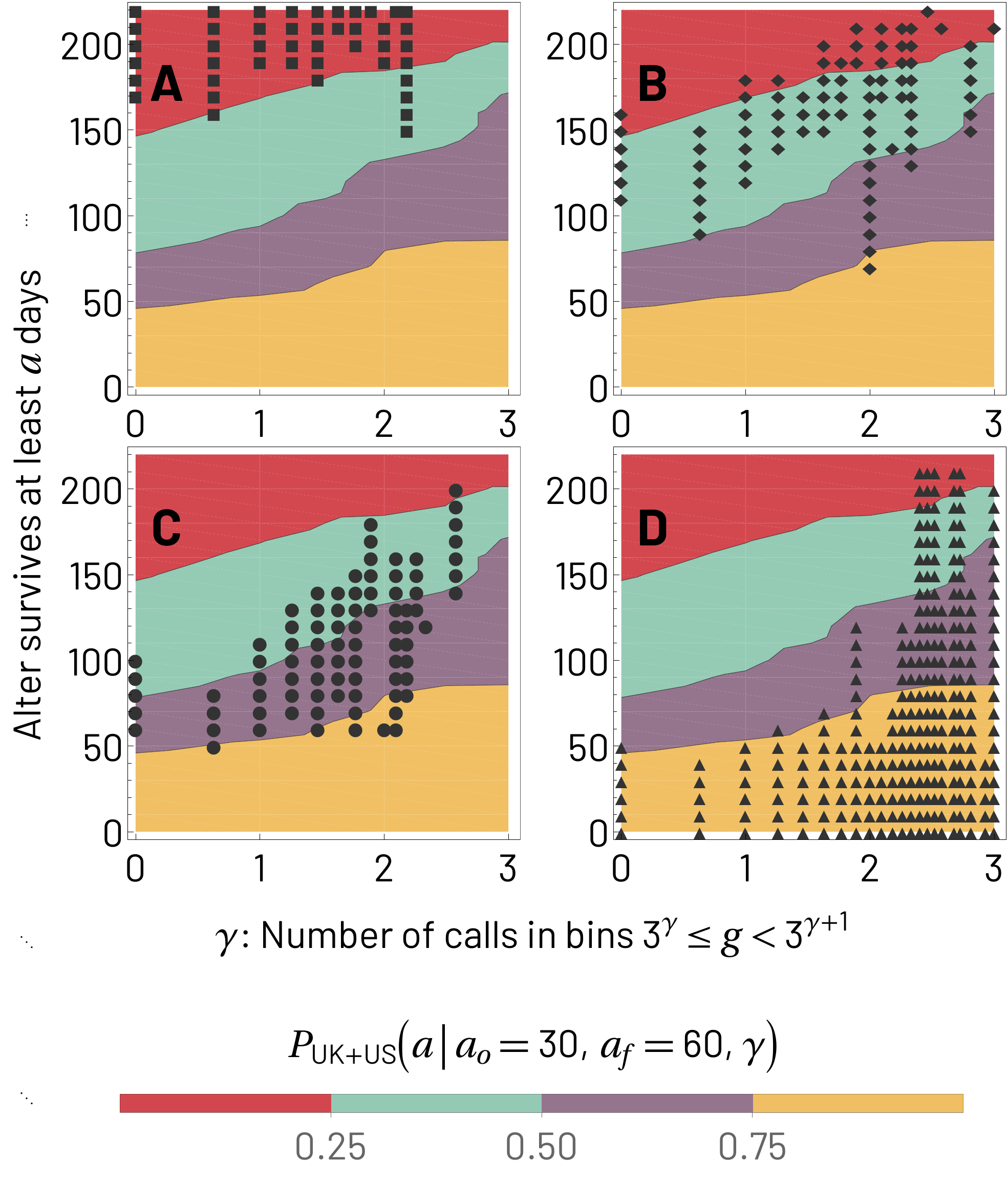}
\caption{Comparison between the survival probabilities $P(a\mid a_o,a_f,\gamma)$ for the combined UK and US data sets (contours) and the Italian data set (symbols). The color background represents ranges of $P_{{\rm UK + US}}(a\mid a_o,a_f,\gamma)$, namely $[0,0.25)$ (red), $[0.25,0.5)$ (teal), $[0.5,0.75)$ (purple), and $[0.75,1]$ (yellow). Panel A shows the symbol $\blacksquare$ for $P_{{\rm IT}}(a\mid a_o,a_f,\gamma)$ in the interval $[0,0.25)$, panel B shows the symbol $\diamond$ for the interval $[0.25,0.5)$, panel C uses the symbol $\bullet$ for the interval $[0.5,0.75)$, and panel D uses the symbol $\blacktriangle$ for the interval $[0.75,1)$. The match in location between the symbols and the colored regions means that the behavior of different cohorts is consistent, supporting the reliability of $g$ as a helpful predictor of $\ell$.}
\label{fig:contour}
\end{figure}

The results of our analysis are shown in Fig.~\ref{fig:contour}, generated as follows: combining the US and UK cohorts, we calculate $P(a\mid a_o,a_f,\gamma)$ for $a$ between $0$ and $\mathcal{L}_{{\rm US}}$ (the smallest value of largest $\ell$ possible among the cohorts in the figure), with the values of the parameters of $a_o,a_f$ and bins $\gamma$ as shown in Fig.~\ref{fig:survival}. Let us call this survival probability $P_{\text{US+UK}}(a\mid a_o,a_f,\gamma)$. The background of Fig.~\ref{fig:contour} is a $2$-dimensional color map version of Fig.~\ref{fig:survival} with $P_{\text{US+UK}}(a\mid a_o,a_f,\gamma)$, where the horizontal axis captures the call volume in the period between $a_o$ and $a_f$ (here, the second month), the vertical axis captures relationship survival up to duration $a$, and the color represents the value of $P_{\text{US+UK}}(a\mid a_o,a_f,\gamma)$ differentiated into four ranges, $[0,0.25)$ (red), $[0.25,0.5)$ (teal), $[0.5,0.75)$ (purple), and $[0.75,1.0]$ (yellow). One way to intuitively understand the construction of the figure is to do a parallel transport out of the page of each of the curves in Fig.~\ref{fig:survival} by an amount proportional to the $\gamma$ associated with call volume between $a_o$ and $a_f$, and then connect the curves along lines of equal probability. These lines of equal probability are the boundaries between colors seen in Fig.~\ref{fig:contour}. To interpret this contour map, note that if we organize the colors in decreasing order of the probability of survival they represent, we obtain the ordered sequence yellow, purple, teal, and red. This order of colors is the same we encounter as we travel the contour map in the direction of increasing $a$, which means that longer lifetimes are less probable. However, note that we can travel along the increasing $a$ direction on a variety of parallel paths each corresponding to a fixed value of $\gamma$. Since the lines that separate the colored regions of the contour map bend upwards as $\gamma$ increases, it means that traveling in the increasing $a$ direction along a line that has a large fixed $\gamma$, the probability of survival decays more slowly with increasing $a$, indicating that lifetime increases with increased calling in the period between $a_o$ and $a_f$.

To understand the connection between the Italian cohort (represented by the symbols in the panels of Fig.~\ref{fig:contour}) and the combined US and UK cohort (represented by the colored background), we test if the survival probabilities for the two cohorts are similar. In symbolic terms, we check if $P_{\text{IT}}(a|a_o,a_f,\gamma)$ is similar to $P_{\text{US+UK}}(a|a_o,a_f,\gamma)$ when the two inputs of these functions, the relationship survival time $a$ and the volume of early communication $\gamma$, are the same. To test this similarity, we divide the values of $P_{\text{IT}}(a|a_o,a_f,\gamma)$ into the same four ranges used for $P_{\text{US+UK}}(a|a_o,a_f,\gamma)$. Concretely, $P_{\text{IT}}(a|a_o,a_f,\gamma)$ can lay in the range $[0,0.25)$ (squares), $[0,25,0.5)$ (diamonds), $[0.5,0.75)$ (circles), or $[0,75,1]$ (triangles). Now, because $a$ and $\gamma$ represent a location in Fig.~\ref{fig:contour}, it means that if the symbols representing a range of $P_{\text{IT}}(a|a_o,a_f,\gamma)$ land in the colored area with corresponding range of $P_{\text{US+UK}}(a|a_o,a_f,\gamma)$, then it means that indeed the probability of survival of alters given a certain amount of early communication volume are similar across the cohorts. For example, if the square symbols land in the red region, it means that the survival probabilities in the range $[0,0.25)$ for both IT and the combined US and UK cohorts occur for the same survival times and amounts of early activity. Going through the four panels, each corresponding to a different range of values of survival probability, the match in location of $P_{\text{IT}}(a\mid a_o,a_f,\gamma)$ and $P_{\text{US+UK}}(a\mid a_o,a_f,\gamma)$ is clearly visible. There is a small discrepancy between the Italian and combined US and UK cohorts for the long lifetimes at the largest values of $\gamma$, but this effect can be explained from Fig.~\ref{fig:bestimation} where we clearly see that given a specific value of volume of communication, lifetimes are longer in Italy than in the US and UK. Notwithstanding this minor discrepancy, the figure shows that indeed the increase in survival time probability of transient relationships for increasing $\gamma$ is a robust phenomenon across countries. In the Supplementary Information, Sec.~S7, we construct alternative combinations of countries and find similar consistency.

\end{spacing}

\section{Discussion}\label{sec:discussion}
\begin{spacing}{1.5}
In this study, we use three mobile phone data sets from the UK, US, and Italy to examine the temporal evolution of communication between an ego and those of its alters that show a considerable communication hiatus - transient relationships. Our results show there is a large range of relationship lifetimes for which communication volume displays no dominant trend, with longer lifetimes associated with larger volumes of communication. One interpretation that emerges from the lack of dominant trends is that relationship end cannot be inferred from a decay in calling. A considerable fraction of relationships begin with a period of more frequent contact before settling into their long-lasting pattern, a result particularly well supported by the UK and IT$_n$ cohorts made up of new alters. Finally, these effects are sufficiently robust that, over the various countries, ages, and life circumstances of our three cohorts, call volume at an early period of communication is found to contain a considerable amount of information about relationship lifetimes even across cohorts.

In terms of how transient relationships may fit into the picture of overall communication, we highlight the following aspects. First, in terms of the mechanics of pursuing relationship communication, the lack of systematic trends reported here for transient relationships is in line with our expectations of communication for long-term contacts~\cite{Dunbar1995}; in such relationships (say, with parents, relatives, or significant others) steady communication is needed. It may be that, \textit{even if subjective evaluation of a relationship may be changing (e.g. decaying)}, it is more economical cognitively, or a better way to attain reciprocity, to have a temporal approach to communication that does not directly imitate the subjective evaluation. Second, as can be appreciated from Figs.~\ref{fig:exb_vertical} and \ref{fig:bestimation}, even relationships that only last $5$ or $6$ months have communication volumes that are substantial (roughly between $1$ to $3$ calls every $15$ days), meaning that transient relationships do not typically constitute meaningless links. This does not, of course, mean that all relationships have such limited lifespans; a few of those with a slighlty larger call volume can last a lifetime, i.e. are non-transient. Saram\"{a}ki et al. \cite{saramaki2014persistence} noted that, among 18-20 year-olds, turnover in friendships could be extremely high: only 40\% of the alters retained their relative rank in terms of communication activity over an 18-month sample. More generally, a 10-15\% of alters left or joined a network in any given year. The data for the US and Italian samples suggest that similarly high rates may be observed in older age cohorts in their later 20s and into their 30s. Longer term studies have also shown a high degree of turnover, with only 27\% of close ties remaining after a decade in Canadian adults \cite{wellman1997}. However, as our own results show, these transient alters take up a \textit{substantial portion} of ego's communication. Overall, this clearly illustrates the fact that contrary to what is often supposed relationship turnover is rather high in human relations, and the explanation of this effect is an important outstanding issue.

Another consideration emerging from our study concerns the complementary nature of objective communication information and subjective measures of relationships. Thus, although the use of Call Detail Records avoids some of the shortcomings that have been previously identified in self-reported patterns of communication~\cite{roberts2011communication, wellman1997, wellman2007challenges,hogan2007visualizing}, including limited time resolution and poor recall effects, questionnaire or interview data are the only sources of subjective relationship measures (e.g. emotional closeness) and are therefore critical. In fact, the discrepancy between subjective relationship intensity's decay over time and the absence of systematic decaying communication trends observed here indicates that \textit{both} approaches are necessary to develop a full picture of an ego's mechanisms in navigating social network creation, maintenance, and modification. To further clarify this, future studies should contemplate dimensions such as face-to-face contact, which has been previously associated with further longevity in relations~\cite{roberts2015}, and sampling of the reasons why people effectively cease to communicate with their alters. Indeed, understanding the interplay between objectively and subjectively measured relationship characteristics may be relevant to understand a variety of aspects of human communication, including how transient and long-term relationships are associated with well-being~\cite{hawkley2010}.

It is reasonable to think that our definition of transient relationships will require further qualitative and quantitative studies because, as research into short- and long-term romantic relationships demonstrates~\cite{buss2019mate}, there are likely to be a variety of reasons why transient relationships exist and why they end. In addition, the monotonic relation between volume of communication and lifetime we observe cannot be absolute, i.e. at some point, an increase in $\ell$ cannot lead to a further increase in $b$ because this would mean that for large enough $\ell$, long transient relationships would in fact take up all available communication time. Thus, it would be important to learn at what point $\ell$ does not lead to further increases in $b$ and, indeed, whether or not $b$ stabilizes or maybe even starts to decrease with very stable (yet still possibly transient) relationships.

At a practical level, our results also have implications in designing research protocols, because they suggest that even relatively short time series of mobile data (say between $100$ and $180$ days, but above the $\ell_s$ limit) are sufficient to distinguish among alters who will go on to have different lifetimes in the network over a longer time period. As participant drop-out is a key issue in longitudinal studies~\cite{ibrahim2009missing, mclean2017explaining}, this finding may enable researchers to design studies that optimize the balance between the length of the study and the likelihood of participant drop-out.

Whilst we found robust relationships between early call volume and lifetime in transient relationships in all three countries, there were some limitations to this study that may have impacted our research findings. First, the focus of this study was on understanding the temporal patterns of communication in transient relationships independent of individual characteristics. Thus, factors such as gender~\cite{ghosh2019quantifying,david2016communication}, personality~\cite{centellegher2017personality,staiano2012friends}, or whether a relationship is between friends or romantic partners~\cite{david2016communication,roberts2011}, may all affect these temporal patterns of communication. Therefore, future research could examine how ego and alter characteristics may modify the patterns we have identified. Second, given our initial motivation for testing whether gradual decay in subjective ratings also translated into objective gradual decay in communication volume, we focused our study on patterns of call volume. However, as has been shown in the context of related questions~\cite{Navarro2017}, different characterizations of temporal signals may be informative. In the future, an expanded exploration of different temporal characterizations of communication in transient relationships may provide further valuable information about how such relationships evolve. Third, the lack of data that couples high temporal resolution subjective ratings with call patterns prevents us from understanding subjective ratings at a level of detail equivalent to that of calling data. Until such data are available, our understanding of the mismatch between objective and subjective measurement of transient relationship temporal behavior will remain unclear.

Another question pertains to patterns of communication as these increasingly shift from mobile calls and texts to messaging platforms and social media sites such as Whatsapp~\cite{bano2019whatsapp}, Twitter, Instagram~\cite{phua2017uses, gonccalves2011modeling, Huberman_Romero_Wu_2008}, and WeChat~\cite{montag2018}. The diversity of these platforms makes collecting communication data more complex than relying solely on mobile data, but the development of applications that passively collect accurate data on mobile application use provides new opportunities for research in this area~\cite{ferreira2015aware, torous2016new, ranjan2019radar}. This variety of platforms and channels is not relevant to the present study due to the time frame when our data were collected (before the widespread use of smartphones in the respective countries). However, based on the fact that communication regularities seen in phone calls also appear in channels such as email~\cite{Godoy} and Facebook~\cite{DunbarArnaboldi}, once the various channels of communication are aggregated, the overall signal may show a great deal of similarity with our present findings.

The connection between early call volume and lifetime of transient relationships may suggest support for a description of the effect of homophily in relationships called the ``Seven Pillars of Friendship.'' This description is made up of a set of seven cultural dimensions that define the individual and the cultural community they belong to~\cite{dunbar2018anatomy}. These dimensions include: dialect, place of origin, career trajectory, hobbies/interests, moral/religious views, musical tastes, and sense of humour. Friendship quality has been shown to depend on the number of these friendship dimensions that an ego and a particular alter share~\cite{Curry2013}, reflecting the extent to which friendships are dominated by homophily -- the tendency for 'birds of a feather to flock together'~\cite{Curry2013,dunbar21_frien,McPherson}. It has been suggested that, after first meeting, dyads initially devote time to checking out each others' respective positions on the seven pillars, and then adjust their rate of contact to that appropriate for the quality of relationship defined by the number of pillars they share~\cite{sutcliffe2012relationships,dunbar21_frien}. Evaluating our results against this proposal, a number of areas of consistency emerge. First, note that call volume measured in the early part of a relationship (say the second month) has predictive power about a relationship’s lifetime (Figs.~\ref{fig:survival} and \ref{fig:contour}). If the lifetime of a relationship was merely a consequence of a continuous evaluation in which, at any point, a relationship could be dissolved, call volume at the early part of a relationship would provide no information about lifetime (for example, Figs.~\ref{fig:survival} and~\ref{fig:contour} would not show differences due to early call volumes). A second consideration that may signal consistency between our results and the Seven Pillars of Friendship is the fact that many cohorts of different lifetime $\ell$ do exhibit a fast very early period of elevated volume of communication followed by a rapid decay, as visible from Fig.~\ref{fig:exb_vertical} and Fig.~S8, near $a\approx 0$. Further study of this possible connection is probably warranted.

In summary, communication volume of egos to groups of similar transient alters is stable, with no signs of a dominant gradual decay of such call volume over lifetime. The volume of calls is associated with greater longevity of a transient relationship. These findings are consistent across three countries and for different demographic groups. Similar to a few other studies, we observe that the volume of communication egos invest in transient alters is far from negligible, suggesting that such relationships are essential. In a broader context, our results uncover a new striking regularity in ego networks that reinforces related findings~\cite{Bernard,dunbar1998social,Zhou,saramaki2014persistence} of regularity and steadiness within the dynamics of communication.
\end{spacing}

\section{Methods}\label{sec:methods}
\begin{spacing}{1.5}
\subsection*{Data}\label{sec:data}
All the analyses are based on three mobile phone data sets: (i) the UK data comes from an 18-month ($T_{{\rm UK}} = 546$) study of 30 students in their final year of secondary school, who were followed as they made the transition from school to university~\cite{roberts2011};
(ii) Friends and Family data set collected phone calls of 130 people from a residential community centered around a university in the US~\cite{socialfmri2011} over a period of $\approx 17$ months ($T_{{\rm US}} = 505$); and finally (iii) the Italy data set, containing phone calls collected from 142 parents with young children aged 0 through 10 years \cite{centellegher2016mobile} over a period of around 22 months ($T_{{\rm IT}} = 699$). For each data set $\mathcal{E}=\{{\rm UK, US, IT}\}$, we limit the ego-alter pairs used to those with a maximum duration of communication $\mathcal{L}_{\mathcal{E}}$ based on the point at which, due to study design and duration of each of the national studies, the percentage of egos with active relationships begins to decay significantly (see Supplementary Information, Sec.~S1.2). The UK data set is further filtered (as explained in Sec.~\hyperref[sec:results]{Results}) to ego-alter pairs that appear only after $6$ months of the study, when participants begin university study. We also exclude relationships with less than $3$ calls since such relations are uninformative. Further filtering is applied to determine transient relationships (see Sec.~\hyperref[sec:altselection]{Transient alter selection}). Finally, the IT${}_n$ cohort is constructed by further filtering relationships to those in the Italian data that do not commence until after a minimum number of days since the entry of the participant. These filters define our four cohorts UK, US, IT, and IT${}_n$. All data sets were collected before smartphones became common and thus capture the bulk of people's non-face-to-face communication.

\subsection*{Transient alter selection}\label{sec:altselection} Each communication event (outgoing phone call) between ego $i$ and alter $x$ occurs on a particular day $a_{ix}$ after their first observed communication, where the first day corresponds to $a_{ix}=0$. From the perspective of when each cohort $\mathcal{E}$ begins, the first observed contact between $i$ and $x$ occurs on day $t^{(1)}_{ix}$ which is a number between $0$ and $T_{\mathcal{E}} - 1$. If there are $n_{ix}$ total observed calls between $i$ and $x$, the last call occurs on day $t^{(n_{ix})}_{ix}$ of the study, which corresponds to $\ell_{ix}=t^{(n_{ix})}_{ix}-t^{(1)}_{ix}$. In our study, we exclude any alter $x$ such that $T_{\mathcal{E}} -t^{(n_{ix})}<\Delta t_w$ where $\Delta t_w$ is an excluded window that provides confidence that a relationship has indeed stopped communicating for a significant amount of time.

\subsection*{$\bar{f}_{i}(a, \ell)$ and $\bar{f}(a, \ell)$ definitions}\label{sec:definitions}
The call volume $f_{ix}(a_{ix},\ell_{ix})$ between $i$ and $x$ captures the evolution of relationship $ix$ over time, but it is a considerably noisy signal, generally with few samples for given values of $a=a_{ix}$ and $\ell=\ell_{ix}$. To address the possibility that egos have a systematic trend over time in communicating with their alters, we average over alters of $i$ in $\mathcal{A}_i(\ell,\Delta\ell)\subset\mathcal{A}_i$, the set of alters $x$ such that $\ell\leq\ell_{ix}<\ell+\Delta\ell$.
The bin size in the main text has been chosen as $\Delta\ell=50$ days, but other values are shown in the Supplementary Information, Sec.~S3.3. From these definitions, as well as a window of $a$ such that $a\leq a_{ix}<a+\Delta a$ (with $\Delta a=15$), we introduce $\bar{f}_i(a,\ell)=\sum_{x\in\mathcal{A}_i(\ell,\Delta \ell)}f_{i,x}(a_{i,x},\ell_{i,x})/\vert\mathcal{A}_i(\ell,\Delta\ell)\vert$. We also introduce $\bar{f}(a,\ell)=\sum_{i}\bar{f}_i(a,\ell)/\sum_{i}\theta(\vert\mathcal{A}_i(\ell,\Delta \ell)\vert)$, where $\theta(\cdot)$ corresponds to the step function ($\theta(x)=1$ if $x>1$, and $0$ otherwise), and $\vert\vert$ produces the cardinality of a set. Note that any trend consistently present in $f_{ix}(a_{ix},\ell_{ix})$ would be inherited by both $\bar{f}(a,\ell)$ and $\bar{f}_i(a,\ell)$.

\subsection*{Kolmogorov-Smirnov test for $\bar{f}_{i}(a, \ell)$}
We study the level of steadiness of $\bar{f}_i(a,\ell)$ as a function of $a$ ego by ego, taking for each time series $\bar{f}_i(a,\ell)$ two parts of equal duration in $a$ that exclude the first ($a=0$) and last ($a=\lfloor\ell/\Delta a \rfloor\Delta a$) points of the time series. These two points are excluded for specific reasons. The first point is affected by initial tendencies to have communication that has not stabilized, as can be seen in Fig.~S8. The last point is excluded because, unless $\ell$ is a perfect multiple of $\Delta a$, the call volume captured by the last time point of the series is likely to have less call volume simply because it is not fully used (there is a period between $\ell$ and $\lfloor\ell/\Delta a\rfloor\Delta a$ with no activity). After excluding these two points, the two resulting ranges of elapsed duration ($\Delta a\leq a< \lfloor(1/2)\left(\lfloor\ell/\Delta a\rfloor -1\right)\rfloor\Delta a$ and $\lfloor(1/2)\left(\lfloor\ell/\Delta a\rfloor -1\right)\rfloor\Delta a\leq a\leq\lfloor\ell/\Delta a \rfloor\Delta a-\Delta a$) generate for each ego two samples of $\bar{f}_i(a,\ell)$ at points in $a$ within each of the periods, and we perform a Kolmogorov-Smirnov test to determine if the values of the two samples come from the same distribution. The result of the Kolmogorov-Smirnov test for each ego is a $p$-value that, the closer it is to $1$, the more likely it is that the series $\bar{f}_i(a,\ell)$ is steady. Let us label the $p$-value obtained for each ego as $p_i$. We conduct these tests for egos with medium and long lifetimes. Fig.~\ref{fig:boxplots}A shows box plots of the $\{p_i\}_{i\in\mathcal{E}}$ obtained from the tests for all cohorts $\mathcal{E}$.

\subsection*{$b(\ell)$, $b_i(\ell)$, and $\ell_s$ computation}\label{sec:bandells}
The determination of $b(\ell)$, $b_i(\ell)$, and $\ell_s$ is made by identifying a \textit{stable region average} of $b(\ell)$. In order to obtain this average, we find the longest range of values of $a$, pivoted around the center of the range, where $\bar{f}(a,\ell)$ or $\bar{f}_i(a,\ell)$ is steady (flat) in $a$. In this description, we label both $\bar{f}(a,\ell)$ and $\bar{f}_i(a,\ell)$ as $u(a)$, where $\ell$ is not written to avoid complicating the notation but it is implied in that $a\leq\ell$. The criterion to determine if $u(a)$ is close to flat is based on whether its average slope oscillates around $0$. This flatness is tested iteratively between two values of $a$, $a_m$ and $a_M$, which must be found by the method. The algorithm starts with $a_m=0$ and $a_M=\lfloor\ell/\Delta a\rfloor\Delta a$, and alternatively and iteratively increases $a_m$ while leaving $a_M$ fixed and, in the next step, decreases $a_M$ while leaving $a_m$ fixed, and so on. The changes in both $a_m$ and $a_M$ are done in increments of $\Delta a$. The algorithm stops when the average slope of $u(a)$ starts to oscillate around $0$, or if no stable region is found. The method takes advantage of the fact that typically when $u(a)$ does reach a stable regime, only the regions near $a=0$ and $a=\ell$ substantially deviate from being flat and are each only a few units of $\Delta a$ in the range of $a$. The concrete application of the method is as follows. Let the values of $a$ for which we calculate $u(a)$ be given by $a=\alpha \Delta a$ with $\alpha$ an integer between $0$ and $\lfloor\ell/\Delta a\rfloor$. Using integer $q$, we calculate the slope
\begin{equation}\label{eq:avg-slope}
    {\rm average\;slope}(q)=\frac{u\left(\lfloor\frac{\ell}{\Delta a}\rfloor\Delta a -\left\lfloor \frac{q+1}{2}\right\rfloor\Delta a\right) -u\left(\left\lfloor\frac{q}{2}\right\rfloor\Delta a\right)}{\left(\lfloor\frac{\ell}{\Delta a}\rfloor-q\right)\Delta a }
\end{equation}
for each value of $q$, starting at $0$, and increasing in increments of $1$ until the sign of the average slope (Eq.~\ref{eq:avg-slope}) first alternates twice in consecutive values of $q$, or until $q=\lfloor\ell/\Delta a\rfloor$ if the alternation condition is never met.
Note that $\lfloor q/2\rfloor\Delta a$ and $\left[\lfloor\ell / \Delta a\rfloor-\lfloor (q+1) / 2\rfloor)\right]\Delta a$ correspond to two values of $a$ roughly equidistant to the center (one to the left and one to the right) of the range of $a$ for $u(a)$. These two values are labelled $a_m(q)=\lfloor q/2\rfloor\Delta a$ and $a_M(q)=\left[\lfloor\ell / \Delta a\rfloor-\lfloor (q+1) / 2\rfloor)\right]\Delta a$ as indicated before. The increase in $q$ one unit at a time increases $a_m$ to $a_m+\Delta a$ in one step while leaving $a_M$ unchanged, and in the next step decreases $a_M$ to $a_M-\Delta a$ while leaving $a_m$ unchanged. This process truncates the two ends of the range of values of $u(a)$ over which the average slope is being calculated. If alternation of the sign of Eq.~\ref{eq:avg-slope} occurs for two consecutive increases of $q$, i.e. when $q$ changes from value $q_x$ to $q_x+1$ and from $q_x+1$ to $q_x+2$, we take $q_x$ as the beginning of the approximately $0$-average slope of $u(a)$. If the average slope sign alternation condition is never met, or if the average slope is always identical to $0$, the algorithm stops when the range of $a$ cannot be truncated any further, which is when $q=\lfloor\ell/\Delta a\rfloor$, and in this case, we make $q_x=2$ which means that we revert to looking at all but the two endpoints of $u(a)$ (although the algorithm is deemed to have failed to converge and we use its results differently). Using the resulting $q_x$ (converging or non-converging), we measure $\bar{u}(a)$, the average of $u(a)$, using all $u(\alpha\Delta a)$ with $\left\lfloor q_x / 2\right\rfloor\leq \alpha \leq \lfloor\ell / \Delta a\rfloor -\left\lfloor (q_x+1) / 2\right\rfloor$. When $u(a)$ corresponds to $\bar{f}(a,\ell)$, then $b(\ell)=\bar{u}$; when $u(a)$ corresponds to $\bar{f}_i(a,\ell)$, then $b_i(\ell)=\bar{u}$.

The method described above also yields the minimum lifetime $\ell_s$ at which stable regions begin to emerge. As noted above, $q_x=2$ when the method does not converge, otherwise, the method converges and therefore, at $a=\lfloor q_x/2\rfloor\Delta a$ a flat region of $u(a)$ begins. Noting that this $a$ is equivalent to the shortest possible value of lifetime, we equate $\ell_s$ with $\lfloor q_x/2\rfloor\Delta a$ and take $u(a)$ to be $\bar{f}_i(a,\ell)$. This produces a sample of $\ell_s$, one for each ego, and provides a statistical picture for the smallest lifetimes that exhibit a steady regime.

\subsection*{$P(a\mid a_o,a_f,\gamma)$ computation}\label{sec:pcomputation} The probability $P(a\mid a_o,a_f,\gamma)$ of a relationship continuing to be active to at least elapsed duration $a$, with a number of calls $g$ that falls in bin $\gamma$ during the window $a_o\leq a\leq a_f$, is calculated over a set of transient ego-alter relationships with $0 \leq \ell \leq \mathcal{L}_{{\rm US}}$ (the smallest value of largest $\ell$ possible among the cohorts in the figure). Concretely, if the total number of alters that receive $g$ calls (falling in bin $\gamma$) in the window between $a_o$ and $a_f$ is $N(a_o,a_f,\gamma)$ and only $N(a;a_o,a_f,\gamma)$ out of those are still communicating at some $a>a_o$, then  $P(a\mid a_o,a_f,\gamma)= N(a;a_o,a_f,\gamma)/N(a_o,a_f,\gamma)$. Bins are exponentially spaced, corresponding to the ranges $[3^0,3^1);\dots;[3^4,3^5]$ and $\gamma$ is the exponent of the minimum power of $3$ that identifies each bin.

\subsection*{Mutual information}\label{sec:mutualinfo} The measurement of mutual information between the random variables $\ell$ and $g$ is performed for all the combined cohorts together and also for individual cohorts. Mutual information $I(\mathbf{X},\mathbf{Y})$ between two random variables $\mathbf{X}$ and $\mathbf{Y}$ is defined as the amount of information one of the random variables contains about the other. Specifically, for discrete random variables,
\begin{equation}
    I(\mathbf{X},\mathbf{Y})=\sum_{x\in\mathbf{X},y\in\mathbf{Y}}{\rm Pr}(\mathbf{X}=x,\mathbf{Y}=y)\log_2\left[\frac{{\rm Pr}(\mathbf{X}=x,\mathbf{Y}=y)}{{\rm Pr}(\mathbf{X}=x){\rm Pr}(\mathbf{Y}=y)}\right],
\end{equation}
where ${\rm Pr}(\mathbf{X}=x,\mathbf{Y}=y)$ is the joint probability to draw $x$ and $y$ simultaneously, ${\rm Pr}(\mathbf{X}=x)$ the marginal probability to draw $x$, and ${\rm Pr}(\mathbf{Y}=y)$ the marginal probability to draw $y$.
$I(\mathbf{X}, \mathbf{Y})$ is measured in \textit{bits}, which we can normalize to a \textit{symmetric uncertainty} $U(\mathbf{X}, \mathbf{Y})$,
\begin{equation}
    \label{symmetric}
    U(\mathbf{X}, \mathbf{Y}) = \frac{2I(\mathbf{X}, \mathbf{Y})}{H(\mathbf{X}) + H(\mathbf{Y})},
\end{equation}
where $H(\mathbf{X})$ and $H(\mathbf{Y})$ are the entropies of $\mathbf{X}$ and $\mathbf{Y}$, respectively, defined as $H(\mathbf{X})=-\sum_{x\in\mathbf{X}}{\rm Pr}(\mathbf{X}=x)\log_2 {\rm Pr}(\mathbf{X}=x)$ (and similarly for $H(\mathbf{Y})$). The advantage of using $U(\mathbf{X}, \mathbf{Y})$ is mostly its interpretation. When the two variables are independent, $U(\mathbf{X}, \mathbf{Y}) = 0$, and when there is complete information about one variable from the other, $U(\mathbf{X}, \mathbf{Y}) = 1$.

\subsection*{Computational and statistical tools used}
In this article, most of the statistical functions employed have been programmed from scratch, using Python 3.10. For some particular uses, the following Python packages were used: for data cleaning (applying all filters described above to identify transient relationships), \texttt{pandas 1.5.1}; for some mathematical functions required to create histograms, \texttt{numpy 1.23}; for KS tests and OLS estimations, \texttt{statsmodels 0.13}; for mutual information tests, \texttt{scikit-learn 1.1}.

\clearpage
\begin{table}[!h]
  \centering
  \begin{tabular}{|llp{0.63\linewidth}|}
    \hline \hline
    Symbol  & Concept & Definition\\
    \hline
    $\mathcal{E}$ & Cohort & Cohort $\mathcal{E}$ which can be US, UK, IT, and IT$_n$.\\
    $\mathcal{A}_i(\ell,\Delta\ell)$ & -- & Set of alters of ego $i$ with lifetimes between $\ell$ and $\ell+\Delta\ell$.\\
    $t^{c}_{i,x}$ & -- & Day of $c$th call from ego $i$ to alter $x$ counted from the start of the data set that $i$ and $x$ belong to.\\
    $n_{i,x}$ & -- & Total calls from ego $i$ to alter $x$\\
    $a_{i,x}$ & Elapsed duration & Observed elapsed duration in days of the relationship between ego $i$ and alter $x$.\\
    $\ell_{i,x}$ & Lifetime & Observed lifetime in days of alter $x$ in ego $i$'s network.\\
    $\Delta t_{s}$ & -- & Exclusion days at the start of IT data to create IT$_n$. If ego calls alter for the first time at or after $\Delta t_{s}$ days, we identify the relationship as \textit{new}.\\
    $\Delta t_{w}$ & -- & Exclusion days before the end of the cohort data. If the last contact between an ego-alter pairs occurs $\Delta t_{w}$ days or more before the end of data in their cohort, we identify a relationship as \textit{transient}.\\
    $f_{i,x}(a_{ix},\ell_{ix})$ & -- & Volume of communication, measured as number of phone calls from ego $i$ to alter $x$ at elapsed duration $a_{ix}$ of their relationship of lifetime $\ell_{ix}$.\\
    $\Delta a$& -- & Time window within which to measure call volume.\\
    $\Delta\ell$ & -- & Time window for the selection of relationship lifetimes.\\
    $\bar{f_{i}}(a, \ell)$ & --  & Average per alter number of phone calls from ego $i$ to its alters with lifetime between $\ell$ and $\ell+\Delta\ell$ at elapsed duration between $a$ and $a+\Delta a$.\\
    $\bar{f}(a, \ell)$ & -- & Average of $\bar{f_{i}}(a, \ell)$ over all egos.\\
    $b(\ell)$ & -- & Steady volume of communication to alters with lifetime between $\ell$ and $\ell+\Delta\ell$.\\
    $p_i$ & -- & $p$-value of the Kolmogorov-Smirnov test between the first and second half of $\bar{f}_i(a,\ell)$.\\
    $g_{i,x}(a_{o}, a_{f})$ & -- & Number of phone calls from ego $i$ to alter $x$ when the relationship is between elapsed durations $a_{o}$ and $a_{f}$.\\
    $P(a \mid a_{o}, a_{f}, \gamma)$ & Survival Probability & Probability that an alter is active at elapsed duration $a$, given that it had activity $3^{\gamma}\leq g<3^{\gamma+1}$ during the interval $[a_{o}, a_{f}]$\\
    $I(\ell, g)$ &  Mutual Information &Mutual information between $\ell$ and $g$. It quantifies the amount of information that can be obtained from one variable by observing the other.\\
    $U(\ell, g)$ & Symmetric uncertainty & Symmetric uncertainty between $\ell$ and $g$. It measures the same as $I(\ell, g)$, but in a scale that goes from 0 (when the variables are independent) to 1 (when the information one variable gives about the other is complete).\\
    \hline \hline
  \end{tabular}
  \caption{All symbols used in this paper, both in the main text and SI. Next to each symbol, there is a brief explanation.}
  \label{tab:symbols}
\end{table}

\begin{table}[!h]
    \centering
    \begin{tabular}{|lrr|r|r|r|}
      \hline\hline
      Cohort                                            & Number of egos    & Alters & Short lifetime & Medium lifetime & Long lifetime\\
      \hline
      UK, IT, and US combined                           & 303               & 7625 & -- & -- & -- \\
      UK                                                & 30                & 920 & 483 & 90 & 76\\
      IT${}_n$                                          & 142               & 2736 & 1102 & 278 & 157\\
      IT                                                & 143               & 4052 & 1369 & 447 & 313\\
      US                                                & 130               & 2653 & 1415 & 399 & 319\\
      \hline \hline
    \end{tabular}
    \caption{Number of transient ego-alter pairs by cohort with $\Delta t_w=60$ days and $\Delta t_s=50$ days (for IT${}_n$ only). The last three columns show the exact number of relationships specifically used in the lifetime groups of Fig.~\ref{fig:exb_vertical} which represent a subset of all the transient relationships contained in the data.}
    \label{tab:numegoalter}
  \end{table}

\section*{Data Availability}
The US data can be accessed through the Reality Commons database (MIT) as the Friends and Family data (\url{http://realitycommons.media.mit.edu/friendsdataset.html}). The UK data relevant to this study has been made available previously in the publication Saram\"{a}ki et al. (2014) ``Persistence of social signatures in human communication'', PNAS 111 (3) 942-947.

The Mobile Territorial lab data used in this study are not freely available on an open repository for privacy reasons. However, they are available upon request by contacting the authors at lepri@fbk.eu. The data will be made available in a timely manner and in compliance with any ethical or legal requirements.
\end{spacing}

\bibliography{references}

\begin{thebibliography}{10}

\bibitem{aharony2011social}
Nadav Aharony, Wei Pan, Cory Ip, Inas Khayal, and Alex Pentland.
\newblock Social fmri: Investigating and shaping social mechanisms in the real
  world.
\newblock {\em Pervasive and Mobile Computing}, 7(6):643--659, 2011.

\bibitem{socialfmri2011}
Nadav Aharony, Wei Pan, Cory Ip, Inas Khayal, and Alex Pentland.
\newblock Social fmri: Investigating and shaping social mechanisms in the real
  world.
\newblock {\em Pervasive and Mobile Computing}, 7(6):643--659, 2011.

\bibitem{Almansoori2012}
Wadhah Almansoori, Shang Gao, Tamer~N. Jarada, Abdallah~M. Elsheikh, Ayman~N.
  Murshed, Jamal Jida, Reda Alhajj, and Jon Rokne.
\newblock Link prediction and classification in social networks and its
  application in healthcare and systems biology.
\newblock {\em Network Modeling Analysis in Health Informatics and
  Bioinformatics}, 1(1):27--36, Jun 2012.

\bibitem{Asikainen}
Aili Asikainen, Gerardo Iñiguez, Javier Ureña-Carrión, Kimmo Kaski, and
  Mikko Kivelä.
\newblock Cumulative effects of triadic closure and homophily in social
  networks.
\newblock {\em Science Advances}, 6(19):eaax7310, 2020.

\bibitem{bano2019whatsapp}
Shehar Bano, Wu~Cisheng, Ali~Nawaz Khan, and Naseer~Abbas Khan.
\newblock Whatsapp use and student's psychological well-being: Role of social
  capital and social integration.
\newblock {\em Children and Youth Services Review}, 103:200--208, 2019.

\bibitem{Barabasi2005}
Albert-L{\'a}szl{\'o} Barab{\'a}si.
\newblock The origin of bursts and heavy tails in human dynamics.
\newblock {\em Nature}, 435(7039):207--211, May 2005.

\bibitem{Bernard}
H.Russell Bernard and Peter~D Killworth.
\newblock On the social structure of an ocean-going research vessel and other
  important things.
\newblock {\em Social Science Research}, 2(2):145--184, 1973.

\bibitem{bidart20_livin}
Claire Bidart.
\newblock {\em Living in networks : the dynamics of social relations}.
\newblock Cambridge University Press, Cambridge, United Kingdom New York, NY,
  2020.

\bibitem{bidart2005}
Claire Bidart and Daniel Lavenu.
\newblock Evolutions of personal networks and life events.
\newblock {\em Social Networks}, 27(4):359–376, Oct 2005.

\bibitem{burt2000}
Ronald~S Burt.
\newblock Decay functions.
\newblock {\em Social Networks}, 22(1):1–28, May 2000.

\bibitem{buss2019mate}
David~M Buss and David~P Schmitt.
\newblock Mate preferences and their behavioral manifestations.
\newblock {\em Annual review of psychology}, 70:77--110, 2019.

\bibitem{centellegher2016mobile}
Simone Centellegher, Marco De~Nadai, Michele Caraviello, Chiara Leonardi,
  Michele Vescovi, Yusi Ramadian, Nuria Oliver, Fabio Pianesi, Alex Pentland,
  Fabrizio Antonelli, et~al.
\newblock The mobile territorial lab: a multilayered and dynamic view on
  parents’ daily lives.
\newblock {\em EPJ Data Science}, 5:1--19, 2016.

\bibitem{centellegher2017personality}
Simone Centellegher, Eduardo L{\'o}pez, Jari Saram{\"a}ki, and Bruno Lepri.
\newblock Personality traits and ego-network dynamics.
\newblock {\em PloS one}, 12(3):e0173110, 2017.

\bibitem{Curry2013}
Oliver Curry and Robin I.~M. Dunbar.
\newblock Do birds of a feather flock together?
\newblock {\em Human Nature}, 24(3):336--347, Sep 2013.

\bibitem{david2016communication}
Tamas David-Barrett, Janos Kertesz, Anna Rotkirch, Asim Ghosh, Kunal
  Bhattacharya, Daniel Monsivais, and Kimmo Kaski.
\newblock Communication with family and friends across the life course.
\newblock {\em PloS one}, 11(11):e0165687, 2016.

\bibitem{dunbar21_frien}
R.~I.~M Dunbar.
\newblock {\em Friends : understanding the power of our most important
  relationships}.
\newblock Little, Brown, London, 2021.

\bibitem{Dunbar1995}
R.~I.~M. Dunbar and M.~Spoors.
\newblock Social networks, support cliques, and kinship.
\newblock {\em Human Nature}, 6(3):273--290, Sep 1995.

\bibitem{DunbarArnaboldi}
R.I.M. Dunbar, Valerio Arnaboldi, Marco Conti, and Andrea Passarella.
\newblock The structure of online social networks mirrors those in the offline
  world.
\newblock {\em Social Networks}, 43:39--47, 2015.

\bibitem{dunbar1998social}
Robin I~M Dunbar.
\newblock The social brain hypothesis.
\newblock {\em Evolutionary Anthropology: Issues, News, and Reviews: Issues,
  News, and Reviews}, 6(5):178--190, 1998.

\bibitem{dunbar2018anatomy}
Robin I~M Dunbar.
\newblock The anatomy of friendship.
\newblock {\em Trends in cognitive sciences}, 22(1):32--51, 2018.

\bibitem{ferreira2015aware}
Denzil Ferreira, Vassilis Kostakos, and Anind~K Dey.
\newblock Aware: mobile context instrumentation framework.
\newblock {\em Frontiers in ICT}, 2:6, 2015.

\bibitem{Fu}
Feng Fu, Christoph Hauert, Martin~A. Nowak, and Long Wang.
\newblock Reputation-based partner choice promotes cooperation in social
  networks.
\newblock {\em Phys. Rev. E}, 78:026117, Aug 2008.

\bibitem{ghosh2019quantifying}
Asim Ghosh, Daniel Monsivais, Kunal Bhattacharya, Robin~IM Dunbar, and Kimmo
  Kaski.
\newblock Quantifying gender preferences in human social interactions using a
  large cellphone dataset.
\newblock {\em EPJ Data Science}, 8(1):9, 2019.

\bibitem{Godoy}
Antonia Godoy-Lorite, Roger Guimerà, and Marta Sales-Pardo.
\newblock Long-term evolution of email networks: Statistical regularities,
  predictability and stability of social behaviors.
\newblock {\em PLOS ONE}, 11(1):1--11, 01 2016.

\bibitem{gonccalves2011modeling}
Bruno Gon{\c{c}}alves, Nicola Perra, and Alessandro Vespignani.
\newblock Modeling users' activity on twitter networks: Validation of dunbar's
  number.
\newblock {\em PloS one}, 6(8):e22656, 2011.

\bibitem{hamed98}
Khaled~H. Hamed and A.~Ramachandra Rao.
\newblock A modified mann-kendall trend test for autocorrelated data.
\newblock {\em Journal of Hydrology}, 204(1-4):182--196, 1998.

\bibitem{hawkley2010}
Louise~C. Hawkley and John~T. Cacioppo.
\newblock Loneliness matters: A theoretical and empirical review of
  consequences and mechanisms.
\newblock {\em Annals of Behavioral Medicine}, 40(2):218–227, Jul 2010.

\bibitem{hill03_social_networ_size_human}
R.~A. Hill and R.~I.~M. Dunbar.
\newblock Social network size in humans.
\newblock {\em Human Nature}, 14(1):53--72, 2003.

\bibitem{hogan2007visualizing}
Bernie Hogan, Juan~Antonio Carrasco, and Barry Wellman.
\newblock Visualizing personal networks: Working with participant-aided
  sociograms.
\newblock {\em Field methods}, 19(2):116--144, 2007.

\bibitem{HOLME201297}
Petter Holme and Jari Saramäki.
\newblock Temporal networks.
\newblock {\em Physics Reports}, 519(3):97--125, 2012.
\newblock Temporal Networks.

\bibitem{holt2010}
Julianne Holt-Lunstad, Timothy~B. Smith, and J.~Bradley Layton.
\newblock Social relationships and mortality risk: A meta-analytic review.
\newblock {\em PLoS Medicine}, 7(7):e1000316, Jul 2010.

\bibitem{Huberman_Romero_Wu_2008}
Bernardo Huberman, Daniel~M Romero, and Fang Wu.
\newblock Social networks that matter: Twitter under the microscope.
\newblock {\em First Monday}, 14(1), Dec. 2008.

\bibitem{Hussain2019pyMannKendall}
Md. Hussain and Ishtiak Mahmud.
\newblock pymannkendall: a python package for non parametric mann kendall
  family of trend tests.
\newblock {\em Journal of Open Source Software}, 4(39):1556, 7 2019.

\bibitem{ibrahim2009missing}
Joseph~G Ibrahim and Geert Molenberghs.
\newblock Missing data methods in longitudinal studies: a review.
\newblock {\em Test}, 18(1):1--43, 2009.

\bibitem{johnson1982}
Michael~P. Johnson and Leigh Leslie.
\newblock Couple involvement and network structure: A test of the dyadic
  withdrawal hypothesis.
\newblock {\em Social Psychology Quarterly}, 45(1):34, Mar 1982.

\bibitem{kendall1955rank}
MG~Kendall.
\newblock {\em Rank Correlation Methods, p 160}.
\newblock Charles Griffin, London, 1955.

\bibitem{Kossinets}
Gueorgi Kossinets and Duncan~J. Watts.
\newblock Origins of homophily in an evolving social network.
\newblock {\em American Journal of Sociology}, 115(2):405--450, 2009.

\bibitem{maccarron16_callin_dunbar_number}
P.~Mac~Carron, K.~Kaski, and R.~Dunbar.
\newblock Calling dunbar's numbers.
\newblock {\em Social Networks}, 47(nil):151--155, 2016.

\bibitem{mann1945nonparametric}
Henry~B Mann.
\newblock Nonparametric tests against trend.
\newblock {\em Econometrica: Journal of the econometric society}, pages
  245--259, 1945.

\bibitem{mclean2017explaining}
Derrick~C McLean, Jeanne Nakamura, and Mihaly Csikszentmihalyi.
\newblock Explaining system missing: Missing data and experience sampling
  method.
\newblock {\em Social Psychological and Personality Science}, 8(4):434--441,
  2017.

\bibitem{McPherson}
Miller McPherson, Lynn Smith-Lovin, and James~M Cook.
\newblock Birds of a feather: Homophily in social networks.
\newblock {\em Annual Review of Sociology}, 27(1):415--444, 2001.

\bibitem{milardo1987changes}
Robert~M Milardo.
\newblock Changes in social networks of women and men following divorce: A
  review.
\newblock {\em Journal of Family Issues}, 8(1):78--96, 1987.

\bibitem{milardo1983}
Robert~M. Milardo, Michael~P. Johnson, and Ted~L. Huston.
\newblock Developing close relationships: Changing patterns of interaction
  between pair members and social networks.
\newblock {\em Journal of Personality and Social Psychology}, 44(5):964–976,
  1983.

\bibitem{miritello2013limited}
Giovanna Miritello, Rub{\'e}n Lara, Manuel Cebrian, and Esteban Moro.
\newblock Limited communication capacity unveils strategies for human
  interaction.
\newblock {\em Scientific reports}, 3(1):1--7, 2013.

\bibitem{miritello2013time}
Giovanna Miritello, Esteban Moro, Rub{\'e}n Lara, Roc{\'\i}o
  Mart{\'\i}nez-L{\'o}pez, John Belchamber, Sam~GB Roberts, and Robin~IM
  Dunbar.
\newblock Time as a limited resource: Communication strategy in mobile phone
  networks.
\newblock {\em Social Networks}, 35(1):89--95, 2013.

\bibitem{MOK2007}
Diana Mok and Barry Wellman.
\newblock Did distance matter before the internet?: Interpersonal contact and
  support in the 1970s.
\newblock {\em Social Networks}, 29(3):430--461, 2007.
\newblock Special Section: Personal Networks.

\bibitem{mollenhorst2014}
Gerald Mollenhorst, Beate Volker, and Henk Flap.
\newblock Changes in personal relationships: How social contexts affect the
  emergence and discontinuation of relationships.
\newblock {\em Social Networks}, 37:65–80, May 2014.

\bibitem{montag2018}
Christian Montag, Benjamin Becker, and Chunmei Gan.
\newblock The multipurpose application wechat: A review on recent research.
\newblock {\em Frontiers in Psychology}, 9, Dec 2018.

\bibitem{munch1997gender}
Allison Munch, J~Miller McPherson, and Lynn Smith-Lovin.
\newblock Gender, children, and social contact: The effects of childbearing for
  men and women.
\newblock {\em American sociological review}, 62(4):509--520, 1997.

\bibitem{Navarro2017}
Henry Navarro, Giovanna Miritello, Arturo Canales, and Esteban Moro.
\newblock Temporal patterns behind the strength of persistent ties.
\newblock {\em EPJ Data Science}, 6(1):31, 2017.

\bibitem{oswald2003}
Debra~L. Oswald and Eddie~M. Clark.
\newblock Best friends forever?: High school best friendships and the
  transition to college.
\newblock {\em Personal Relationships}, 10(2):187–196, Jun 2003.

\bibitem{link-prediction}
Wang Peng, Xu~BaoWen, Wu~YuRong, and Zhou XiaoYu.
\newblock Link prediction in social networks: the state-of-the-art.
\newblock {\em SCIENCE CHINA Information Sciences}, 58:011101:1–011101:38,
  January 2015.

\bibitem{phua2017uses}
Joe Phua, Seunga~Venus Jin, and Jihoon~Jay Kim.
\newblock Uses and gratifications of social networking sites for bridging and
  bonding social capital: A comparison of facebook, twitter, instagram, and
  snapchat.
\newblock {\em Computers in human behavior}, 72:115--122, 2017.

\bibitem{Rand}
David~G. Rand, Samuel Arbesman, and Nicholas~A. Christakis.
\newblock Dynamic social networks promote cooperation in experiments with
  humans.
\newblock {\em Proceedings of the National Academy of Sciences},
  108(48):19193--19198, 2011.

\bibitem{ranjan2019radar}
Yatharth Ranjan, Zulqarnain Rashid, Callum Stewart, Pauline Conde, Mark Begale,
  Denny Verbeeck, Sebastian Boettcher, Richard Dobson, Amos Folarin, RADAR-CNS
  Consortium, et~al.
\newblock Radar-base: Open source mobile health platform for collecting,
  monitoring, and analyzing data using sensors, wearables, and mobile devices.
\newblock {\em JMIR mHealth and uHealth}, 7(8):e11734, 2019.

\bibitem{roberts2015}
Sam B.~G. Roberts and R.~I.~M. Dunbar.
\newblock Managing relationship decay.
\newblock {\em Human Nature}, 26(4):426–450, Oct 2015.

\bibitem{roberts2011communication}
Sam~GB Roberts and Robin~IM Dunbar.
\newblock Communication in social networks: Effects of kinship, network size,
  and emotional closeness.
\newblock {\em Personal Relationships}, 18(3):439--452, 2011.

\bibitem{roberts2011}
Sam~G.B. Roberts and Robin~I.M. Dunbar.
\newblock The costs of family and friends: an 18-month longitudinal study of
  relationship maintenance and decay.
\newblock {\em Evolution and Human Behavior}, 32(3):186--197, 2011.

\bibitem{rozer2015}
Jesper~Jelle Rözer, Gerald Mollenhorst, and Beate Volker.
\newblock Romantic relationship formation, maintenance and changes in personal
  networks.
\newblock {\em Advances in Life Course Research}, 23:86–97, Mar 2015.

\bibitem{saramaki2014persistence}
Jari Saram{\"a}ki, Elizabeth~A Leicht, Eduardo L{\'o}pez, Sam~GB Roberts, Felix
  Reed-Tsochas, and Robin~IM Dunbar.
\newblock Persistence of social signatures in human communication.
\newblock {\em Proceedings of the National Academy of Sciences},
  111(3):942--947, 2014.

\bibitem{staiano2012friends}
Jacopo Staiano, Bruno Lepri, Nadav Aharony, Fabio Pianesi, Nicu Sebe, and Alex
  Pentland.
\newblock Friends don't lie: inferring personality traits from social network
  structure.
\newblock In {\em Proceedings of the 2012 ACM conference on ubiquitous
  computing}, pages 321--330, 2012.

\bibitem{sutcliffe2012relationships}
Alistair Sutcliffe, Robin Dunbar, Jens Binder, and Holly Arrow.
\newblock Relationships and the social brain: integrating psychological and
  evolutionary perspectives.
\newblock {\em British journal of psychology}, 103(2):149--168, 2012.

\bibitem{torous2016new}
John Torous, Mathew~V Kiang, Jeanette Lorme, and Jukka-Pekka Onnela.
\newblock New tools for new research in psychiatry: a scalable and customizable
  platform to empower data driven smartphone research.
\newblock {\em JMIR mental health}, 3(2):e16, 2016.

\bibitem{wellman2007challenges}
Barry Wellman.
\newblock Challenges in collecting personal network data: The nature of
  personal network analysis.
\newblock {\em Field Methods}, 19(2):111--115, 2007.

\bibitem{wellman1997}
Barry Wellman, Renita Yuk-lin Wong, David Tindall, and Nancy Nazer.
\newblock A decade of network change: Turnover, persistence and stability in
  personal communities.
\newblock {\em Social Networks}, 19(1):27–50, Jan 1997.

\bibitem{Zhou}
W.-X. Zhou, D.~Sornette, R.~A. Hill, and R.~I.~M. Dunbar.
\newblock Discrete hierarchical organization of social group sizes.
\newblock {\em Proceedings of the Royal Society B: Biological Sciences},
  272(1561):439--444, 2005.

\end{thebibliography}
\bibliographystyle{plain}

\section*{Acknowledgments}
VVH acknowledges support from Chile's National Agency for Research and Development (ANID) /  Scholarship Program / DOCTORADO BECAS CHILE/2018 - 72190510. BL and SC acknowledge The Mobile Territorial Lab (MTL), a joint initiative created by TIM - Telecom Italia, Fondazione Bruno Kessler, MIT Media Lab and Telefonica Research. BL and SC also thank all the MTL study participants. RD and SR acknowledge support from UK EPSRC/ESRC research grant number EP/D052114/2.

\newpage
\appendix
\renewcommand{\thetable}{S\arabic{table}}
\renewcommand{\thefigure}{S\arabic{figure}}
\renewcommand{\thesection}{S}
\setcounter{figure}{0}

\section{Supplementary Information}\label{sec:SM}
\begin{spacing}{1.5}

As a general note, in this document the term relationship between ego-alter pairs refers to \textit{transient relationships}. In many instances, we simply write relationship for brevity, but this should always be understood to mean transient relationships. The \textit{only exception} to this rule is encountered in Sec.~\ref{sec:L}. and its only figure (Fig.~\ref{fig:egosbya}).

\subsection{Construction of cohorts for each study}\label{sec:construction}
A succinct explanation of the data is provided in the main text (Sec.~Methods). These data sets have been fully described in previous articles, and the corresponding citations can be found below in the respective subsections.

Here we expand on some details of the studies that generated these data, namely the timing of entry of participants into the study and the life circumstances of the participants in each study. These details affect the way in which we choose the transient relationships we analyze here. After these descriptions, we elaborate on the filters we apply to arrive at the final cohorts in this study.

\subsubsection{Detailed information about each National study}

\paragraph{UK study}
The UK dataset was collected between 2007 and 2008, with all participants (egos) starting the experiment simultaneously. The timing of data collection was chosen to start observing participants during the last few months of secondary school and then continue to observe them for a time period that would capture an entire first year of university study. All participants were recruited from the same cohort in one school. The transition from secondary school to university occurs around six months after the start of data collection. With this design, participants begin the study while interacting generally with well-established network members (alters), and after six months, participants start to engage with a host of contacts that can be taken to be newly met alters (for details see~\cite{roberts2011}). Of the total cohort of $30$, only $2$ participants did not transition to university but their networks were still deeply disrupted due to the loss of alters and new personal circumstances.

\paragraph{US study}
The US dataset was collected between 2010 and 2011, with a pilot phase lasting $6$ months, and then a second phase of $12$ months in which an additional larger pool of participants is recruited at the beginning of this phase. This means that the egos did not all start simultaneously (for details see~\cite{aharony2011social}). Effectively, this produced a sample of approximately $17$ months.

The circumstances of the participants (egos) in the US data set are generally steady in time, i.e. the participants were not intentionally recruited to capture a particularly large change in their circumstances. Furthermore, egos do not generally enter synchronously in the phases of the study.

\paragraph{Italian study}
The Mobile Territorial Lab experiment recruited participants in two groups, one beginning their participation in early 2013 and the second in early 2014~\cite{centellegher2016mobile}. As with the US dataset, egos were not selected to be in a particularly dynamic stage of their lives where large changes to their circumstances could be foreseen. As with the US data set, egos in this study do not begin synchronously.

\subsubsection{Bounds on lifetimes by cohort}\label{sec:L}
The numbers of ego-alter relationships in all data sets decrease as a function of observed lifetimes. These decreases begin at a steady rate for all data sets, but as the lifetimes start to reach values that resemble the duration of the study, and in most cases considerably less time, the number of ego-alter pairs drop off at increased rates.

Fig.~\ref{fig:egosbya} shows the situation for all three national studies used. In each study, if an ego has at least one relationship that is still active at elapsed duration $a$ (horizontal axis), we count that ego into the proportion of active egos in the study; otherwise, the ego does not count. The proportion is displayed on the vertical axis. To choose the longest lifetimes $\mathcal{L}_{\mathcal{E}}$ used in each of the studies, we locate the value of $a$ at which the proportion of egos with at least one relationship active at $a$ starts to decay at an increased rate. This filtering prevents the possibility of attempting to build statistical inferences about lifetimes (particularly those with values close to $\mathcal{L}_{\mathcal{E}}$), on the basis of one or two ego-alter pairs per ego. The specific $\mathcal{L}_{\mathcal{E}}$ values are indicated in the plot by vertical dashed lines, and correspond to $\mathcal{L}_{\mathcal{E}}$ of the respective country $\mathcal{E}$. The values in days, reported in the plot, are $\mathcal{L}_{{\rm UK}}=270$ days, $\mathcal{L}_{{\rm US}}=220$ days, and $\mathcal{L}_{{\rm IT}}=365$ days. For the US, we choose to take a slightly larger $\mathcal{L}_{{\rm US}}$ than strictly supported by the plot (specifically $20$ days) because otherwise long lifetimes for the US become comparable to medium lifetimes of other cohorts on the basis of the rules we have chosen for selecting lifetime groups (see main text, discussion of Fig.~1). However, this adjustment is quite minor, and robustness checks presented below in Sec.~\ref{sec:robustness} show that any error that this may lead to is not appreciable.

With all the values of $\mathcal{L}_{\mathcal{E}}$ on hand, the ego-alter pairs we study are those that satisfy
\begin{equation}
    \ell_{ix}\leq \mathcal{L}_{\mathcal{E}}, i\in\mathcal{E}.
\end{equation}

\subsubsection{Calculation of initial relationship times for asynchronous ego entry}
The US and Italian studies add participants in an asynchronous way, i.e. not all participants become active at the same time. This becomes relevant for some measurements presented in the main text and in this supplementary document.

To deal with this effect, for each ego $i$ in both the Italian and US data sets, we define an entry time $\epsilon_i$, equal to the day $t$ of the study that $i$ belongs to when this ego is seen to make the first contact with any of its alters $\mathcal{A}_i$ (defined in the Methods, main text). Then, for ego-alter pair $ix$, we generate a relative start day $\tau_{o,ix}=t^{(1)}_{ix}-\epsilon_i$, where $t^{(1)}_{ix}$ is the day of first contact between $i$ and $x$ measured from the first day of the study to which $i$ belongs. Thus, $\tau_{o,ix}$ measures the number of days since ego $i$ entered his/her study before $i$ and $x$ were observed to begin contact.

We use this information in two different ways. First, we use it in determining how many transient relationships in each national study begin at or after a certain number of days from the entry of an ego into the study (see Sec.~\ref{sec:ego-entry-exit}). The second way we use this information is the construction of the IT${}_n$ cohort, described next.

\subsubsection{Construction of cohorts}
We develop four cohorts. Two of the cohorts (UK and IT${}_n$) are meant to characterize transient relationships where there is a high chance the initial contact is observed within the data. Two other cohorts (IT and US) are not adjusted to specifically try to capture the initial contact of ego-alter pairs but, as we see in Sec.~\ref{sec:ego-entry-exit}, this occurs for most contacts due to chance.

All the cohorts explained next satisfy the following: i) each ego-alter pair $ix$ has at least 3 contacts, so as to avoid studying meaningless relationships, ii) no lifetimes $\ell_{ix}$ are larger than the $\mathcal{L}_{\mathcal{E}}$, where $i\in\mathcal{E}$, and iii) ego-alter pairs comply with the transient relationship filter $\Delta t_w$, explained in the main text.

The Italian (IT) and US cohorts are fully defined by the three filters just mentioned. The other two cohorts satisfy additional conditions.

\paragraph{Construction of UK cohort}
Beyond the conditions stated above, the UK cohort is generated by only using ego-alter pairs that become active after 6 months or more from the start of the study. All egos have activity and therefore the cohort has all egos that enter the study from the beginning. Alters seen entering ego networks at that point are believed to be almost all new.

\paragraph{Construction of IT${}_n$ cohort}
This sub-cohort of the Italian cohort, in addition to conditions i, ii, and iii, includes a filter such that an ego-alter pair $ix$ is used only if $\tau_{o,ix}\geq\Delta t_s$, where $\Delta t_s$ is an exclusion window at the start of a participant's time in the study. This condition means that if pair $ix$ is active before the ego has been in the study at least $\Delta t_s$ days, the relationship is ignored. This filter is a way to reduce the number of ego-alter pairs analyzed for IT${}_n$ that may have been active before the actual start of the study. As we show in Sec.~\ref{sec:ego-entry-exit}, a large portion of transient relations actually begin after the start of the study and therefore, the filter introduced by using $\Delta t_s$ further lessens the likelihood of using an ego-alter pair that was in communication before the start of the study.

\bigskip

\noindent \textit{The result of the application of all the filters stated above is cohorts where the sample size is indicated in Table~2 of the main text.}

\subsection{Starting and ending times of relationships inside each study}\label{sec:ego-entry-exit}
As explained in the main text, even though in the US and Italian studies, some ego-alter pairs may have been active before the start of the study, the majority of transient relationships begin well after an ego enters his/her respective study (see next). This goes a long way in explaining why the various analyses we undertake in this study work similarly well when using UK and IT${}_n$ or US and IT.

To provide evidence for this interpretation, we present the cumulative distribution for $\tau_{o}$ (that is, the random variable associated with the individual $\tau_{o,ix}$ for concrete $ix$ pairs), separated by cohorts (Fig.~\ref{fig:totf}A). Each curve shown provides, per cohort, the percentage of ego-alter pairs that are first seen to be active on or before day $\tau_{o}$. Clearly, although many relationships are first observed for small values of $\tau_o$, they are by no means the majority. For instance, in most studies, $40\%$ of relationships require that $\tau_o$ reaches a value of $\approx 50$ days. The US is the only exception, starting at $\approx 40\%$ when $\tau_o$ is still quite small. By contrast, IT${}_n$ requires well over $100$ days to reach $40\%$ of transient relationships.

A similar analysis can be carried out with regards to the final contact between ego $i$ and alter $x$, which takes place $T_{\mathcal{E}}-t^{(n_{ix})}_{ix}$ days before the end of each study. This is relevant when assessing how close to the filter $\Delta t_w$ each pair $ix$ gets. As Fig.~\ref{fig:totf}{\color{blue}B} shows, this filter is not commonly reached. Thus, for example, in most cohorts only about $40\%$ of ego-alter pairs remain in touch when there still over $100$ days before the end of the study.

These two analyses provide support for the consistency seen in the results coming from cohorts UK, IT${}_n$, IT, and US because, although for the first two $a,\ell$ more strictly measure the actual elapsed duration and lifetime of transient relationships than the latter two, in practice many transient relationships are well contained inside the time boundaries of the studies, effectively making all four cohorts similar.

\subsection{Robustness check for $\bar{f}(a,\ell)$}\label{sec:robustness}
This section is concerned with robustness checks for the results and interpretation of $\bar{f}(a,\ell)$, presented in Fig.~1 of the main text. In summary, the results in the next subsections support the robustness of our conclusions about $\bar{f}(a,\ell)$, i.e. that for lifetimes of enough duration (see Sec.~\ref{sec:ego-tests} regarding this point) $\bar{f}(a,\ell)$ is indeed steady over $a$ until it ceases (\textit{\textbf{the steadiness feature}}), and that $\bar{f}(a,\ell)$ generally increases as a function of $\ell$ (\textbf{\textit{the monotonicity feature}}). After showing the standard error for each point of $\bar{f}(a, \ell)$, each subsection focuses on testing the effect of each particular measurement parameter ($\Delta t_w, \Delta \ell, \Delta a$) on $\bar{f}(a,\ell)$, with an additional subsection that tests $\bar{f}(a,\ell)$ against $\Delta t_s$ in IT${}_n$.

\subsubsection{Standard error of each value of $\bar{f}(a, \ell)$}
The results presented in Fig.~1 in the main text consider the stable volume of communication for the aggregated alters and egos of a given lifetime range. In order to show that variation exists at the individual level, we support our results with those from Fig.~3B in the main text. Additionally, as an alternative visualization for the variation among individual egos, we show one standard error for each point of $\bar{f}(a, \ell)$ in Fig.~\ref{fig:fig1_sem} as shaded regions in a color corresponding to their lifetime group.

\begin{figure}
  \centering
  \includegraphics[width=0.95\textwidth]{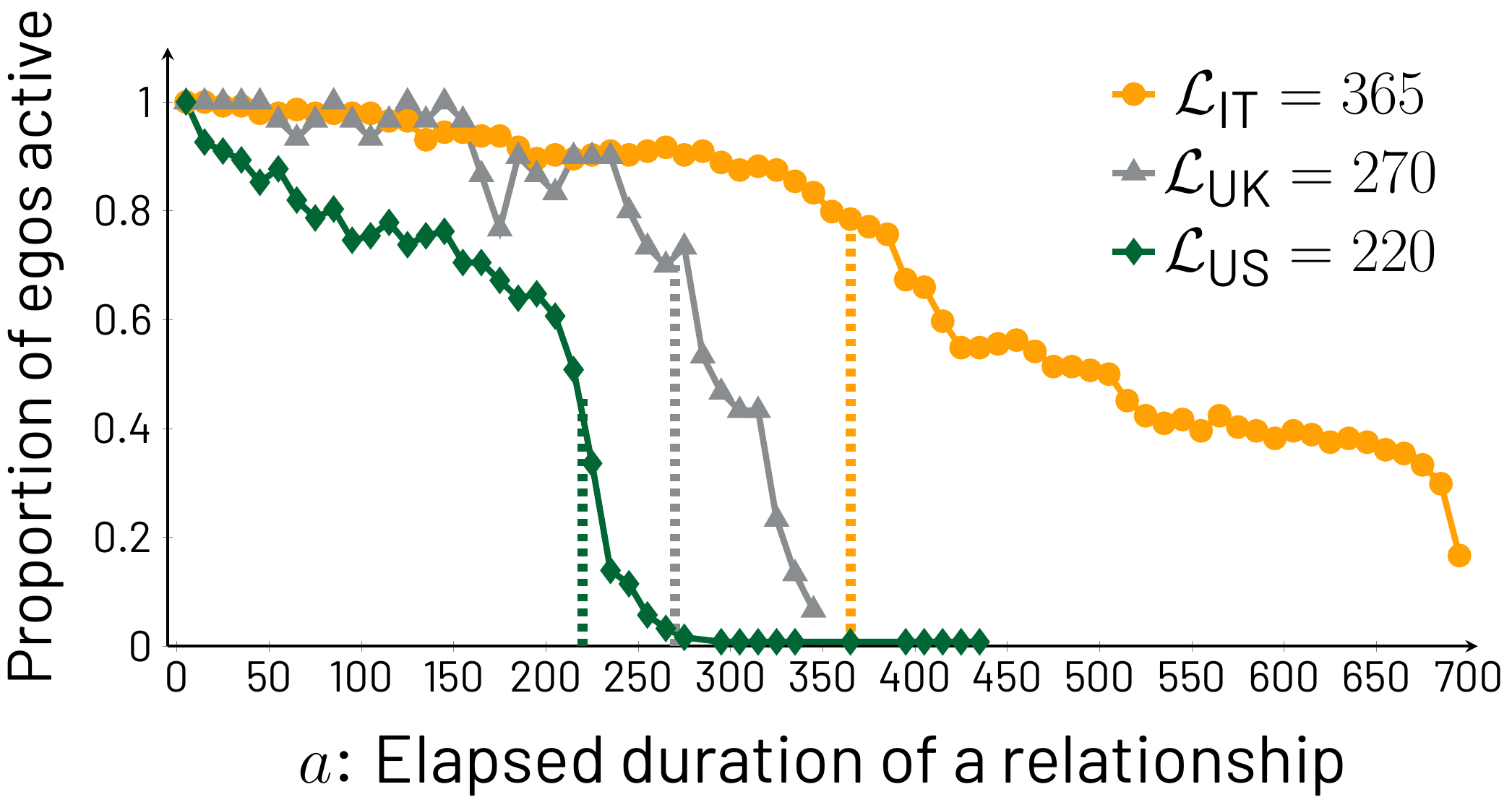}
  \caption{Proportion of egos with at least one active relationship at or above elapsed duration $a$, country by country. These relationships are only filtered by having a minimum of $3$ contacts throughout each data set; in other words, these relationships are not necessarily transient. Vertical dashed lines show the chosen values for $a=\mathcal{L}_{\mathcal{E}}$, for each $\mathcal{E}=\{{\rm UK,IT,US}\}$. These values are located at the point where the number of egos still having active alters begins to decay rapidly.}
  \label{fig:egosbya}
\end{figure}

\begin{figure}
    \centering
    \includegraphics[width=0.95\textwidth]{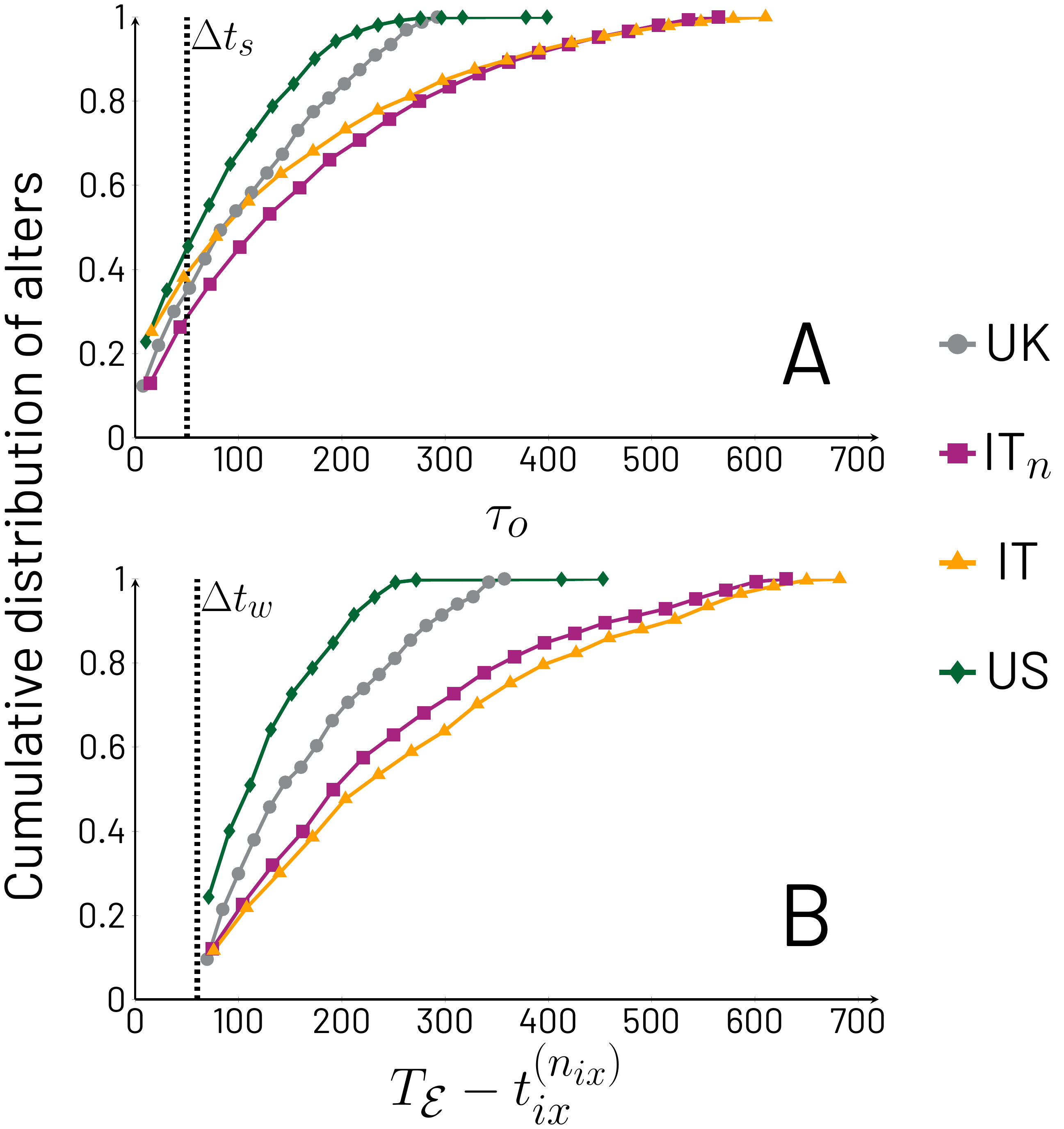}
    \caption{Panel A: cumulative distribution of $\tau_o$, the random variable of the variates $\tau_{o,ix}$ which corresponds to the starting day of relationship $ix$ compared to the beginning of ego $x$ in its respective cohort. These $\tau_{o,ix}$ come from transient relationships. Panel B: cumulative distribution of the number of days between the last phone call between transient ego-alter pair $ix$ and the last day of the cohort, cohort by cohort.}
    \label{fig:totf}
  \end{figure}

\begin{figure}
    \centering
    \includegraphics[width=\textwidth]{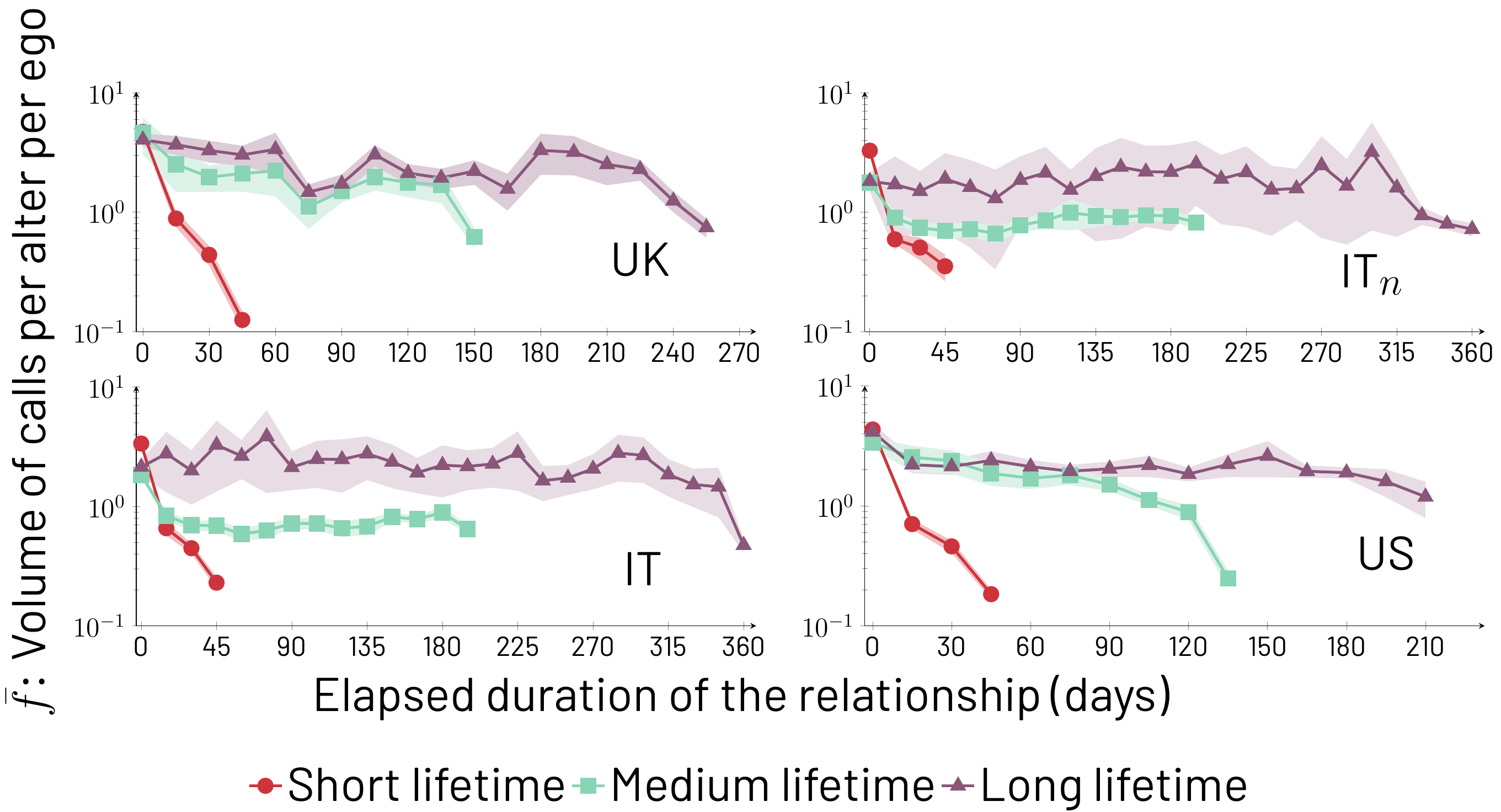}
    \caption{Version of Fig.~1 of the main text with the addition of one standard error (above and below) for each point of $\bar{f}(a, \ell)$, represented as a shaded region with color corresponding to those used for each lifetime group in the paper. All other parameters are the same as those presented in the main text, Fig.~1.}
    \label{fig:fig1_sem}
\end{figure}

For all cohorts, as lifetime increases, so does the standard error. This is due to the fact that the number of transient alters decreases with $\ell$, as seen in Fig.~\ref{fig:egosbya}.

\subsubsection{Decay in $\bar{f}(a, \ell)$ for $a$ between $\ell$ and $\ell+\Delta\ell$}
Our method for measuring $\bar{f}(a,\ell)$, although effective in terms of generating a reliable estimate of the per ego per alter calling volume of each ego to their alters, also has the unintended consequence of frequently generating a fast decaying tail for $\bar{f}(a,\ell)$ for $\ell\leq a\leq\ell+\Delta\ell$ visible in medium and long lifetimes (Fig.~1 of the main text as well as figures of the robustness checks of the current section). Here, we explain the origin of this effect, which is \textit{not} a behavioral feature of egos, but rather a statistical nuisance effect.

As defined in the main text, $\bar{f}(a,\ell)$ is given by
\begin{equation}\label{eq:barf}
    \bar{f}(a,\ell)=\frac{\sum_{i}\bar{f}_i(a,\ell)}{\sum_{i}\theta(\vert\mathcal{A}_i(\ell,\Delta \ell)\vert)},
\end{equation}
where $\theta(\cdot)$ corresponds to the step function ($\theta(x)=1$ if $x>1$, and $0$ otherwise), and $\vert\vert$ produces the cardinality of a set. In the range of $a$ starting with $\ell$ and ending at $\ell+\Delta\ell$, there is a progressive reduction of the numerator of Eq.~\ref{eq:barf} that occurs because not all individual egos have alters until $a=\ell+\Delta\ell$. Instead, any given ego typically has activity until a value of $a$ somewhere in the middle of the range between $\ell$ and $\ell+\Delta\ell$. Let us assume that the ego in question is $i$. If the last active alter with lifetime between $\ell$ and $\ell+\Delta\ell$ in ego $i$'s network stops activity at $a^{\text{(end)}}_{i}$, then the time series $\bar{f}_i(a,\ell)=0$ for $a>a^{\text{(end)}}_{i}$. However, in Eq.~\ref{eq:barf}, the denominator is unchanging, which means that between $a^{\text{(end)}}_{i}$ and $\ell+\Delta\ell$, $\bar{f}(a,\ell)$ is calculated with the same denominator but a diminished numerator with no contributions from ego $i$. Crucially, the distribution of end-times for each of the time series $\bar{f}_i(a,\ell)$ occurs all throughout the range between $\ell$ and $\ell+\Delta\ell$.

\begin{figure}
    \centering
    \includegraphics[width=0.9\textwidth]{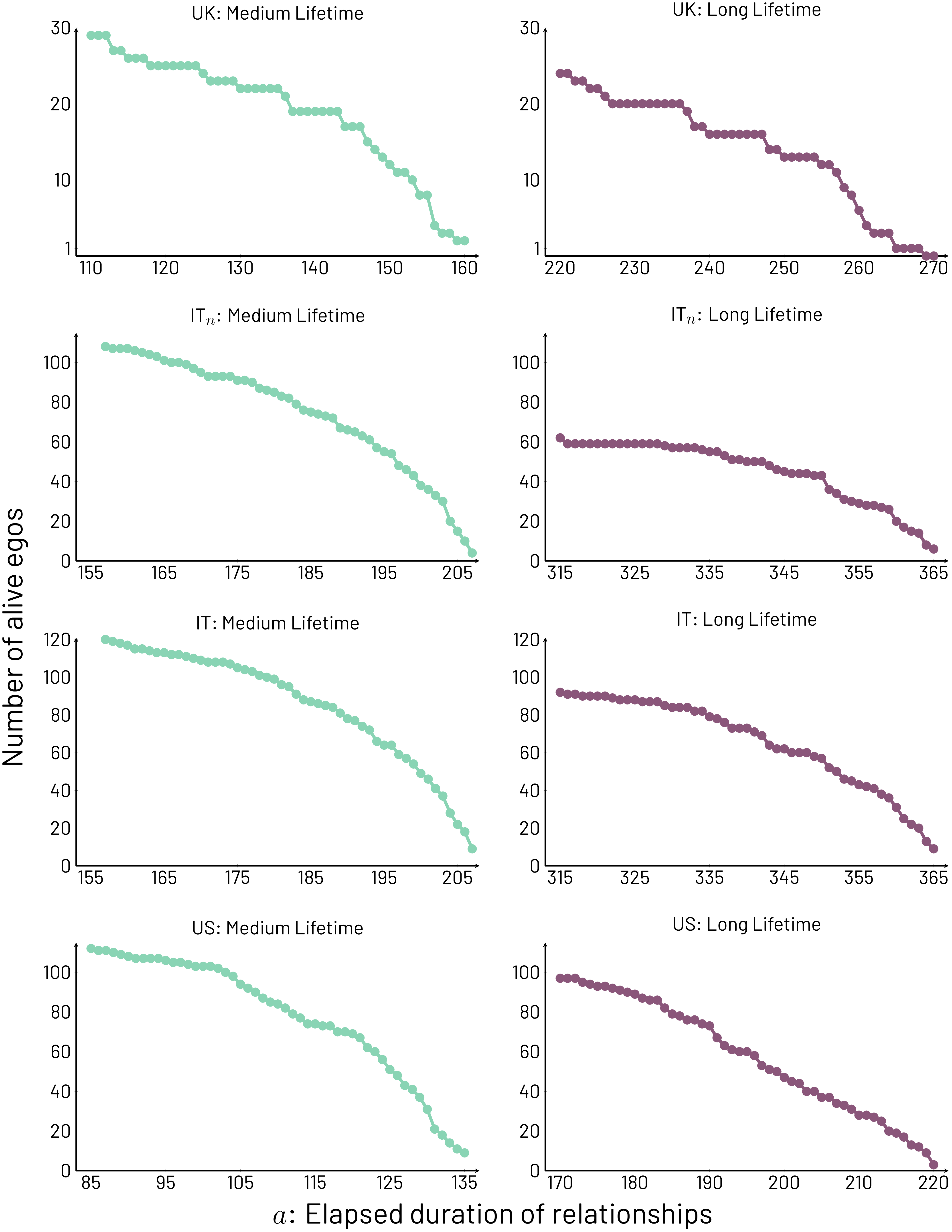}
    \caption{Number of egos with active alters at elapsed durations in the ranges $\ell\leq a\leq\ell+\Delta \ell$ for all countries and lifetime cohorts shown in Fig.~1, main text. Left column shows medium lifetimes, the right columns shows long lifetimes. The colors consistent with the main text, Fig.~1. From these plots, one can observe how the number of active egos decays steadily within the window between $\ell$ and $\ell+\Delta\ell$, generating the fast decaying effect seen in Fig.~1, a purely statistical effect of the definition of $\bar{f}(a,\ell)$.}
    \label{fig:aliveegos}
  \end{figure}

In Fig.~\ref{fig:aliveegos}, we show the number of egos still active in the range between $\ell$ and $\ell+\Delta\ell$ respective to each of the cohorts shown in Fig.~1 of the main text. As it is clear from these plots, the number of active alters decays rapidly from a value of $\vert\mathcal{A}_i(\ell,\Delta \ell)\vert$ to $0$ causing $\bar{f}(a,\ell)$ to also decay within this temporal range. This effect leads to the generation of the fast drops seen in most of the curves in Fig.~1. However, we should note that the decay can be partially attenuated if, by random chance, a group of egos in some lifetime group $\ell$ to $\ell+\Delta\ell$ remains active until closer to $\ell+\Delta\ell$ and/or the total call volume among those egos near the end of the time series fluctuates upwards (see e.g. medium lifetime for Italy in Fig.~1).

\subsubsection{Robustness in $\Delta t_{w}$}
In our study, we exclude any alter $x$ such that $T_{\mathcal{E}}-t_{ix}^{(n_{ix})}\leq\Delta t_w$ which means that we study ego-alter pairs that stop communicating at some point in the study and remain without communication until the end of the study and for a minimum of at least $\Delta t_w$ days. This is effectively our transient relationship operational criterion.

\begin{figure}
    \centering
    \includegraphics[width=0.95\textwidth]{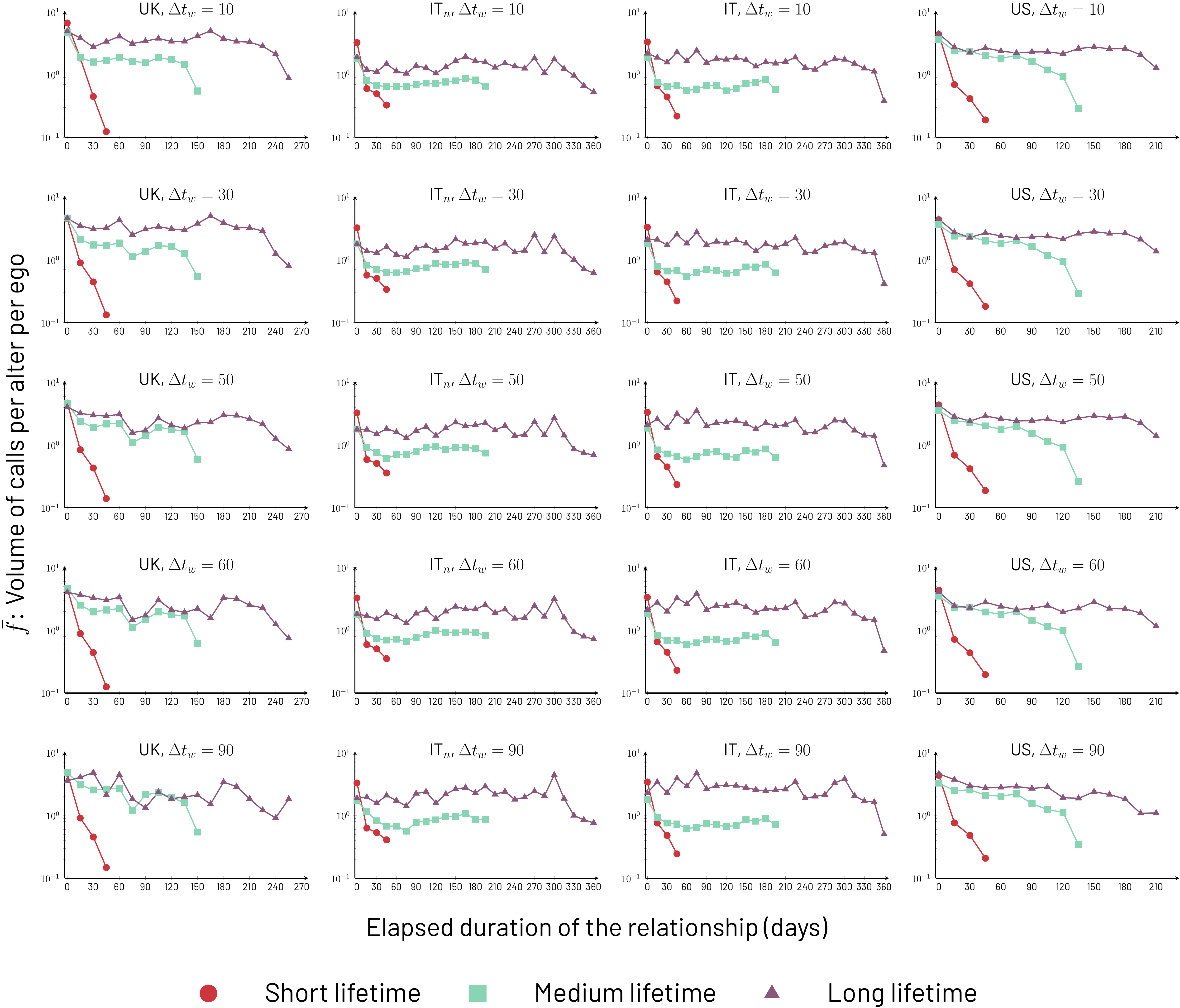}
    \caption{Robustness check of Fig.~1 from the main text with respect to changes in the choice of $\Delta t_w$. Each column corresponds to a cohort and each row to a value of $\Delta t_w$, all indicated in each plot. The main text uses $\Delta t_w=60$ days. The qualitative behavior of $\bar{f}(a,\ell)$ is consistent through the choices of $\Delta t_w$. Although as $\Delta t_w$ increases, sampling decreases appreciably and leads to more fluctuations in the signal, the conclusions drawn about Fig.~1 of the main text remain valid.}
    \label{fig:Deltatw}
  \end{figure}

The main text presents results with $\Delta t_{w} = 60$ (fourth row in Fig.~\ref{fig:Deltatw}). Here, we test robustness by also checking $\Delta t_{w}=10,30,50,90$. As Fig.~\ref{fig:Deltatw} shows, using different values of $\Delta t_w$  does not affect either the steadiness nor the monotonicity features.

\subsubsection{Robustness in $\Delta\ell$}
\begin{figure}
    \centering
    \includegraphics[width=0.95\textwidth]{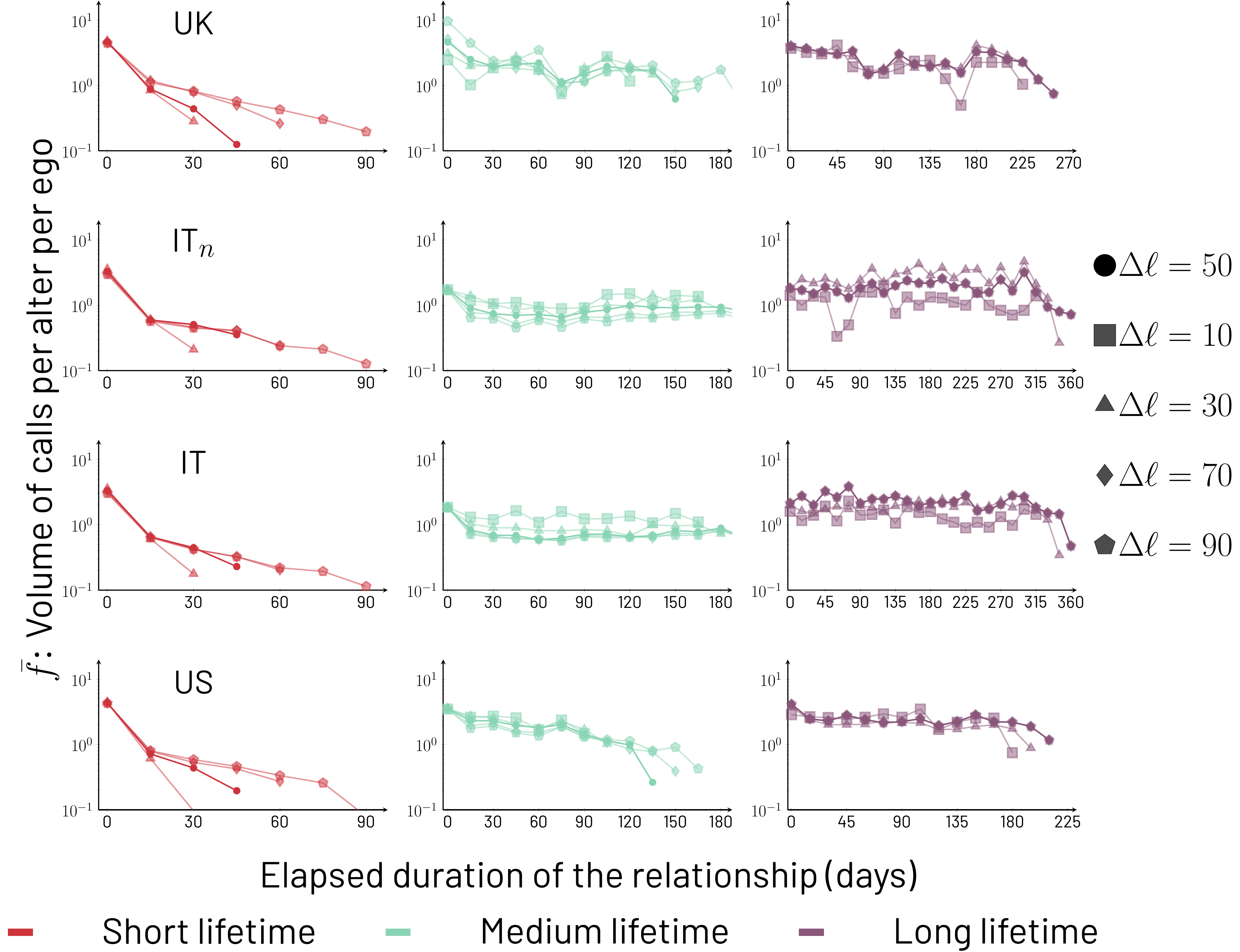}
    \caption{Robustness checks of $\bar{f}(a,\ell)$ with respect to the choice of $\Delta \ell$, separated by lifetime group (short, medium, long). In the main text, $\Delta\ell=50$ days is used. Each row corresponds to a cohort, and each column to a lifetime group. For small $\Delta\ell$, as expected, the signal fluctuates more. The only qualitative change observed across all the plots is in the column for short lifetimes: as $\Delta\ell$ increases, we observe the gradual emergence of weak plateaus, specially as $\Delta\ell\to\ell_s$, the minimum value at which steadiness emerges for $\bar{f}(a,\ell)$.}
    \label{fig:Deltaell}
  \end{figure}

The value of $\Delta\ell$ of each lifetime group in Fig.~1 of the main text has been chosen as $\Delta\ell=50$ days (circles in Fig.~\ref{fig:Deltaell}). Testing $\Delta \ell = 10$, $\Delta \ell = 30$, $\Delta \ell = 70$, and $\Delta \ell = 90$ leads to consistent results for medium and long lifetimes in terms of steadiness and monotonicity of $\bar{f}(a,\ell)$. An interesting observation also emerges for short lifetimes where $\ell$ is below the threshold value $\ell_s$ (see Sec.~\ref{sec:ells}) for steady behavior: as $\Delta \ell$ increases, $\bar{f}(a,\ell)$ begins to change from a decaying behavior to one that develops a steadier range over values of $a$, signalling a trend towards steadiness.

\subsubsection{Robustness in $\Delta a$. Estimation of $a_s$}
\begin{figure}
    \centering
    \includegraphics[width=0.95\textwidth]{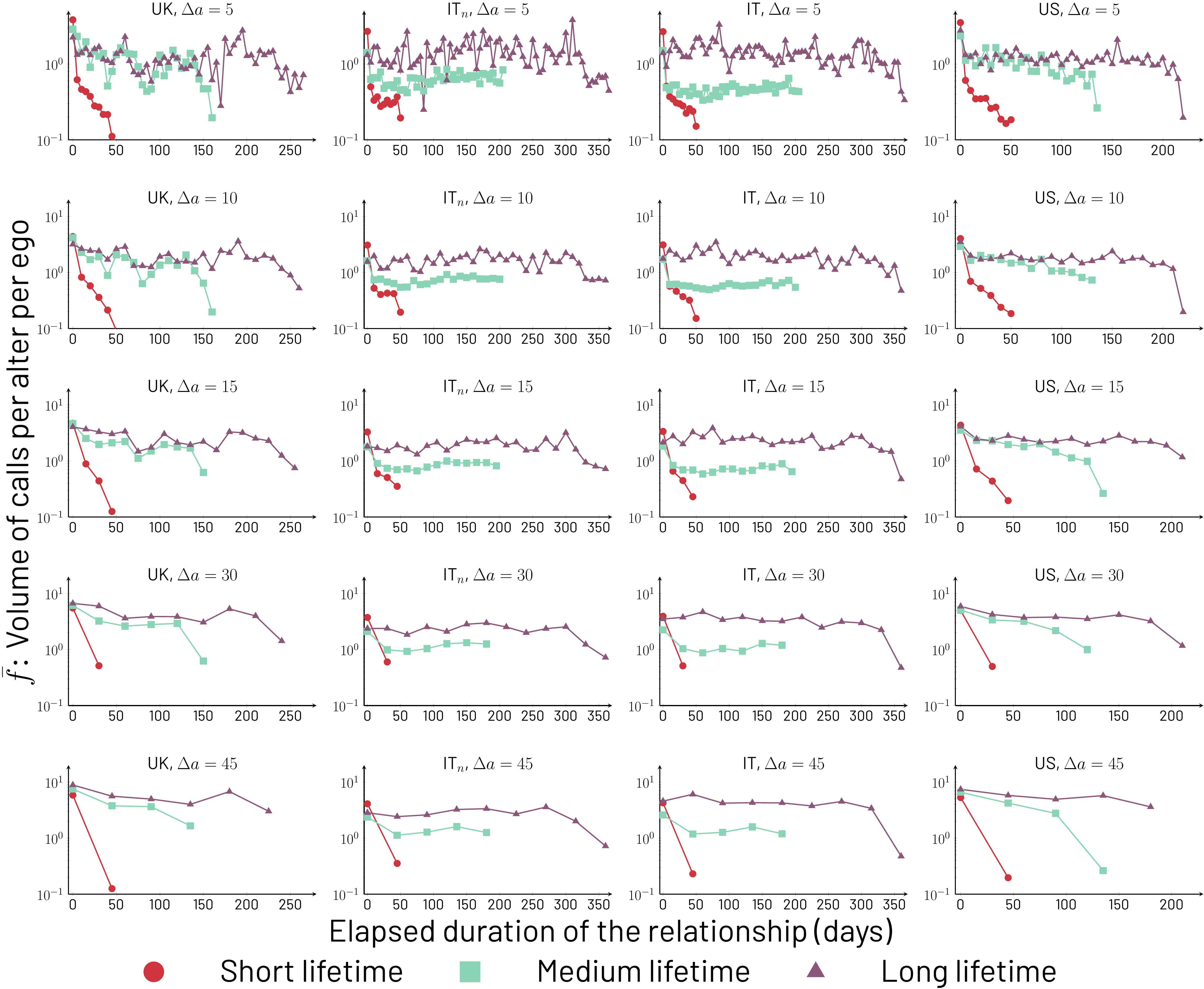}
    \caption{Robustness check of Fig.~1 from the main text with respect to changes in the choice of $\Delta a$. Each column corresponds to a cohort and each row to a value of $\Delta a$, all indicated in each plot. The main text uses $\Delta a=15$ days. The qualitative behavior of $\bar{f}(a,\ell)$ is consistent through the choices of $\Delta a$. Smaller values of $\Delta a$ display more fluctuations as expected, but the conclusions drawn about Fig.~1 of the main text remain valid.}
    \label{fig:Deltaa}
  \end{figure}

In the main text, Fig.~1 uses $\Delta a = 15$. Fig.~\ref{fig:Deltaa} shows different values of $\Delta a$ ($\Delta a$ = 5, 10, 15, 30, 45). Small $\Delta a$ leads to $\bar{f}(a,\ell)$ with more fluctuations, while large $\Delta a$ exhibits very steady features. In all cases, the qualitative features of $\bar{f}(a,\ell)$ are preserved, in terms of the steadiness of communication to alters with medium and long lifetimes.

\begin{figure}
    \centering
    \includegraphics[width=0.95\textwidth]{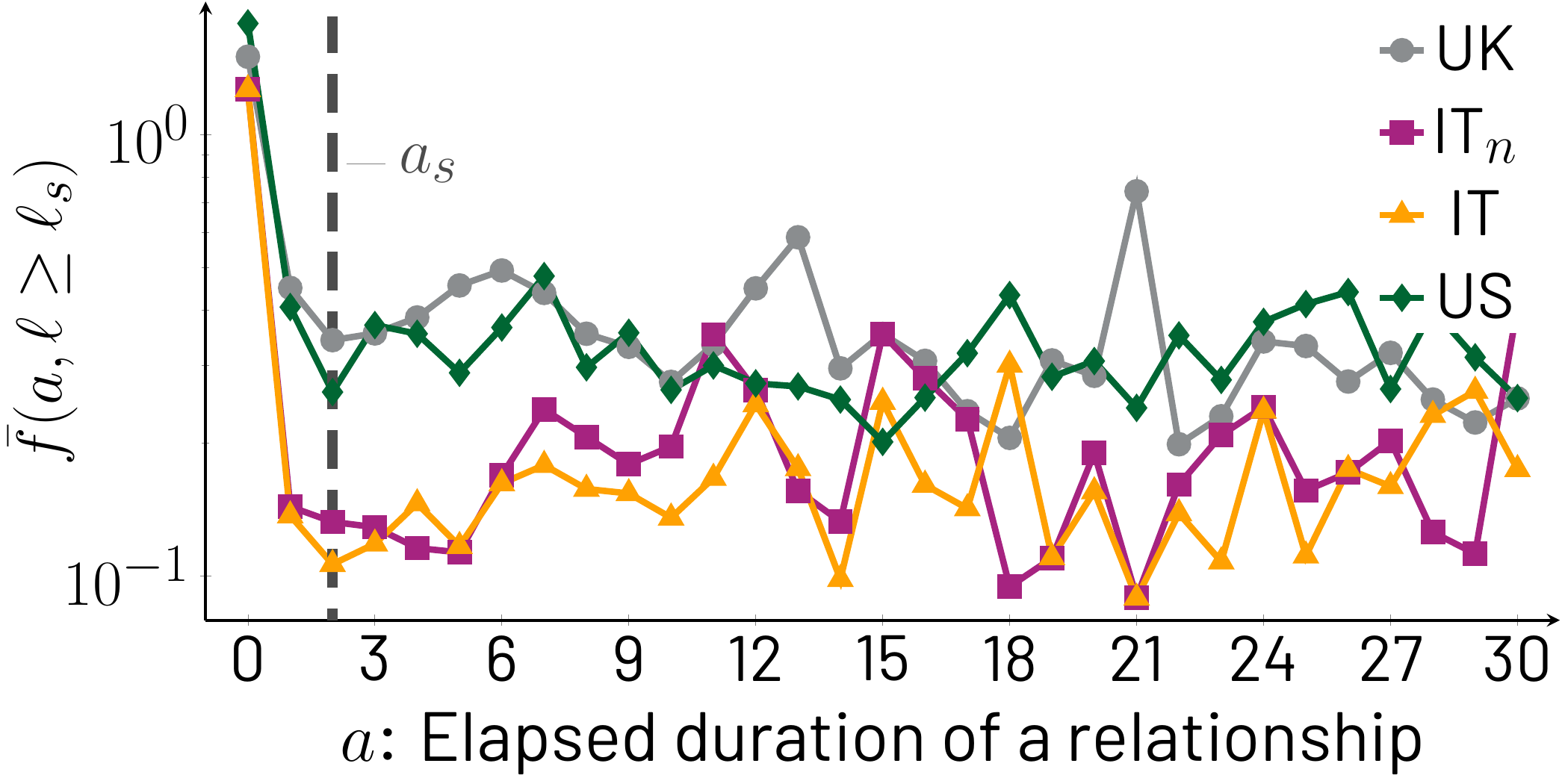}
    \caption{Time series of $\bar{f}(a, \ell \geq \ell_s)$ for all cohorts with $\Delta a=1$. In all curves, the derivative of $\bar{f}(a, \ell \geq \ell_s)$ is negative for $a \leq 2$. Thus, we identify $a_s = 2$ as the minimum value of $a$ for which Eq.~1 of the main text is valid. For each country, $\ell_s$ is determined by direct averaging as described in Sec.~\ref{sec:ells}}
    \label{fig:faells}
  \end{figure}

In addition to the $\Delta a$ above, we also apply $\Delta a=1$ to estimate the value $a_s$. However, given that this quantity appears across values of $\ell$, we use $\bar{f}(a, \ell \geq \ell_{s})$. This has the additional advantage of improving our sample size. As Fig.~\ref{fig:faells} shows, the first set of points on the plot, $a=0,1$, and $2$, all show a steady decreasing trend before the curve begins to stabilize. Therefore, we believe that $a_s=2$ constitutes a lower bound for the applicability of Eq.~1 in the main text.

\subsubsection{Robustness in $\Delta t_s$ for IT${}_n$}
\begin{figure}
    \centering
    \includegraphics[width=0.95\textwidth, height=0.9\textheight]{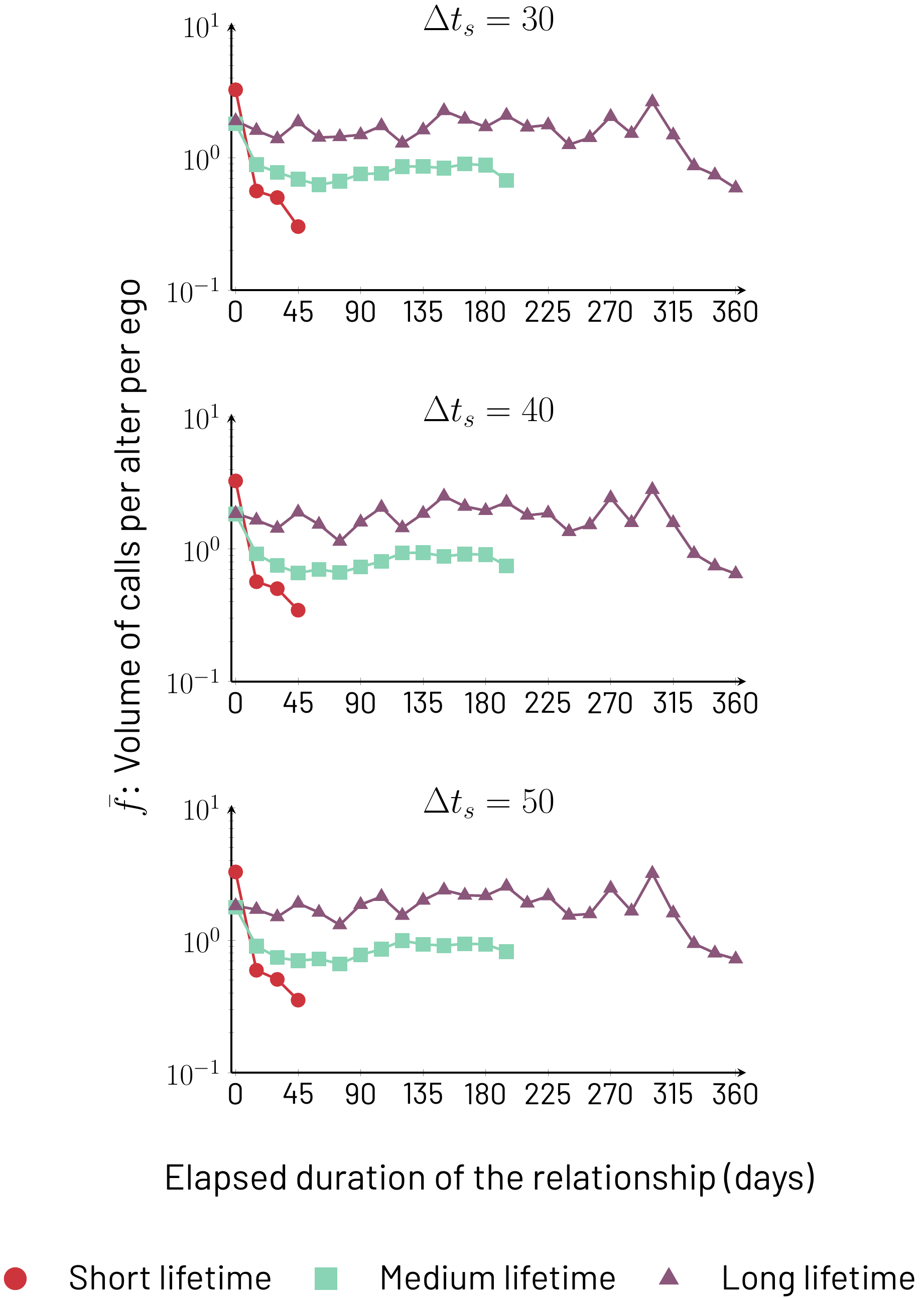}
    \caption{Robustness check for $\bar{f}(a, \ell)$ for different values of $\Delta t_{s}$ for the IT${}_{n}$ cohort. The main text shows $\Delta t_s=50$ days. The results are consistent across the range of values.}
    \label{fig:Deltats}
  \end{figure}

To construct IT${}_n$ in the main text, we use $\Delta t_s=50$. Although most of our subsequent analysis in the main text as well as in this supplementary document (Sec.~\ref{sec:ego-entry-exit}) supports the idea that $\bar{f}(a,\ell)$ is robust even if the start of a transient relationship is not captured, we nevertheless test for this robustness. In Fig.~\ref{fig:Deltats}, we present $\bar{f}(a,\ell)$ calculated for $\Delta t_s=30,40,50$ and find consistent results, supporting the steadiness and monotonicity features. Further filtering of $\Delta t_s$ reduces the sample considerably and thus becomes unreliable for values $\Delta t_s \geq 50$.

\subsubsection{Volume and duration of calls}
\begin{figure}
    \centering
    \includegraphics[width=0.95\textwidth]{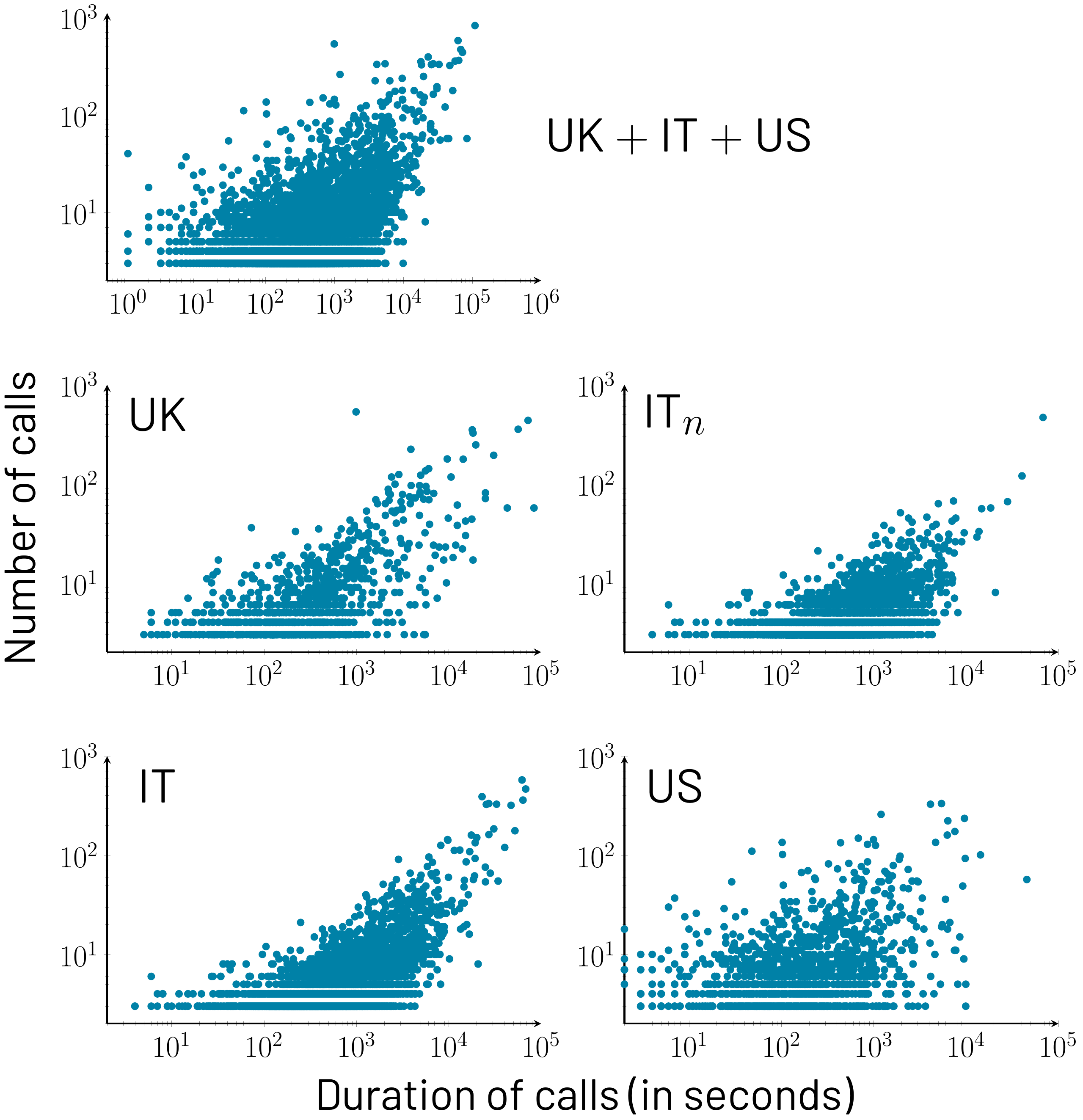}
    \caption{Scatter plots showing the association between the added duration of phone calls in a relationship, and their total number. Results are presented a log-log scale and show a strong positive correlation, with the coefficients presented in the supplementary text.}
    \label{fig:ndur}
  \end{figure}

To study the temporal signal of communication, one can study numbers of calls or call durations. However, these two choices are known to be correlated for the UK data we use here~\cite{saramaki2014persistence}. In order to provide a full picture of this correlation focused on transient relationships, we measure the Person correlation of total numbers of calls and total time spent communicating between each ego-alter pair in all the our cohorts, and obtain the following results: the combined cohort shows a correlation of $r = 0.6544$, while the cohorts coefficient are: $r_{\rm{UK}} = 0.5739$; $r_{\rm{IT}_{n}} = 0.8238$; $r_{\rm{IT}} = 0.8769$; and $r_{\rm{US}} = 0.2332$. Fig.~\ref{fig:ndur} shows scatter plots for each of the cohorts, in which each point corresponds to the total number of calls and total time ego spent talking to one of its transient alters. These results support our choice of only focusing on number of calls as a useful metric, as call duration would produce redundant analysis.

\subsection{Determination of $b(\ell)$, and $b_i(\ell)$}\label{sec:bell}
The height of the plateaus of each $\bar{f}(a,\ell)$ associated with a set of alters of a given range of lifetimes $\ell$ to $\ell+\Delta \ell$ in a cohort $\mathcal{E}$ is measured by $b(\ell)$. Similarly, the set of $b_i(\ell)$ captures the heights of the plateaus of individual egos' $\bar{f}_i(a,\ell)$. In this section, we discuss an alternative method to obtain $b(\ell)$ and $b_{i}(\ell)$ to the \textit{stable region average} presented in the main text, Methods section.

\subsubsection{Mann-Kendall method to identify $b(\ell)$}\label{sec:bmethods}

As an alternative to the \textit{Stable Region Average} method presented in the main text, here we  give an alternative calculation of $b(\ell)$, using the Mann-Kendall test~\cite{kendall1955rank, mann1945nonparametric}, explained further in~\cite{hamed98}, to detect trends in the data. We use the \texttt{Python} implementation provided by~\cite{Hussain2019pyMannKendall}. The basic intuition of the test can be understood as a simplification first proposed by Mann~\cite{mann1945nonparametric} of the Kendall rank-correlation test~\cite{kendall1955rank}. In particular, for a signal such as $\bar{f}(a,\ell)$ or $\bar{f}_i(a,\ell)$ (which for generality we denote as $u(a)$), one defines a test statistic $S=\sum_{a}{\rm sign}(u(a+\Delta a)-u(a))$, where ${\rm sign}(u(a+\Delta a)-u(a))=1$ if $u(a+\Delta a)>u(a)$, $=0$ if $u(a+\Delta a)=u(a)$, and $=-1$ if $u(a+\Delta a)<u(a)$. The null hypothesis is for there to be no trend, in which case $S$ is a normally distributed random variable with mean $0$. Normality of the input random variable is not a requirement of this non-parametric test. To apply the test, we truncate the range of $a$ from the left and from the right systematically as in the stable region test (see subsection~``$b(\ell)$, $b_i(\ell)$, and $\ell_s$ computation'' in the main text), stopping when no trend is detected (slope tends to $0$), i.e. when the null hypothesis of the test can no longer be rejected at a significance level of $0.05$. This procedure identifies the steady regions of $u(a)$.

\subsubsection{Results from application of the Mann-Kendall method}

\begin{figure}
    \centering
    \includegraphics[width=0.95\textwidth]{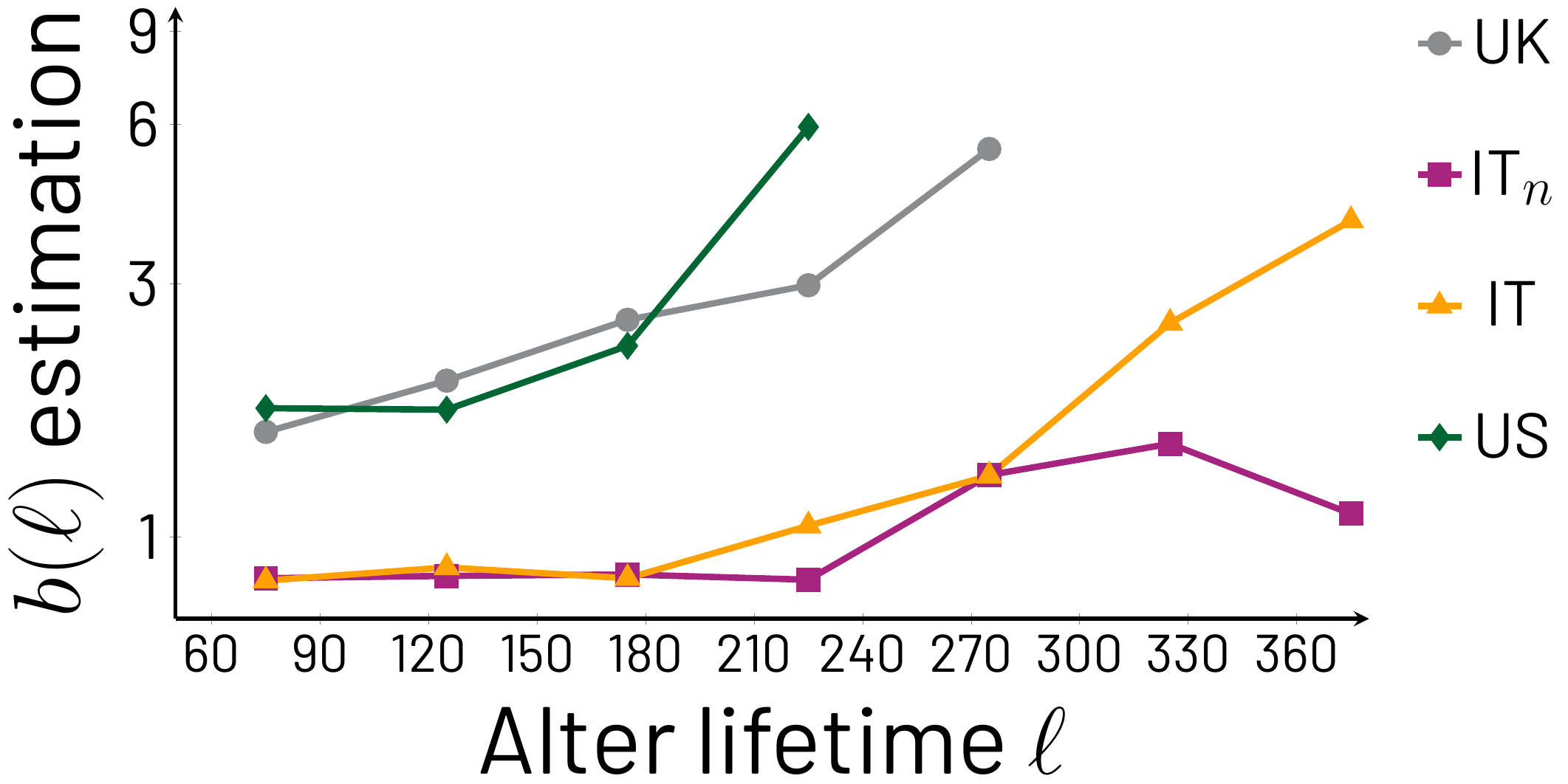}
    \caption{$b(\ell)$ as a function of $\ell$ obtained through the Mann-Kendall method, cohort by cohort. The vertical axis is in logarithmic scale. Clearly, $b(\ell)$ has an increasing trend with respect to $\ell$, with minor exceptions. This result, remarkably similar to Fig.~2 in the main paper, highlights the consistency between the Mann-Kendall and stable region average methods.}
    \label{fig:bestimationMK}
  \end{figure}

In this section, we show results for $b(\ell)$ done across the values of $\ell$ with the Mann-Kendall method described above. This result can be seen in Fig.~\ref{fig:bestimationMK}. From the plot, we see that $b(\ell)$ increases with $\ell$ almost universally, with the exception of minuscule fluctuations early in IT and IT${}_n$, and again for the longest $\ell$ for IT${}_n$. This last deviating point occurs because the introduction of $\Delta t_s$ effectively eliminates a great deal of the ego-alter samples that are available for equivalent lifetimes of IT, thus reducing statistical sampling. Overall, the trends are clear and highly consistent across cohorts (see Fig.~2 in the main text).

The fact that $b(\ell)$ is increasing with $\ell$ also supports the claim made in the main text that the selection of the medium and long lifetimes used in Figs.~1 and~3 is mostly arbitrary and for the purposes of illustrating the behavior of $\bar{f}(a,\ell)$ for concrete values of $\ell$. However, these choices of $\ell$ are not restrictive and in fact one can work with values of $\ell$ from $\ell_s$ and up. %

One last observation is that, while the trends of $b(\ell)$  are increasing, there are differences among the cohorts, with the US and UK showing a more rapid growth than the Italian cohorts, which start roughly steady and then begin their marked increase for larger values of $\ell$. This may have implications in terms of how effectively one can distinguish medium lifetimes in Italian ego-alter pairs in comparison to the other cohorts on the basis of early phone call activity. This will require further research.

Results for $b_i(\ell)$ and $\ell_s$ are presented in Sec.~\ref{sec:ego-tests} as they pertain to ego-level features, the subject of that section.

\subsection{Individual ego tests}\label{sec:ego-tests}
The features captured by $\bar{f}(a,\ell)$ are also shared by $\bar{f}_i(a,\ell)$ for individual egos. This is supported in the main text through the results displayed in Fig.~3, and further tested in other parts of the main manuscript, namely, those that check if the increase of $\bar{f}(a,\ell)$ with $\ell$ has predictive power such as Figs.~4 and~5. In this section, we complement this evidence by showing primary analyses that allows us to construct the results presented in the main text, as well as additional robustness checks for the main text results.

\subsubsection{Visual inspection of random sample of $\bar{f_{i}}$}
\begin{figure}
    \centering
    \includegraphics[width=0.95\textwidth]{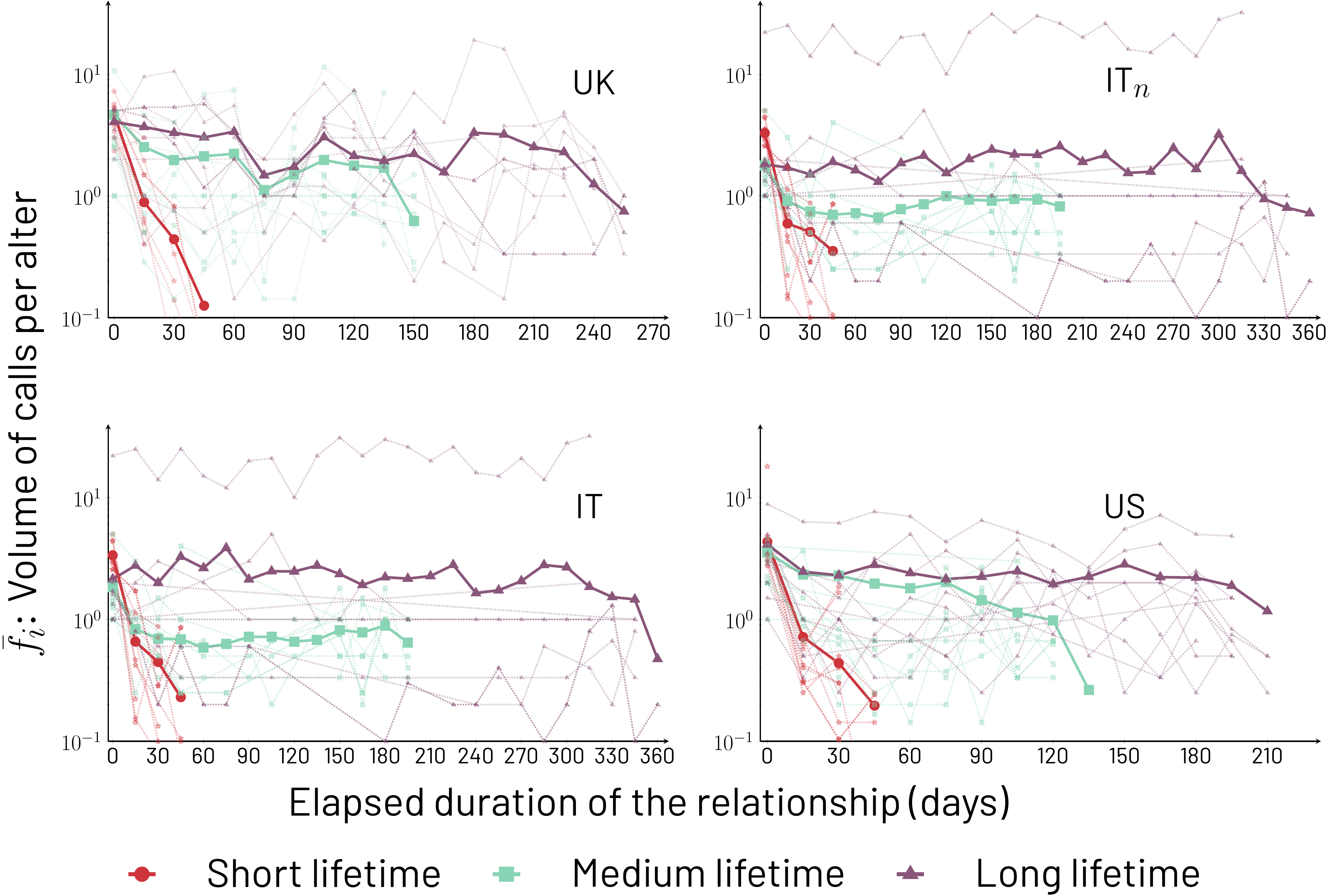}
    \caption{Visual comparison of a sample of $\bar{f_{i}}(a,\ell)$ curves (light color) with their respective cohort average $\bar{f}(a,\ell)$ (dark color). Cohorts are indicated in each plot. As expected, individual egos exhibit larger fluctuations than their cohort average, yet the fluctuations are generally centered around the averages, providing evidence that the general behavior of individual egos is qualitatively similar to that of the cohort average with respect to monotonicity with respect to $\ell$, and steadiness with respect to $a$. More quantitative evidence for these features being present in the $\bar{f}_i$ is provided by the Kolmogorov-Smirnov test shown in the main text, Fig.~3, as well as below in Fig.~\ref{fig:biavg}}
    \label{fig:fiandf}
  \end{figure}

A simple and illuminating check for the consistency between $\bar{f}_i$ for individual egos and the aggregate result $\bar{f}$ is to plot the series together. Fig.~\ref{fig:fiandf} shows the $\bar{f}$ reported in the main text (dark curves), Fig.~1, as well as $\bar{f}_{i}$ for a random sample of 10 egos in each cohort (light-colored curves). While the results from individual egos are noisier, as expected, the steadiness and monotonicity features are still present at the level of individual egos. Thus, $\bar{f}_i(a,\ell)$ for different egos are generally steady through a large range of values of $a$, and generally increase with $\ell$.

\subsubsection{Distribution of $b_{i}$}
\begin{figure}
    \centering
    \includegraphics[width=0.9\linewidth]{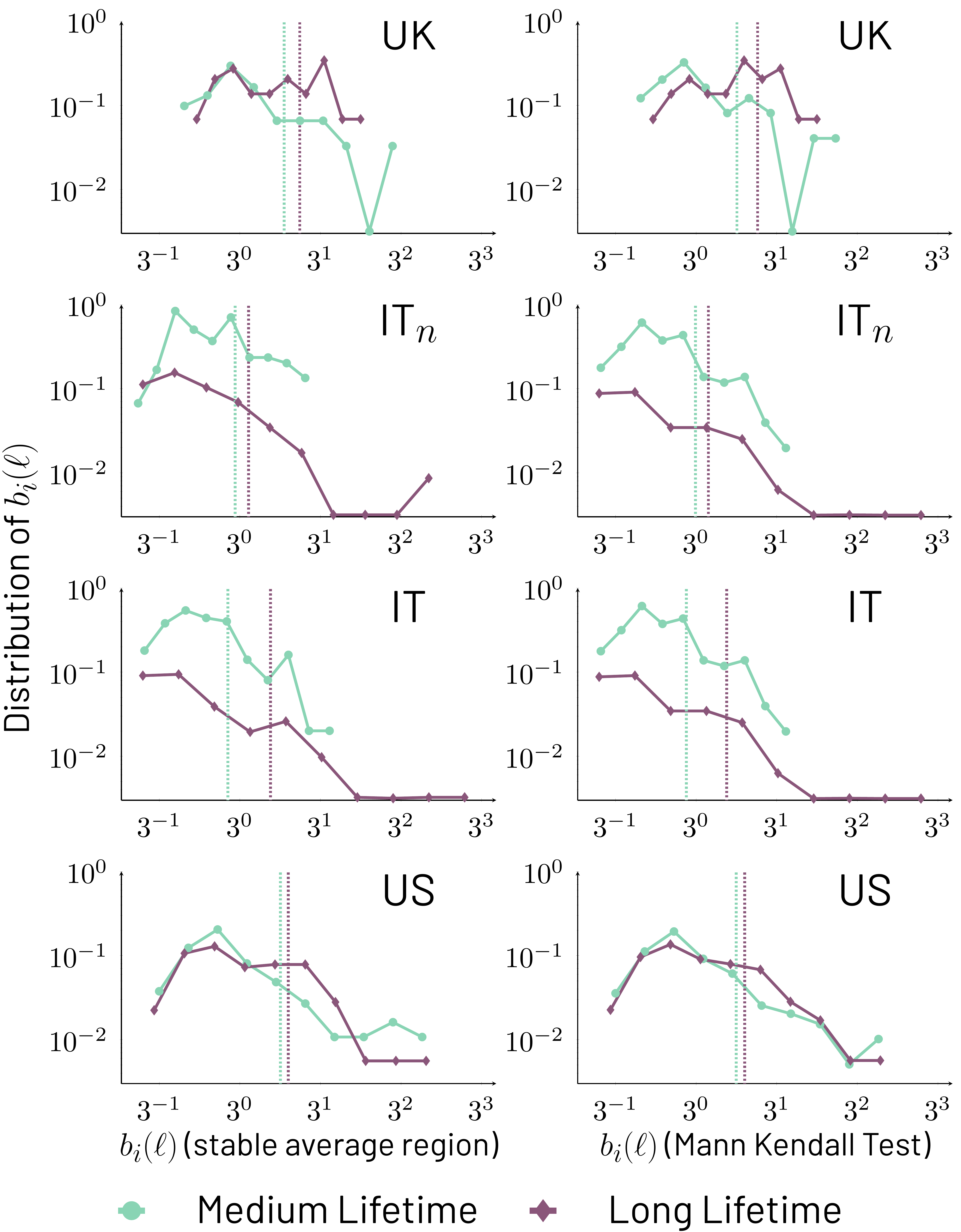}
    \caption{Distribution of $b_{i}(\ell)$ per cohort (indicated in each plot), with both axes in logarithmic scale. These distributions show that the average stable call volume of egos to their alters does not vary much given a lifetime, supporting the point that $b(\ell)$ and $b_i(\ell)$ are similar given $\ell$. In order to show all information in logarithmic scale, all values with frequency 0 are shown with frequency $10^{-2.5}$. Also, since no ego has a value of $b_i(\ell) = 0$, no information was lost due to the scaling of the horizontal axes in any of the plots. All plots to the left use the stable average region method of estimation of $b_i(\ell)$, and the right column shows the estimation using the Mann-Kendall test for trends. The averages of $b_i(\ell)$ over the egos of a cohort are displayed for each cohort and lifetime group with a vertical dashed line.}
    \label{fig:bdistribution}
  \end{figure}

The methods discussed in Sec.~\ref{sec:bmethods} allow us to determine the stable regimes of communication of each $\bar{f}_i(a,\ell)$, along with their associated stable volumes of communication $b_i(\ell)$. Fig.~\ref{fig:bdistribution} shows the probability distributions of values of $b_i$ for medium and long lifetimes (as defined in Fig.~1 of the main text) of each of the cohorts; the left column of plots shows the results of using the stable region averages and the right column plots show the results from the Mann-Kendall method. The plots also represent the averages of each of the distributions through vertical dotted lines. The color scheme representing lifetime groups is consistent with that of the main text.

The general characteristics of the distributions are very similar over all the plots. First, they show a rapid decay as $b_i$ increases, signaling that in general the values of $b_i$ are distributed over narrow ranges. Second, the long lifetime groups display a slower decay than the medium lifetime groups consistently across all cohorts, in agreement with the monotonic behavior of communication volume with lifetime. Third, both the stable average and Mann-Kendall methods lead to very similar distributions of $b_i$ cohort by cohort, indicating that the results are robust.

\begin{figure}
    \centering
    \includegraphics[width=0.95\textwidth]{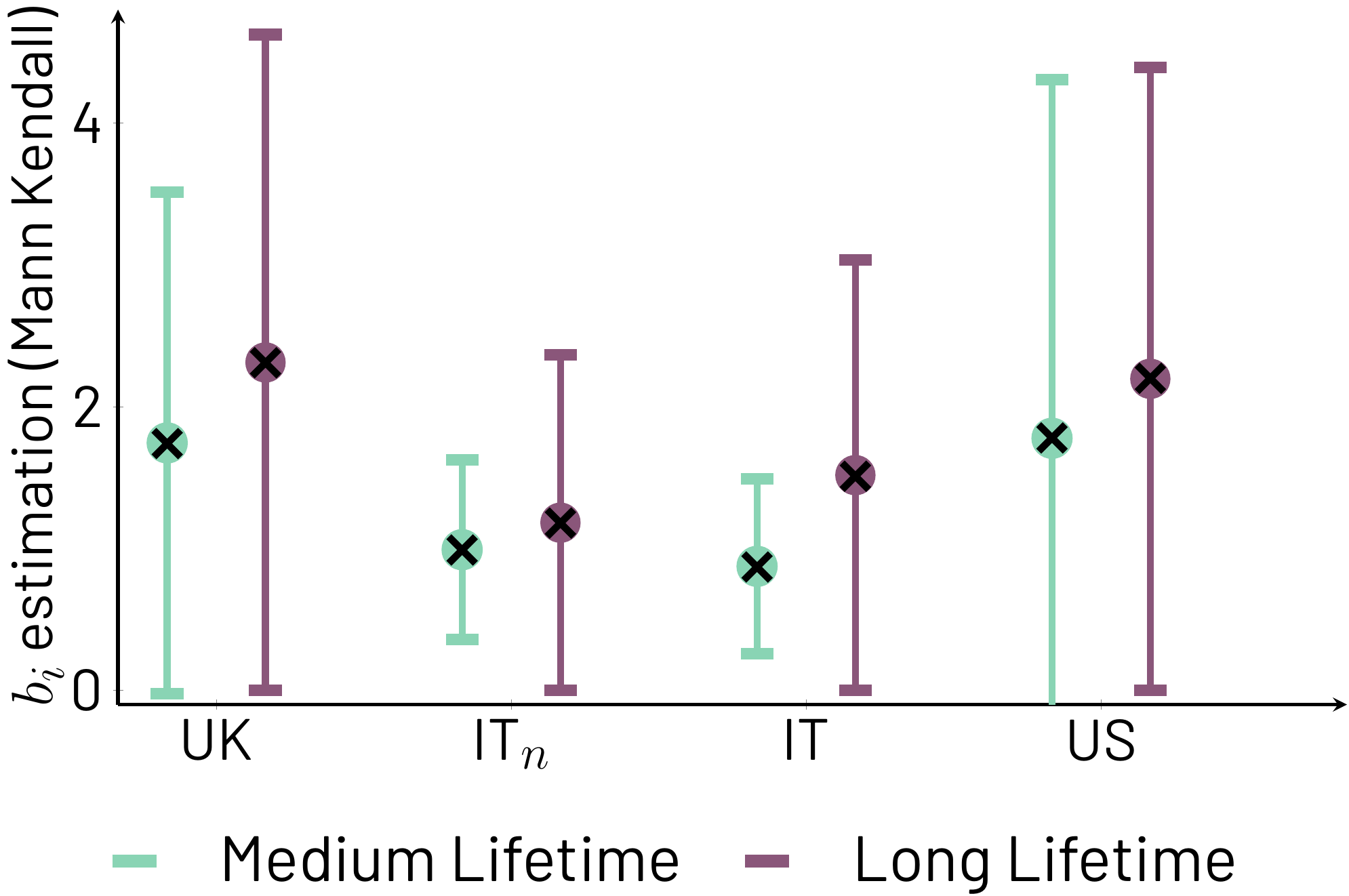}
    \caption{Averages (circles) and standard deviations (lines-whiskers) of $b_{i}(\ell)$ over egos, determined through the Mann-Kendall method, separated by cohorts and lifetime groups (medium and long). This figure is equivalent to Fig.~3B of the main text, generated with the alternative plateau finding method. The $\times$ symbol represents the value of $b(\ell)$ per cohort and lifetime group. The results support both the point that volume of calls increases with $\ell$, and that $b(\ell)$ is a reasonable approximation for the behavior of individual ego patterns of communication with their alters of a given lifetime group.}
    \label{fig:biavg}
  \end{figure}

For completeness, we also present a version of Fig.~3B of the main text using the Mann-Kendall method (Fig.~\ref{fig:biavg}). The figure presented here and the one in the main text have the same qualitative features, including good agreement between the average value of $b_i(\ell)$ and the corresponding $b(\ell)$ for cohort and lifetime $\ell$.

\subsubsection{Distribution of $p$-values from the Kolmogorov-Smirnov test}\label{sec:KS-test}
As explained in the main text, we study the level of steadiness of $\bar{f}_i(a,\ell)$ as a function of $a$ ego by ego, taking for each time series $\bar{f}_i(a,\ell)$ two parts of equal duration in $a$ around the mid-point of the time series that exclude the first ($a=0$) and last ($a=\lfloor\ell/\Delta a \rfloor\Delta a$) points. The two ranges of elapsed duration ($\Delta a\leq a< \lfloor(1/2)\left(\lfloor\ell/\Delta a\rfloor -1\right)\rfloor\Delta a$ and $\lfloor(1/2)\left(\lfloor\ell/\Delta a\rfloor -1\right)\rfloor\Delta a\leq a\leq\lfloor\ell/\Delta a \rfloor\Delta a-\Delta a$) generate for each ego two samples of $\bar{f}_i(a,\ell)$ at points in $a$ within each of the periods, and we perform a Kolmogorov-Smirnov test to determine if the values of the two samples come from the same distribution. The result of the Kolmogorov-Smirnov test for each ego is a $p$-value that, the closer it is to $1$, the more likely it is that the series $\bar{f}_i(a,\ell)$ is steady. Let us label the $p$-value obtained for each ego as $p_i$. We conduct these tests for egos with medium and long lifetimes.

\begin{figure}
    \centering
    \includegraphics[width=0.72\textwidth]{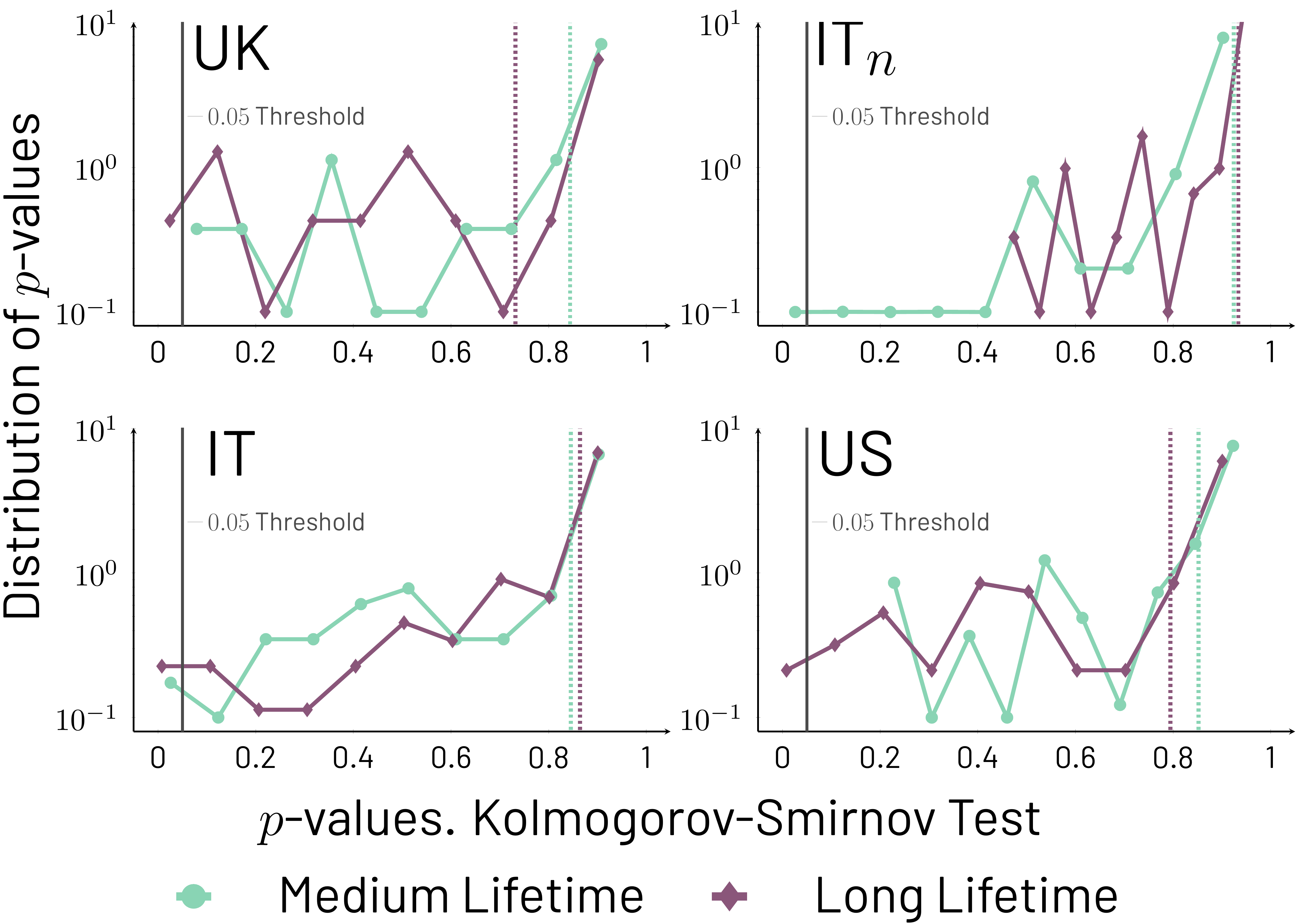}
    \caption{Distribution of $p$-values obtained from applying the Kolomogorov-Smirnov test to each individual curve $\bar{f_i}$. Each $p$-value is generated by dividing $\bar{f_i}$ into two equal-sized ranges of $a$ and comparing the values of $\bar{f_i}(a,\ell)$ between the two halves. Medium and long lifetime groups are chosen per cohort and are consistent with those chosen in the main text (cohorts are indicated in the plots). The distributions shown use $10$ equal-sized bins for the $p$-values. Due to the logarithmic scale in the vertical axis of each plot, in order to show all information, values with frequency 0 are shown with frequency $10^{-1}$. In all plots, purple and teal dashed lines represent the average of the distribution for alters with medium and long lifetimes, respectively, calculated from the raw data that generates the distributions (values reported in Sec.~\ref{sec:KS-test}). The black vertical line represents $p = 0.05$. The significance of the plots is that they show that the time series $\bar{f}_i(a,\ell)$ for the different cohorts and medium and long lifetimes of Fig.~1 do not have decaying or increasing trends.}
    \label{fig:pksdistribution}
  \end{figure}

In the main text, we show box plots of the $\{p_i\}_{i\in\mathcal{E}}$ obtained from the tests (Fig~3A) for all cohorts $\mathcal{E}$. Here, we present the probability distributions of these $\{p_i\}_{i\in\mathcal{E}}$ (Fig.~\ref{fig:pksdistribution}) over the cohorts and, in addition, the average values of the distributions (vertical dashed lines). The specific averages for each cohort and, respectively, medium and long lifetimes are: for UK $0.84$ and $0.73$, for IT${}_n$ $0.92$ and $0.93$, for IT $0.84$ and $0.86$, and for US $0.85$ and $0.79$. The color scheme is consistent with the main text (Figs.~1 and~3) regarding lifetimes. For reference, we also show with a black dashed line the $0.05$ statistical significance level.

As is clearly visible from these results, the distributions concentrate toward $p_i\to 1$ for all cohorts and their averages tend to the same limit, i.e. $1$. In the binning used here, a few outlying egos fall under the threshold $0.05$. Specifically, for alters with medium lifetimes, the proportion of egos under this threshold of 0.05 are: UK 0\%, IT${}_n$ 1\%, IT 1\%, and US 0\%. For alters with long lifetimes, the proportions per cohort that fall under the threshold are: UK 4\%, IT${}_n$ 0\%, IT 2\%, and US 1\%. Thus, the large majority of alters with medium and long lifetimes exhibit steadiness in their communication patterns.

\subsubsection{Analysis of $\ell_{s}$}\label{sec:ells}
The functions $\bar{f}_i(a,\ell)$ do not always stabilize to a flat region. This almost always occurs because $\ell$ is too small, i.e. when lifetimes are \textit{short} as described in the main text (small fractions of $\bar{f}_i(a,\ell)$ do fail the Kolmogorov-Smirnov test for medium or long lifetimes too, but at rates between $0$\% and $4$\% over the different cohorts, thus a negligible effect). As explained in Sec.~\ref{sec:bmethods}, the stable region and Mann-Kendall methods may \textit{fail to converge}, which means they never find a region in which the average slope of $\bar{f}_i(a,\ell)$ is close to $0$. On the other hand, when $\ell$ starts to become large, if a stable region is found, we track the values of the smallest $a$ at which such stable regions begin for each ego. In the methods described in Sec.~\ref{sec:bmethods}, these values are labelled $a_m$. At the threshold between $\ell$ being too small, not showing a steady regime, and starting to show stability, $a_m$ and $\ell$ are very similar as $a_M$ is not too far above $a_m$. Therefore, as a conservative approximation, we equate $a_m$ to the smallest lifetimes at which $\bar{f}_i(a,\ell)$ for a given $\ell$ can become stable.

\begin{figure}
    \centering
    \includegraphics[width=0.95\textwidth]{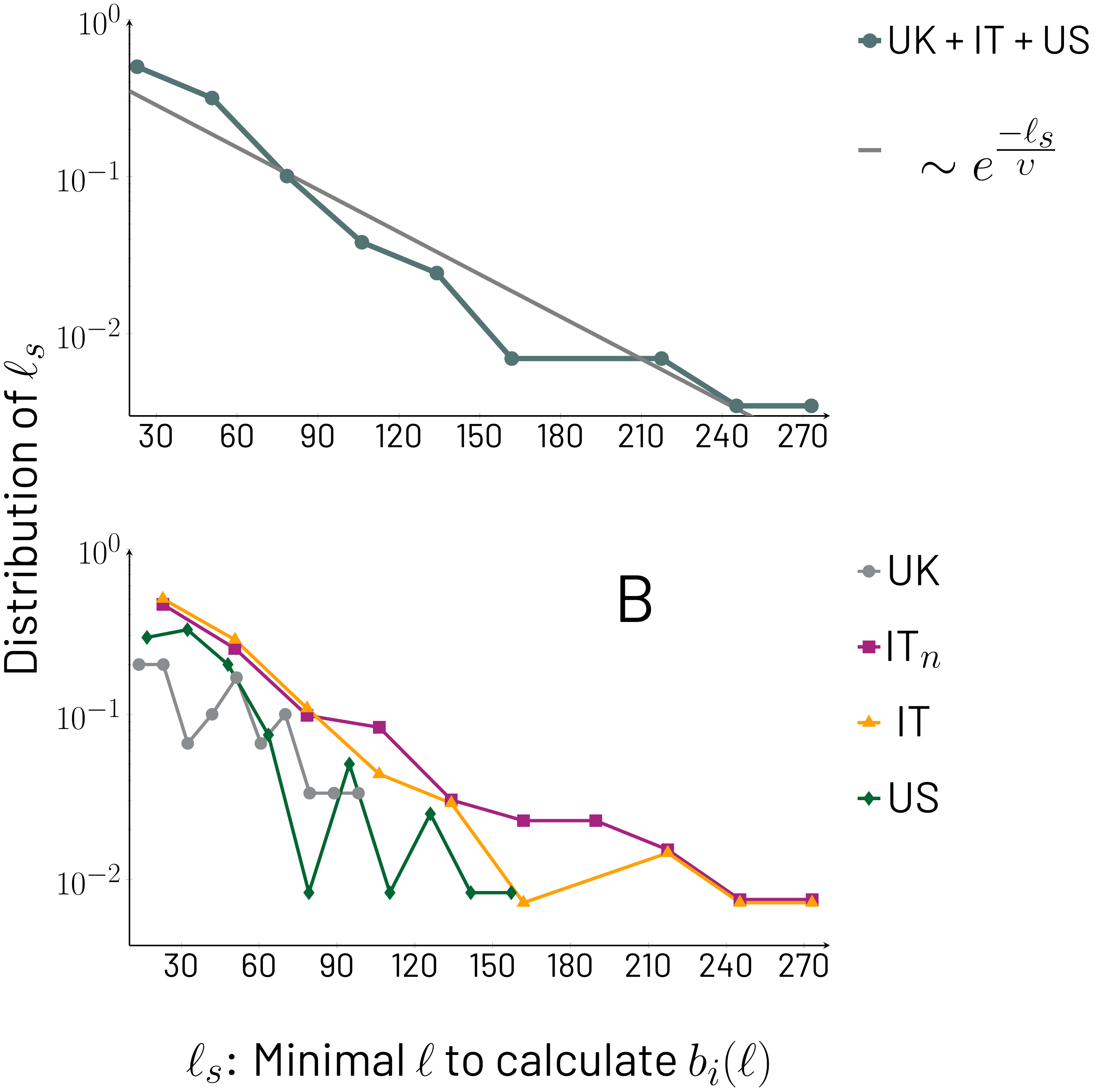}
    \caption{Distribution of $\ell_{s}$ plotted with logarithmic vertical scale. Panel A corresponds to the distribution of $\ell_s$ of all combined ego-alter pairs over all cohorts. The distribution is displayed with symbols, and the line crossing the distribution is the least-squares fit line for Eq.~\ref{eq:Pells-log}. Estimations of the average $\ell_s$ under the assumption that Eq.~\ref{eq:Pells-log} is a good approximation as well as the average calculated directly from the data are given within the text of this Supplementary Information and in Tab.~\ref{tab:ells} (which also contains per cohort averages from panel B of this figure). Panel B shows the distributions of $\ell_s$ separated by cohort. For the individual cohorts as well as the combined cohort, $\ell_s$ is obtained for all relationships for which the calculation of $b_i(\ell)$ converges.}
    \label{fig:ells}
  \end{figure}

Most $\bar{f}_i(a,\ell)$, as $\ell$ is increased, eventually exhibit a stable region starting at some $a_m(q_x)=\lfloor q_x/2\rfloor\Delta a$. We collect all such values over egos of a cohort and label them $\ell_s$, the minimum lifetime for stable communication. In Fig.~\ref{fig:ells} we present distributions of $\ell_s$ for each of the cohorts, the vertical scale is logarithmic and the horizontal scale is linear. The shape of these plots resembles exponential distributions which suggest a narrow set of possible values for $\ell_s$.

To provide estimates for the values of $\ell_s$ at which, generally, $\bar{f}_i(a,\ell)$ becomes steady, we take two approaches. First, we directly calculate the averages of $\ell_s$ of each of the cohort distributions. These averages can be found in Table~\ref{tab:ells}. We also create a single combined cohort that produces an average $\ell_s$ of $55.94$. A second approach is to assume that the distributions indeed are well approximated by the exponential form $\sim e^{-\ell_s/\upsilon}$, and estimate $\upsilon$. In turn, $\upsilon$ can be used to provide estimates for $\ell_s$. This second approach is well supported by the similarity of the distributions for the different cohorts (Fig.~\ref{fig:ells}B), which suggests that $\ell_s$ has similar quantitative properties across cohorts. This is a surprising result given the diversity of the egos.

To perform this second approach based on curve fitting we assume ${\rm Pr}(\ell_s)=Ce^{-\ell_s/\upsilon}$ where $C$ is the normalization constant. The range of values of $\ell_s$ can be limited on the left if desired, with a value we call $\ell_{s,\min}$. The average of the distribution requires we determine $\upsilon$. For this purpose, we use the information in Fig.~\ref{fig:ells}A and perform a least-squares regression of the points using a logarithmic transformation of the vertical axis first but leaving the horizontal scale linear. This gives the equation
\begin{equation}\label{eq:Pells-log}
    \log {\rm Pr}(\ell_s)=-\frac{1}{\upsilon}\ell_s +\log C
\end{equation}
and the slope $1/\upsilon$ is obtained from the least-squares regression. This provides the value $\upsilon\approx 48.7$. Finally, the average of an exponential distribution satisfying ${\rm Pr}(\ell_s)=Ce^{-\ell_s/\upsilon}$ with $\ell\geq \ell_s$ is given by $\upsilon +\ell_{s,\min}$. For the combination of cohorts, the smallest $\ell_s$ is $\ell_{s\min} = 14$, providing a final estimate of $62.7$ for average $\ell_s$, similar in value to the estimate above based on directly computing the average of ${\rm Pr}(\ell_s)$, namely, $55.94$.

\subsection{Information about lifetimes of transient relationships obtained from early communication volume}
Figs.~4 and~5 in the main text show that the survival probabilities of alters increases as a function of activity early in the relationship. In this section we provide further evidence of this, by performing robustness checks on those figures. First, we provide justification for the choice of binning used in Figs.~4 and~5. Following that, we show that the behavior seen in Fig.~4 of the main text, which combines the UK, IT, and US cohorts, is also visible in each one of the cohorts separately, including IT${}_n$. To conclude the section, we explore the effect of changing the choices of the window of time in elapsed duration used to assess the relation between early call activity and ultimate lifetime of an ego-alter pair.

\subsubsection{Overall ego-alter number of calls and exponential binning for $\gamma$}
For Figs.~4 and~5 of the main text we define $g$, the communication volume between an ego and alter pair in a specific range of $a$, selected to be at the initial stages of an observed ego-alter relationship. When we use this parameter to provide estimates for the lifetime $\ell$ of a relationship, we also group the values of $g$ into exponentially sized bins, labelled by the variable $\gamma$. Here we provide the background evidence that such binning choice is necessary given the distribution of numbers of calls seen across ego-alter pairs in our datasets.

\begin{figure}
    \centering
    \includegraphics[width=0.85\textwidth]{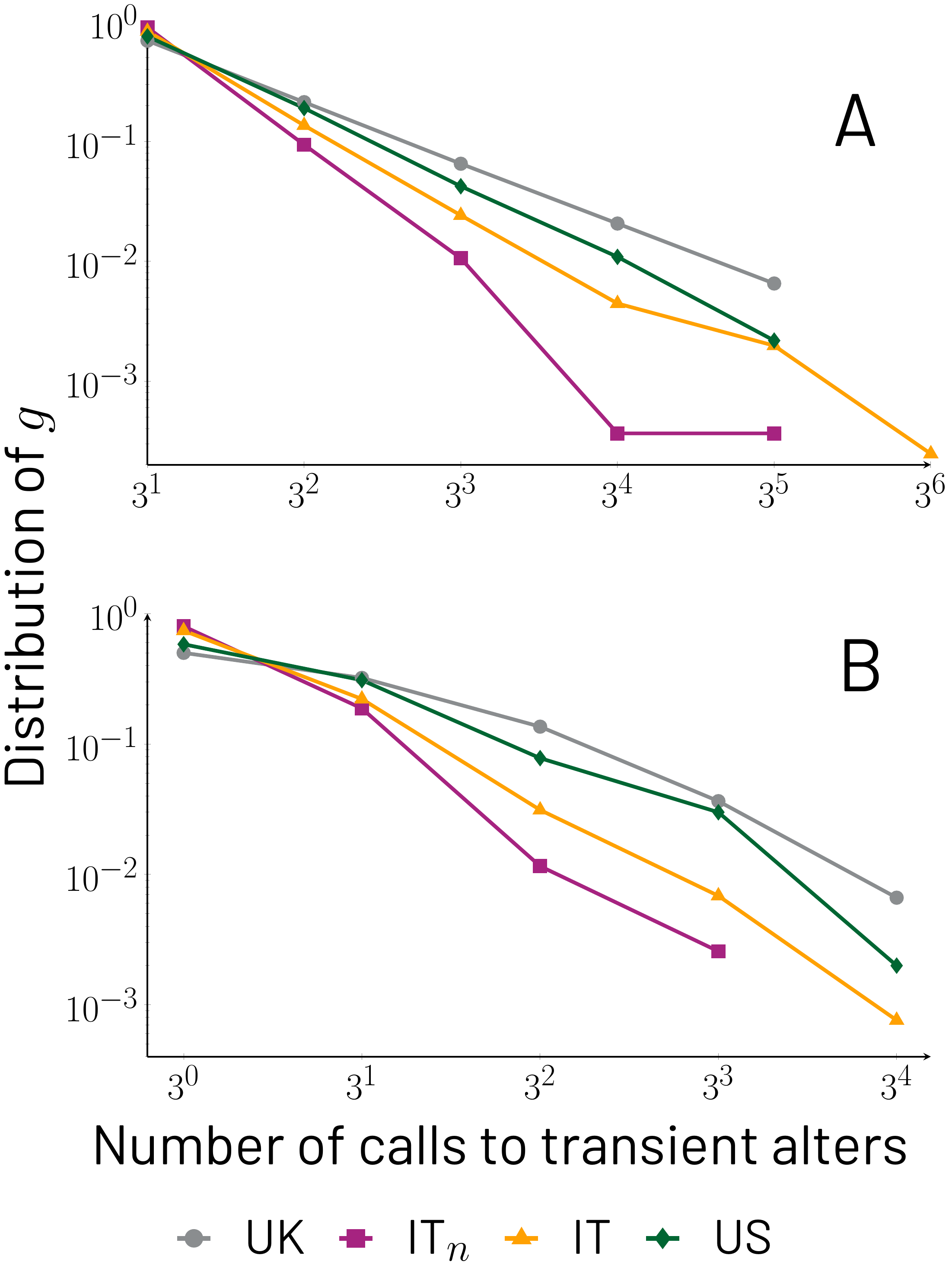}
    \caption{Distribution of $g$ for transient alters, measured between a starting and ending elapsed duration of relationships $a_o$ and $a_f$. Panel A uses $a_o=0$ and $a_f=\mathcal{L}_{\mathcal{E}}$ which means all calls for each ego-alter transient relationship with the $\mathcal{L}_{\mathcal{E}}$ is taken into account; Panel B uses $a_o=30$ and $a_f=60$, the values used in Figs.~3 and~4 of the main text.}
    \label{fig:gdistribution}
  \end{figure}

Fig~\ref{fig:gdistribution} shows the probability distribution of the number of calls between each ego-alter pair for the lifetime of the relationship, separated by cohorts (panel A). Panel B shows the distribution of the number of calls fixed to the period of time of relationships with $30\leq a\leq 60$. As is clearly visible from both panels, the exponential sizes of the bins are well suited to plot the slowly decaying probability distributions, justifying their use in Figs.~4 and~5 of the main text. We find that $3$ is a good base for the exponential bins, capturing clearly the distinction between different levels of call activity captured by $g$.

\subsubsection{Probability of relationship survival as a function of $\gamma$ per country}
\begin{figure}
    \centering
    \includegraphics[width=0.95\textwidth]{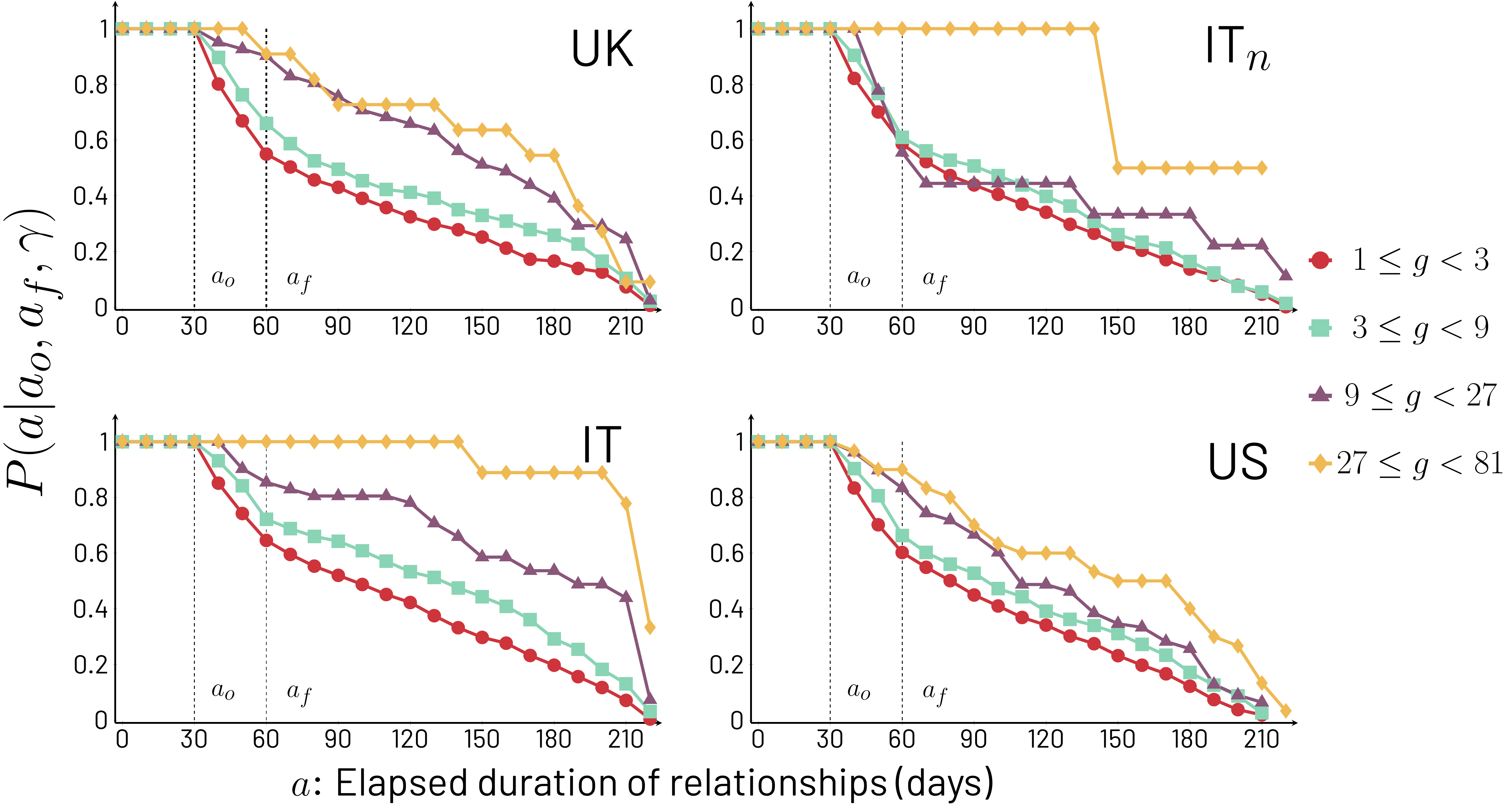}
    \caption{Survival probabilities $P(a\mid a_o,a_f,\gamma)$ of transient alters to duration of at least $a$ for different bins $\gamma$ of amount of mobile phone calls between $a_o=30$ and $a_f=60$ days. Each plot corresponds to a single cohort (indicated in each plot), in contrast to the main text which includes UK, IT, and US. The smaller samples that make up each cohort do lead to noisier results as well as step-wise jumps on the plots. The bins represented by $\gamma$ as the exponent in $3^{\gamma}\leq g<3^{\gamma+1}$ are $\gamma=0,1,2,3$. As $\gamma$ increases, and even though the plots display noisier behavior than in Fig.~4 of the main text, the probability of survival also increases in all cohorts, i.e. for $\gamma'>\gamma$, $P(a\mid a_o,a_f,\gamma') > P(a\mid a_o,a_f,\gamma)$, consistent with the conclusions of the main text.}
    \label{fig:survivalbycountry}
  \end{figure}

To provide a robustness check for the relationship survival probabilities $P(a\mid a_o,a_f,\gamma)$ presented in the main text, we show the same probability here for each separate cohort (Fig.~\ref{fig:survivalbycountry}). The results display the same consistent behavior, i.e. that $P(a\mid a_o,a_f,\gamma)$ increases with $\gamma$ for fixed value of $a$, indicating that more relationships survive longer the greater the amount of early activity seen.

\subsubsection{Using different values of $a_{o}$ and $a_{f}$}
The definition of $g$, as indicated above, admits an arbitrarily chosen range of $a$. Our choice of $a_o=30$ to $a_f=60$ in the main text is driven by the fact that this period is long enough ($30$ days) to capture sufficient activity to provide information about a relationship, and finally is not too long so it remains a \textit{limited} measurement of communication activity rather than the measurement of a large proportion of the activity that, in some sense, one wants to predict.

\begin{figure}[t]
    \centering
    \includegraphics[width=0.857\textwidth]{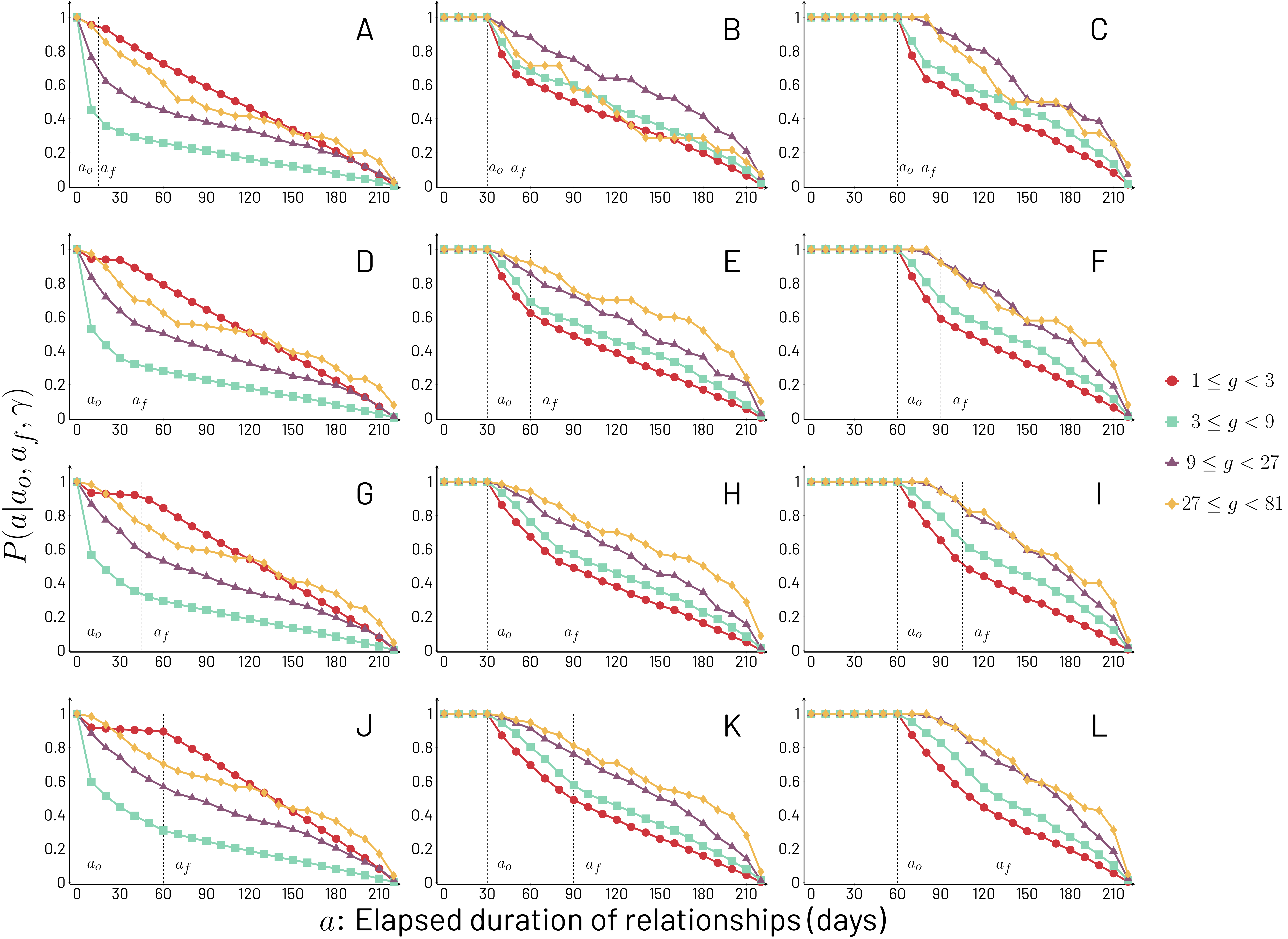}
    \caption{Exploration of survival probabilities $P(a\mid a_o,a_f,\gamma)$ of transient alters to duration of at least $a$ for different bins $\gamma$ of amount of mobile phone calls between $a_o$ and $a_f$ days. To explore the effects of $a_o$ and $a_f$, we proceed in a systematic way: each row of plots fixes the value of $a_f-a_o$, while increasing $a_o$; while each column fixes $a_o$ while increasing $a_f-a_o$. The values of $a_f-a_o$ in order of rows, from top to bottom, are $15$, $30$, $45$, and $60$ days; the values of $a_o$ in order of columns, from left to right, are $0$, $30$, and $60$ days. We use the combined data for UK, Italy and US, and therefore, we only look at relationships active for $\ell<\mathcal{L}_{{\rm US}}=220$ days or less, in order to include data for all three cohorts. The bins represented by $\gamma$ as the exponent in $3^{\gamma}\leq g<3^{\gamma+1}$ are $\gamma=0,1,2,3$. The most important effects observed are that if $a_o$ is chosen early in the relationship (say $a_o=0$), survival curves are closer together and even show inconsistency for the smallest call bin $\gamma=0$. Also, curves with increasing $\gamma$ do not separate as broadly. Larger $a_f-a_o$, on the other hand, leads to greater separation between curves of increasing $\gamma$, although increasing the window of observation is somehow antithetical to the idea of using $g$ measured in a small time window to predict relationship lifetime. In any case, for reasonable $a_o$ (one that is not too small) we still find, as in the main text, that as $\gamma$ increases the probability of survival also increases, i.e. for $\gamma'>\gamma$, $P(a\mid a_o,a_f,\gamma') > P(a\mid a_o,a_f,\gamma)$.}
    \label{fig:fig3avariations}
  \end{figure}

Fig.~\ref{fig:fig3avariations} shows an extensive check of the effect of varying $a_o$ and $a_f$ on $P(a\mid a_o,a_f,\gamma)$, for the combined cohorts UK, IT, and US; organized in matrix form to study the change systematically. Along the horizontal direction, $a_f-a_o$ is kept fixed while $a_o$ increases, thus exploring the effect of using progressively later windows of observation. Along the vertical direction, $a_f-a_o$ increases while $a_o$ stays fixed.

The first general observation is the remarkable robustness of the behavior of $P(a\mid a_o,a_f,\gamma)$ with respect to $a$ for the variety of choices of $a_o,a_f$ we test. A few additional features emerge worth mentioning. First, if $a_o$ is chosen early in the relationship (say $a_o=0$), survival curves are closer together and even show inconsistency for the smallest call bin $\gamma=0$. This is explained by the fact that very early in relationships, communication patterns have not totally settled and therefore longer and shorter lifetime relationships are still somewhat indistinguishable. Second, as $a_f-a_o$ increases, the survival curves separate from each other and call volume $g$ has a greater predictive power. However, a very early choice of $a_o$ dominates over a larger window $a_f-a_o$, indicating the unreliability of trying to predict relationship duration at an extremely early point of the relationship. As can be seen when the choice of $a_o$ becomes $30$, reliability returns, even for shorter windows of observation, even down to $15$ days. Third, the curves for $\gamma < 2$ appear to be smoother than those for higher values of $\gamma$. This occurs due to the number of alters in these bins, as shown in Table~\ref{tab:altersbygamma}. Finally, when $a_o$ and $a_f-a_o$ both increase, the curves $P(a\mid a_o,a_f,\gamma)$ separate very widely as functions of $\gamma$, signalling the much greater ability of $g$ to predict lifetime. The caveat to this is that, since the purpose of using early call volume within a limited time window is mostly to provide some early estimates of $\ell$, it is not practical to increase both $a_o$ and $a_f-a_o$ because in that case the measurement of $g$ in fact amounts to a full measurement of relationship call volume.

\subsection{Consistency between countries}

\subsubsection{Variations on contour plots}
We also test the robustness of the results of Fig.~5 to cohort selection. In that figure of the main text, we combined data from the UK and US to explore how well they predict the Italian cohort. Here, we change our cohort selection in order to test how robust these results are.

\begin{figure}[h!]
    \centering
    \includegraphics[width=0.9\textwidth]{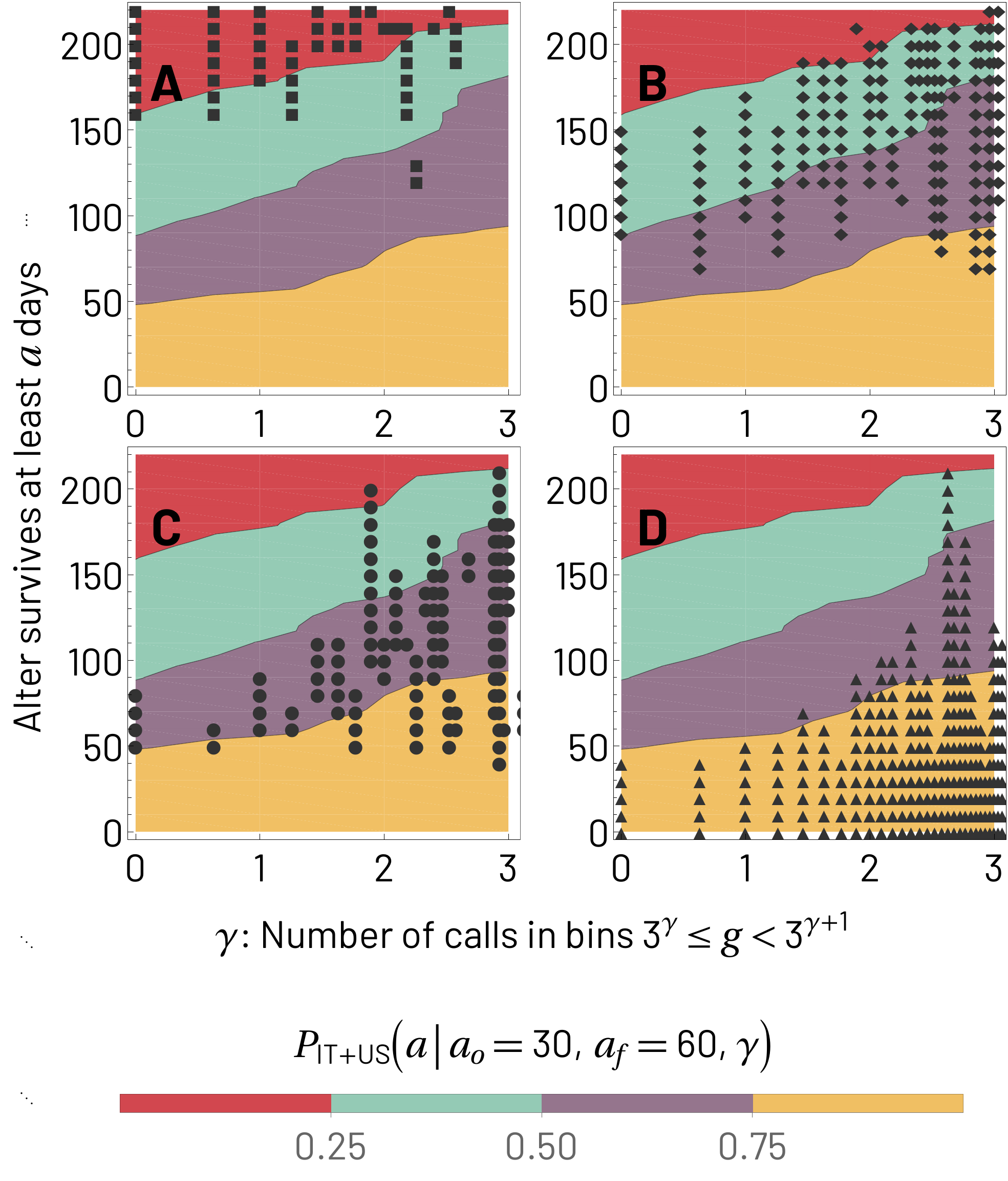}
    \caption{Version of Fig.~5 from the main paper with contours created using US and Italian cohorts and the points correspond to the UK cohort. In the spirit of the main text, the color background represents ranges of $P_{{\rm IT + US}}(a\mid a_o,a_f,\gamma)$, namely $[0,0.25)$ (red), $[0.25,0.5)$ (teal), $[0.5,0.75)$ (purple), and $[0.75,1]$ (yellow). Panel A shows the symbol $\blacksquare$ for $P_{{\rm UK}}(a\mid a_o,a_f,\gamma)$ in the interval $[0,0.25)$, panel B shows the symbol $\diamond$ for the interval $[0.25,0.5)$, panel C uses the symbol $\bullet$ for the interval $[0.5,0.75)$, and panel D uses the symbol $\blacktriangle$ for the interval $[0.75,1)$ The match in location between the symbols and the colored regions is reasonable throughout values of the probability. However, quality is degraded slightly in comparison to the main text, Fig.~5, as the sample for the UK cohort is considerably smaller than the one for Italy (which forms the symbols of the main text). Overall, the qualitative trend of the results presented is still consistent with those of the main text.}
    \label{fig:contourSb}
  \end{figure}

\begin{figure}[h!]
    \centering
    \includegraphics[width=0.9\textwidth]{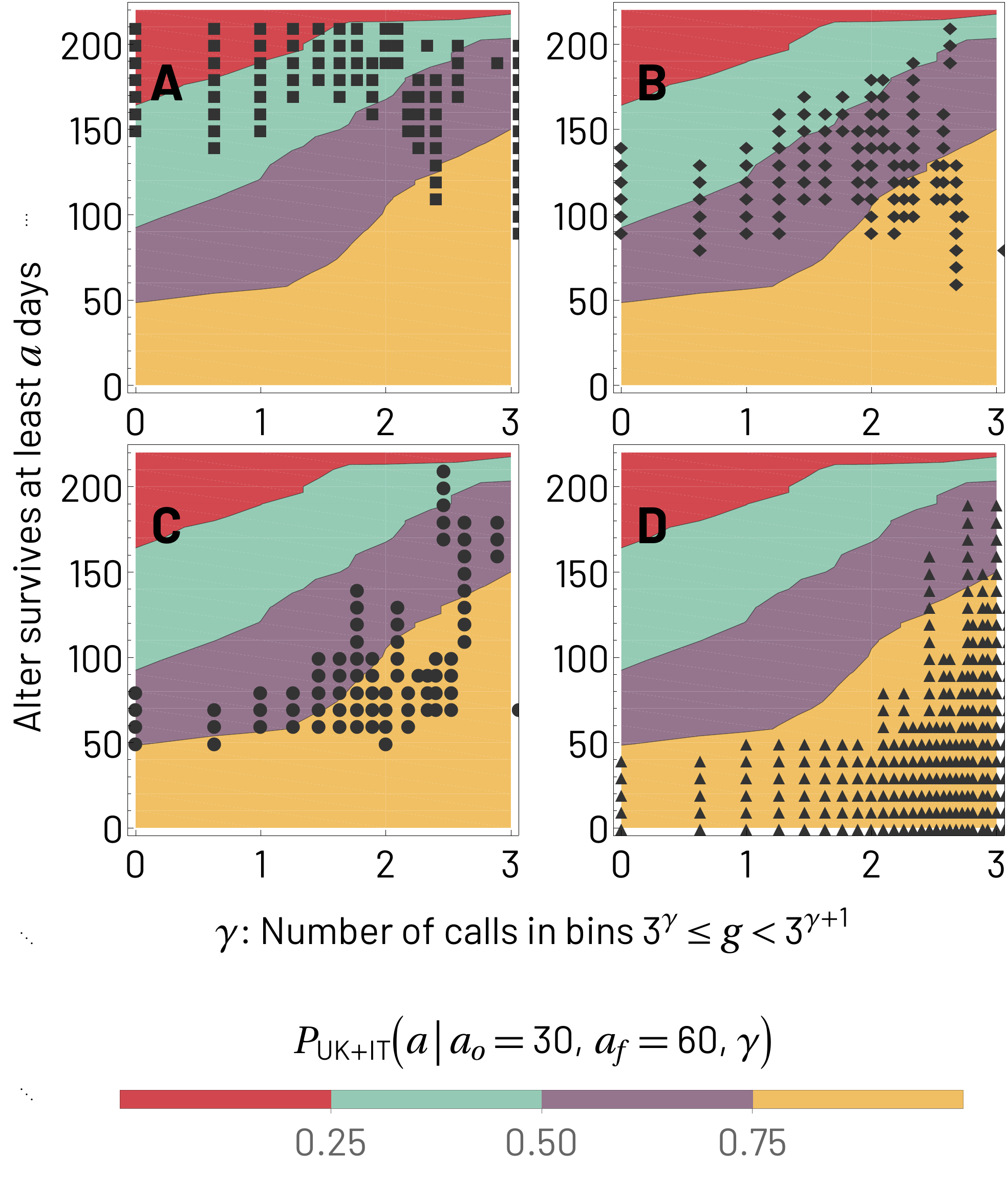}
    \caption{Version of Fig.~5 from the main paper with contours created using UK and Italian cohorts and the symbols correspond to the US cohort. In the spirit of the main text, the color background represents ranges of $P_{{\rm UK + IT}}(a\mid a_o,a_f,\gamma)$, namely $[0,0.25)$ (red), $[0.25,0.5)$ (teal), $[0.5,0.75)$ (purple), and $[0.75,1]$ (yellow). Panel A shows the symbol $\blacksquare$ for $P_{{\rm US}}(a\mid a_o,a_f,\gamma)$ in the interval $[0,0.25)$, panel B shows the symbol $\diamond$ for the interval $[0.25,0.5)$, panel C uses the symbol $\bullet$ for the interval $[0.5,0.75)$, and panel D uses the symbol $\blacktriangle$ for the interval $[0.75,1)$. The match in location between the symbols and the colored regions is best achieved for largest probabilities, i.e. dark yellow and purple regions. For lower probability regions, the match is not as good although it has the correct trend of dependence of survival with respect to $g$, namely, more calling means longer survival. The smaller size of the US cohort plays a role. Overall, the qualitative trend of the results presented is still consistent with those of the main text.}
    \label{fig:contourSa}
  \end{figure}

In Figs.~\ref{fig:contourSb} and~\ref{fig:contourSa} we show two combinations of two countries used to produce the contour plots shown, and the third country is contrasted against those contours. In all cases, as in the main text, the third country's behavior is reasonably predicted. The limited quality in comparison with that of the Fig.~5 in the main text is that the Italian cohort is, by far, the best sampled one leading to a cleaner match of symbols and contours in Fig.~5. Both the UK and US cohorts have limited statistics and therefore produce somewhat coarser results. Nevertheless, their qualitative trends are consistent with those seen in the main text, supporting our conclusions. While other details about the cohorts may play a role, that exploration beyond the scope and goals of the current project.
\end{spacing}

\newpage

\clearpage

\clearpage

\clearpage

  \clearpage

  \clearpage

  \clearpage

  \clearpage

\clearpage

\clearpage

\clearpage

\clearpage

\clearpage

\clearpage

\clearpage

\clearpage

\clearpage

\begin{table}
      \centering
      \begin{tabular}{|lrcc|}
      \hline\hline
      Cohort                        & average $\ell_s$ & &\\
      \hline

      UK                            & 51.13 & & \\
      IT${}_n$                      & 66.14 & &\\
      IT                            & 56.49 & &\\
      US                            & 56.51 & &\\
      UK, Italy, and US combined    & 55.94 & &\\
       & & &\\
      \hline
      Exponential fit          & $\ell_{s, \min{}}$ & $\upsilon$ & Estimated average $\ell_{s}$\\
      \hline
      UK, Italy, and US combined   &  14                   &  48.70        & 62.38 \\
      \hline\hline
      \end{tabular}
      \caption{Calculation/estimation of average $\ell_s$ by cohort, all cohorts combined, and exponential fit for all cohorts combined (Eq.~\ref{eq:Pells-log}). When all cohorts are combined, results are obtained from an ordinary least square (OLS) estimation (Fig.~\ref{fig:ells}A). For the individual cohorts as well as the combined cohort, averages are obtained directly from the $\ell_s$ of all relationships for which the calculation of $b_i(\ell)$ converges. Their distributions are shown in Fig~\ref{fig:ells}B.
      }
      \label{tab:ells}
    \end{table}
\clearpage

\clearpage

\clearpage

\begin{table}[h!]
\centering
\begin{tabular}{|lrrrrrrrrrrrr|}
\hline\hline
$\gamma$ & \textbf{A} & \textbf{B} & \textbf{C} & \textbf{D} & \textbf{E} & \textbf{F} & \textbf{G} & \textbf{H} & \textbf{I} & \textbf{J} & \textbf{K} & \textbf{L}\\
\hline
0 & 3820 & 1186 & 964 & 3053 & 1624 & 1335 & 2531 & 1902 & 1588 & 2094 & 2032 &1716\\
1 & 3462 & 381 & 275 & 4058 & 638 & 500 & 4425 & 850 & 704 & 4764 & 1039 &857\\
2 & 299 & 107 & 52 & 431 & 151 & 94 & 546 & 183 & 138 & 625 & 228 &179\\
3 & 38 & 10 & 12 & 72 & 41 & 33 & 102 & 61 & 45 & 113 & 75 &55\\
\hline\hline
\end{tabular}
\caption{Number of alters used at each series $\gamma$ in Fig.~\ref{fig:fig3avariations}. Every column corresponds to a panel in Fig.~\ref{fig:fig3avariations}. There is a noticeable decrease in the number of alters, as $\gamma$ increases. Just as for Fig.~\ref{fig:fig3avariations}, a combination of the three countries was used for this Table}
\label{tab:altersbygamma}
\end{table}
\clearpage

\clearpage

\end{document}